UNIVERSIDADE FEDERAL DO PARÁ

INSTITUTE OF TECHNOLOGY

GRADUATE PROGRAM IN ELECTRICAL ENGINEERING

DEVELOPMENT AND MODELING OF A LOW-VOLTAGE DC DISTRIBUTION NANOGRID WITH DISTRIBUTED GENERATION SYSTEMS

PEDRO FERREIRA TORRES

DM 36 / 2019

UFPA / ITEC / PPGEE

Guamá University Campus

Belém-Pará-Brasil

2019

UNIVERSIDADE FEDERAL DO PARÁ

INSTITUTE OF TECHNOLOGY

GRADUATE PROGRAM IN ELECTRICAL ENGINEERING

PEDRO FERREIRA TORRES

DEVELOPMENT AND MODELING OF A LOW-VOLTAGE DC DISTRIBUTION NANOGRID WITH DISTRIBUTED GENERATION SYSTEMS

DM 36 / 2019

UFPA / ITEC / PPGEE

Guamá University Campus

Belém-Pará-Brasil

2019

UNIVERSIDADE FEDERAL DO PARÁ

INSTITUTE OF TECHNOLOGY

GRADUATE PROGRAM IN ELECTRICAL ENGINEERING

PEDRO FERREIRA TORRES

DEVELOPMENT AND MODELING OF A LOW-VOLTAGE DC DISTRIBUTION NANOGRID WITH DISTRIBUTED GENERATION SYSTEMS

> Dissertation submitted to the Examining Board of the Graduate Program in Electrical Engineering of Universidade Federal do Pará in fulfillment of the requirements for the degree of Master of Science in Electrical Engineering in the area of Electrical Energy Systems.

UFPA / ITEC / PPGEE

Guamá University Campus

Belém-Pará-Brasil

2019

UNIVERSIDADE FEDERAL DO PARÁ
INSTITUTE OF TECHNOLOGY
GRADUATE PROGRAM IN ELECTRICAL ENGINEERING

"DEVELOPMENT AND MODELING OF A LOW-VOLTAGE DC DISTRIBUTION NANOGRID WITH DISTRIBUTED GENERATION SYSTEMS"

AUTHOR: **PEDRO FERREIRA TORRES**

MASTER DISSERTATION SUBMITTED TO THE EXAMINING BOARD APPROVED BY THE COLLEGATE OF THE GRADUATE PROGRAM IN ELECTRICAL ENGINEERING, BEING SUITABLE FOR OBTAINING THE MASTER OF SCIENCE DEGREE IN ELECTRICAL ENGINEERING, IN THE AREA OF ELECTRICAL ENERGY SYSTEMS

APPROVED IN: 23/10/2019

EXAMINING BOARD:

______________________________________________________________
**Prof. Wilson Negrão Macêdo, Dr.**
(Supervisor – PPGEE/UFPA)

______________________________________________________________
**Prof. Marcos André Barros Galhardo, Dr.**
(Internal Program Evaluator – PPGEE/UFPA)

______________________________________________________________
**Prof. João Tavares Pinho, Dr. Ing-**
(External Evaluator – IEE/USP)

______________________________________________________________
**Prof. Luiz Antônio Corrêa Lopes, PhD**
(External Evaluator – PEER/Concordia University)

______________________________________________________________
**Prof. Samuel James Williamson, PhD**
(External Evaluator – EEMG/University of Bristol)

ATTESTED BY:

______________________________________________________________
**Prof.ª Maria Emília de Lima Tostes, Dr.**
(Program Coordinator PPGEE/ITEC/UFPA)

# ACKNOWLEDGMENT

I thank God for all the opportunities and for the beauty and harmony of life.

To my parents, Amaury and Paula, for the purest love, affection and support in this walk.

To my sisters, Ana and Heloisa, for understanding, companionship and good times of brotherhood.

To my paternal grandmother, Vanda, and maternal grandparents, Otacilio and Joaquina, for being sources of love and wisdom.

To Rafaella, for all the patience and good times shared in these years.

To the friends of GEDAE, present and past, who make the work environment a pleasant place and conducive to the development of various activities, and who fundamental in the development of this work.

To my advisor, Prof. Wilson Negrão Macêdo, for the patience of everyday life, shared wisdom and opportunities granted.

To CAPES for the scholarship fundamental for the full accomplishment of this work.

To the National Institute of Renewable Energy and Energy Efficiency Science and Technology of the Amazon, INCT-EREEA, for their financial support.

# SUMMARY

# LIST OF ILLUSTRATIONS

# LIST OF TABLES

# LIST OF ABBREVIATIONS AND ACRONYMS

| | |
|---|---|
| AC | Alternating current |
| ACDM | AC Distribution minigrid |
| ADC | Analog-digital converter |
| AM | Air mass |
| ANEEL | National Electrical Energy Agency |
| BB | Battery bank |
| BTU | British thermal unit |
| CCSA | Chinese Communication Standards Association |
| DC | Direct current |
| DG | Distributed Generation |
| DCDM | DC distribution minigrid |
| DCDN | DC distribution nanogrid |
| ETSI | European Telecom Standard Institute |
| GEDAE | Group of Studies and Development of Energy Alternatives |
| GSS | Generation and storage system |
| HVDC | High voltage DC |
| IEC | International Eletrotechnical Comission |
| IEEE | Institute of Electrical and Electronic Engineers |
| IEEE-SA | Institute of Electrical and Electronic Engineers Standard Association |
| ITU | International Telecommunication Union |
| LB | Load bank |
| LED | Light emitting diode |
| LVD | Low voltage directive |
| LVDC | Low voltage DC |
| MIGDI | Isolated microsystems of energy generation and distribution |
| MOSFET | Metal-oxide silicon field effect transistor |
| MPPT | Maximum power point tracking |
| NBR | Brazilian Standards Association |
| NEC | National Electrical Code |
| NOCT | Nominal operating cell temperature |
| NR | Newton-Raphson algorithm |
| P&O | Perturb and observe algorithm |
| PID | Proportional, integral and derivative |
| PLC | Power line communication |
| PV | Photovoltaic |
| PVG | Photovoltaic generator |
| PWM | Pulse Width Modulation |
| QEE | Quality of electrical energy |
| QGD | Main switchboard for distribution, measurement and command |
| RC | Reserve capacity |
| SELV | Separated extra-low voltage |
| SLH | Sunlight hours |
| SoC | State of charge |
| SPD | Surge protection device |
| SPI | Serial peripheral interface |
| STC | Standard test conditions |
| UFPA | Universidade Federal do Pará |
| UV | Ultraviolet |
| VB | Voltage balancer |
| Vcc | DC volts |

# LIST OF SYMBOLS

| | |
|---|---|
| $A_{cap}$, $B_{cap}$, $C_{t,coef}$ | Battery model constants for instantaneous capacity in Ampere-hours. |
| $a_{cmt}$, $b_{cmt}$ | Model constants for battery charge efficiency calculation time constant |
| $A_{fonsc}$, $B_{fonsc}$ | Battery Model Constants Regarding Saturation Voltage |
| $A_{gas}$, $B_{gas}$ | Battery Model Constants for Gasification Voltage |
| $A_n$ | Diode ideality factor |
| $A_{\tau sc}$, $B_{\tau sc}$, $C_{\tau sc}$ | Battery model constants related to voltage time constant in the overcharge region |
| $C$ | Battery capacity in Ampere-Hour |
| $C^*_B$ | Rated battery bank capacity given in watt-hours |
| $C_{10}$ | Battery bank Capacity at 10 h Discharge rate, given in Ampere-hour |
| $C_n$ | Battery maximum capacity in Ampere-Hour |
| $C_{nominal}$ | Nominal battery capacity in Ampere-hours |
| $D$ | Duty cycle |
| $D_{MP}$ | Deviation between model and maximum power measurement |
| $D_{OC}$ | Deviation between model and open circuit voltage measurement |
| $D_{SC}$ | Deviation between model and short circuit current measurement |
| $E_{BB}$ | Battery Bank Net Daily Energy measured |
| $E_{BB,SIM}$ | Battery Bank Net Daily Energy simulated |
| $E_{BC}$ | Daily measured energy consumed by the load bank |
| $E_{BC,SIM}$ | Daily simulated energy consumed by the load bank |
| $E_{c.a.,e}$ | AC power consumption of electronic equipment |
| $E_{c.a.,i}$ | Inverter-type air conditioner AC power consumption |
| $E_{c.a.,m}$ | AC power consumption of motor loads |
| $E_{c.c.,e}$ | DC power consumption of electronic equipment |
| $E_{c.c.,i}$ | Inverter-type air conditioner DC power consumption |
| $E_{c.c.,m}$ | Direct current power consumption of motor loads |
| $E_g$ | Energy, in electron-volt, of PV cell diode bandgap |
| $E_{GFV}$ | Daily measured energy produced by photovoltaic generator |
| $E_{GFV,SIM}$ | Daily simulated energy produced by photovoltaic generator |
| $E_{PERDAS}$ | Daily energy dissipated in grid and power conversion processes |
| $f_{sw}$ | Converter switching frequency |
| $G$ | Conductance matrix |
| $G_i$ | Global irradiance incident on photovoltaic generator plane |
| $G_{i,STC}$ | STC Global irradiance incident on photovoltaic generator plane |
| $G_{i,TNOC}$ | TNOC Global irradiance incident on photovoltaic plane |
| $I_{adc}$ | Battery self discharge current |
| $I_{bat}$ | Battery current |
| $I_{FV}$ | Photovoltaic module or PV generator current |
| $I_{MP}$ | Current at maximum power point of PV module or generator |
| $I_{nominal}$ | Battery current for rated discharge |
| $I_o$ | Non-ideal Buck converter output current |
| $I_{o,ideal}$ | Ideal buck converter output current |
| $i_{ph}$ | Photovoltaic cell photogenerated current |
| $i_{Rp}$ | Photovoltaic cell parallel resistance current |
| $i_s$ | Photovoltaic cell diode reverse saturation current |
| $i_{s,STC}$ | Reverse diode saturation current of the photovoltaic cell model obtained under standard test conditions |
| $I_{SC}$ | Photovoltaic module or generator short circuit current |
| $I_{SC,e}$ | Photovoltaic module short circuit current obtained experimentally |
| $I_{SC,m}$ | Photovoltaic module short circuit current obtained from model |
| $I_{SC,STC}$ | Photovoltaic module or generator short circuit current under standard test conditions |
| $i_{sc,STC}$ | Photovoltaic cell short circuit current under standard test conditions |
| $I_\delta$ | Battery current threshold used in the model to enter the transition region between charge and discharge |
| $J$ | Jacobian Matrix |
| $k$ | Boltzmann Constant |
| $k_{AJS}$ | Safety factor for isolated PV generator sizing |
| $L_{c.c.}$ | Daily consumption of energy in DC |
| $L_{c.c.,cor}$ | Daily power consumption corrected by battery charge and discharge efficiency |
| $L_{COND}$ | Conductor Length |
| $LoE$ | Level of Energy |
| $n$ | Battery discharge time in hours to achieve rated capacity |
| $N$ | Grid's number of busses |
| $N_D$ | Number of battery bank days of autonomy |
| $N_s$ | Number of cells associated in series in a photovoltaic module |
| $P_{1c}$, $P_{2c}$, $P_{3c}$, $P_{4c}$, $P_{5c}$ | Constants of the battery model regarding the constructive aspects that influence the internal resistance of the battery operating in the charging region. |
| $P_{1dc}$, $P_{2dc}$, $P_{3dc}$, $P_{4dc}$, $P_{5dc}$ | Constants of the battery model regarding the constructive aspects that influence the internal resistance of the battery operating in the discharge region. |
| $P_{auto}$ | Charge controller self-consumption |
| $P_{D,MAX}$ | Maximum allowable depth of discharge of the battery bank |
| $P_{DT}$ | Power dissipated by dead time |
| $P_{FV}$ | PV module or generator power |
| $P_{FV}^*$ | Photovoltaic generator reference power, obtained by MPPT algorithm |
| $P_{MP}$ | PV module or generator rated power |
| $P_{MP,e}$ | Experimentally obtained rated power of photovoltaic module |
| $P_{MP,m}$ | Rated power of photovoltaic module obtained by simulation |
| $P_{MP,STC}$ | Photovoltaic generator rated power in STC |
| $P_{NOM}$ | Rated power of photovoltaic generator |
| $P_{nom,l}$ | Rated lamp operating power |
| $P_{r,n}$ | Power on $n^{th}$ grid bus |
| $P_{Rds\_on}$ | Power dissipated at MOSFET drain-source resistance |
| $P_{RL}$ | Power dissipated in the internal resistance of the charge controller inductor |
| $P_{SW}$ | Power dissipated by switching |
| $q$ | Elemental electric charge |
| $R_{bat}$ | Internal battery resistance on Thévenin equivalent model |
| $r_{c.c.}$ | Linear Resistance per conductor kilometer |
| $R_{ds\_on}$ | MOSFET Drain Source Resistance |
| $R_G$ | Equivalent grounding resistance |
| $R_L$ | Internal charge controller inductor resistance |
| $R_{lamp}$ | Lamp resistance |
| $R_{n,m}$ | Line resistance between grid nodes n and m |
| $R_P$ | Parallel resistance of the standard photovoltaic cell model |
| $R_S$ | Series resistance of standard photovoltaic cell model |
| $R_{vent}$ | Fan resistance |
| $SLH_{min}$ | Daily average number of sunlight hours for worst-irradiation month |
| $SoH$ | Battery state of health |
| $T_a$ | Ambient temperature |
| $T_{a,TNOC}$ | Ambient temperature at nominal conditions of photovoltaic cell operation |
| $T_{bat}$ | Battery Operating Temperature |
| $T_c$ | Photovoltaic cell temperature |
| $T_{c,k}$ | Photovoltaic cell temperature in absolute value (K) |
| $T_{c,STC}$ | Photovoltaic cell temperature in STC |
| $T_{c,STC,k}$ | Photovoltaic cell temperature in STC in absolute value (K) |
| $T_{COND}$ | Conductor Temperature |
| $t_d$ | MOSFET Switching Fall Time |
| $T_{ref,s}$ | Reference temperature for battery health factor calculation |
| $t_s$ | MOSFET Switching Rise Time |
| $t_{tempo\_morto}$ | Converter dead time |
| $U_L$ | Coefficient of thermal loss of PV module at room temperature |
| $U_{L,TNOC}$ | Coefficient of thermal loss of PV module at nominal operating temperature of the cell |
| $V_{abs}$ | Battery voltage at absorption stage, regulated by charge controller |
| $V_{bat}$ | Voltage at battery terminals |
| $V_{boc}$, $K_{boc}$ | Battery model constants for construction aspects that influence the internal voltage of the battery operating in the charging region |
| $V_{bodc}$, $K_{bodc}$ | Battery model constants refer to the constructional aspects that influence the internal voltage of the battery operating in the discharge region. |
| $V_c$ | Battery voltage in charge region |
| $V_{cdc}$ | Battery voltage in the transition region between charge and discharge |
| $v_d$ | Diode bias voltage of photovoltaic cell model |
| $V_{dc}$ | Battery voltage in discharge, deep discharge and exhaust regions |
| $V_{DESC}$ | Battery Undervoltage Charge Controller Disconnect Voltage |
| $V_{ec}$ | Battery Saturation Voltage |
| $V_{eq}$ | Voltage at equipment terminals |
| $V_{flt}$ | Battery voltage at fluctuation stage, regulated by charge controller |
| $V_{FV}$ | Voltage of PV module or generator |
| $V_g$ | Battery Gasification Voltage |
| $V_{int}$ | Internal battery voltage on Thévenin equivalent model |
| $V_{MP}$ | Voltage at the maximum power point of the PV module or generator |
| $V_{MP,e}$ | Voltage at the maximum power point of the experimentally obtained photovoltaic module |
| $V_{MP,m}$ | Voltage at the maximum power point of the PV module obtained from the model |
| $V_n$ | Rated battery voltage |
| $V_{nom,l}$ | Rated lamp operating voltage |
| $V_o$ | Buck converter output voltage |
| $v_{OC}$ | Photovoltaic cell open circuit voltage |
| $V_{OC}$ | Open circuit voltage of PV module or generator |
| $V_{oc,bat}$ | Battery open circuit voltage |
| $V_{OC,e}$ | Photovoltaic module open circuit voltage obtained experimentally |
| $V_{OC,m}$ | PV module open circuit voltage obtained from the model |
| $V_{r,n}$ | Voltage in grid's $n^{th}$ bar |
| $V_{REC}$ | Charge controller reconnect voltage after battery voltage recovery |
| $V_{sc}$ | Battery voltage in saturation and overcharge regions |
| $V_{th}$ | Diode thermal voltage |
| $Y_{FV}$ | Daily Productivity of Photovoltaic Generator given in kWh / kWp |
| $\alpha_c$ | First-order temperature coefficient relative to battery capacity |
| $\alpha_f$ | Photovoltaic cell thermal absorption coefficient |
| $\alpha_{fc}$ | Thermal coefficient of variation of battery voltage in saturation state |
| $\alpha_{gas}$ | Thermal coefficient of variation of battery voltage in gasification state |
| $\alpha_r$ | Thermal coefficient of variation of conductor resistance |
| $\alpha_{rc}$ | Temperature coefficient for varying internal battery resistance in the charging region |
| $\alpha_{rdc}$ | Temperature coefficient for variation of internal battery resistance in discharge region |

| $\alpha_{sc}$ | Temperature coefficient of short circuit current |
| --- | --- |
| $\alpha_T, \beta_T$ | Temperature coefficients to obtain health factor |
| $\beta_c$ | Second order temperature coefficient relative to battery capacity |
| $\beta_{MP}$ | Thermal coefficient of variation of maximum power voltage |
| $\Delta T_b$ | Battery operating temperature range from reference value |
| $\Delta T_f$ | PV module operating temperature range from reference value |
| $\eta_c$ | Photovoltaic cell energy efficiency |
| $\eta_{c.a\text{-}c.c.}$ | Grinding Efficiency |
| $\eta_{c.c\text{-}c.a.}$ | Inversion efficiency |
| $\eta_{c.c\text{-}c.c.}$ | DC/DC conversion efficiency |
| $\eta_{c10}$ | Coefficient of effective reduction of battery capacity |
| $\eta_{cdc}$ | Overall battery charge and discharge efficiency |
| $\eta_{ch}$ | Battery charging efficiency |
| $\eta_{DCDN}$ | Energy efficiency of load power supply |
| $\eta_{MPPT}$ | Efficiency of the maximum power point tracking algorithm |
| $\eta_q$ | Battery self-discharge coefficient |
| $\eta_{Tf}$ | Battery health factor as a function of operating temperature |
| $\eta_{wz}$ | Battery health factor depending on operating zone |
| $\xi$ | Thermal Transmittance Coefficient of Glass |
| $\tau$ | Time constant for transition between overcharge and exhaustion regions |

## ABSTRACT

The concept of direct current distribution minigrids has been gaining ground in academia and industry regarding the development of distribution grid applications with high penetration of distributed energy sources and storage systems. The adoption of a direct current distribution system facilitates the integration of sources such as photovoltaic and wind generation and storage systems such as batteries, as these technologies in general operate intrinsically in direct current. This work presents the development of a direct current distribution nanogrid (DCDN) installed in the laboratory of the Grupo de Estudos e Desenvolvimento de Alternativas Energéticas (GEDAE), of the Universidade Federal do Pará. The developed grid is composed of three photovoltaic generation and storage systems in battery banks and three load banks, distributed over the 200 m grid in a ring topology, with a 24 Vdc bus. Two simulation methodologies were developed and can reproduce the nanogrid's operational behavior under static and dynamic conditions, allowing the evaluation of the grid performance over a day of operation. Tests are also presented with measurements at strategic points of the grid to evaluate the system behavior under specific operating conditions, being normal or under contingency. The results attest the nanogrid's ability to reliably supply the loads, as long as it respects the limitations of the implemented power generation and storage capacities. In addition, it was found that the characteristics related to the topology of the commercial charge controller that is used to form the DC distribution nanogrid benefits the power quality for the developed grid size and topology.

**Keywords: DC Nanogrid, distributed generation, DC Microgrids, photovoltaic generation, energy storage.**

# INTRODUCTION

The pursuit of environmentally friendly power generation systems, as well as the rising costs of energy from fossil fuel combustion, has led to a worldwide growth in the number of power generation applications from renewable sources, especially the photovoltaic (PV) and wind (EPE, 2018; REN21, 2019). The insertion of intermittent generation sources into distribution bars, traditionally designed to operate only as load, is accompanied by an increase in system complexity. Negative impacts related to the increased participation of distributed generation have already been investigated and followed by new mitigation proposals (RAZAVI et al., 2019; SINGH; MUKHERJEE; TIWARI, 2015).

In this context, the concept of distribution microgrids proposes to make traditional distribution systems more flexible. Assuming that the complexity of a system is directly related to the number of bars present in it, the subdivision of a large distribution grid by smaller independent groups facilitates the adoption of control strategies, protection, among others, making the system more reliable and robust. In Ton & Smith (2012), a microgrid is defined as "an interconnected group of charges and energy sources with clearly defined electrical boundary, that function for the grid as a single controllable entity that can connect to or disconnect from the grid". Thus, the adoption of distribution microgrids in isolated locations is an alternative to the use of individual generation systems and may present greater reliability and robustness, since generation and energy storage systems are shared among consumers.

Low voltage direct current (LVDC) distribution microgrid (DCDM) systems can be a more efficient alternative to traditional alternating current systems. The increased participation of distributed generation sources (largely by photovoltaic systems) and the use of more efficient direct current electric charges justify the development of this type of microgrid, as they reduce the number of conversion stages, making the system more efficient, reliable and with lower cost (ELSAYED; MOHAMED; MOHAMMED, 2015).

The advantages of implementing a distribution system in DCDM include: less complexity and greater control over distributed generation sources; reduction in transmission losses (there is no reactive power transmission, as well as the better use of the conductor, since there is no skin effect in direct current systems); besides the reduced number of conversion stages (both in generation and demand) (KUMAR; ZARE; GHOSH, 2017).

Several DCDM facilities are already operating in various parts of the world, whether in pilot projects for research institutions, industrial and commercial facilities and for electrification of isolated communities. According to Rodriguez-Diaz et. al. (2015), one of the main applications already consolidated in the industry is in telecommunications. The adoption of 380 V / 48 V direct current distribution considerably increases the overall energy efficiency of the installation as well as reduces the high cooling expenditure on this type of installation as heat dissipation in sources and conductors is reduced (KUMAR; ZARE; GHOSH, 2017).

Industrial manufacturer LLC Robert Bosch installed, in 2014, several DCDM systems in its US industrial plants in order to validate expected efficiency gains and verify the safe and reliable operation of DCDM. To carry out the comparison studies, DC and AC systems with similar ratings were installed, both connected to the electric grid, partially supplied by photovoltaic systems and supplying loads with the same end use. The loads used in both DC as for AC are for lighting (high-bay LED) and ventilation (industrial ceiling fans). It has been found that the direct current system allows the energy harvested from the photovoltaic system to be 6% to 8% higher than the alternating current system, considering different scenarios of operating conditions, location and design parameters (FREGOSI et al., 2015). Other examples of DCDM in industrial facilities are presented in (ARDA POWER, 2018) (NEXTEKPOWER, 2019).

Among the projects conducted by research institutions, stands-out those capable of emulating the use of direct current loads considering typical use in various consumption scenarios, as well as the integration of power generation sources and energy storage systems. Díaz et al (2015) developed two laboratory structures that demonstrate the use of DCDM in residential and industrial systems, in Denmark and China, respectively. The proposal is to use the facilities to carry out design studies, modeling, control, coordination, communications and management applied to dc distribution networks. Other examples of pilot projects already in operation in research institutions are presented in (WORLEYPARSONS, 2019). (ZHANG et al., 2015) (MISHIMA et al., 2014).

In Jhunjhunwala, Lolla & Kaur (2016) is presented a case study of the application of a 48 V DCDM for electrification of a small community in India. In this application, although the utility grid is available, the use of DCDM in parallel proved to be more financially advantageous, considering the current energy consumption rates in place. For this project, two models of converters to control battery bank charge and discharge, PV generation control and connection to the mains grid were developed, with capacities to serve up to 24 households.

Many of the difficulties of implementing direct current systems are due to the lack of standardization, regulation and availability of equipment. Many devices, although internally operating on direct current, are powered by alternating current, which is rectified internally at the power supply of the device. The lack of commercially available equipment to operate in direct current is largely due to the absence of standards that seek to standardize this type of system. An example of this is the voltage level that a direct current distribution system must adopt. There are several practical examples in which LVDC systems are applied without following a standardization, especially regarding the voltage level, as presented in the previous references.

Standardization efforts are being made by institutions such as: IEC (International Electrotechnical Commission), which already has standards such as IEC 62040-5-3, IEC 61643-3 and IEC 61643-311, and has recently approved the formation of strategic group SG4, for evaluation of low voltage direct current distribution systems up to 1500 V; IEEE-SA (The Institute of Electrical and Electronics Engineering Standard Association), which currently has a number of activities related to this topic, such as WG946 (Best practices in dc auxiliary systems design using lead acid batteries), P2030.10 (design, operation and maintenance of DCDM for rural or remote applications) and IEEE DC @ Home (investigate the use of LVDC in homes); Emerge Alliance, which recently launched two standards, one for residential and commercial 24 V systems and one for datacenter and 380 V telecommunication centers; In addition to other institutions focusing on telecommunications systems such as ETSI (European Telecom Standard Institute), ITU (The International Telecommunication Union) and CCSA (Chinese Communication Standards Association).

Other difficulties are related to the implementation of protection devices, which differ from the devices in alternating current systems based on the passage of current/voltage through zero - a situation that does not exist in direct current systems. In this case, the abrupt disconnection of loads is more prone to arcing. In the field of photovoltaic systems there are commercially available devices such as circuit breakers, fuses and disconnect switches designed to operate on direct current. Other devices, such as specific arcing suppression sockets, have been developed (TAN; HUANG; MARTIN, 2014).

Despite the many efforts being made around the world, there is still little practical experience evaluating the implementation of DCDM under real operating conditions, especially in equatorial climate regions, such as the Amazon region, which has several environmental and load usage profile peculiarities. In this sense, the main objective of this work is to experimentally evaluate a small direct current distribution grid under different real operating conditions. The physical grid is installed in the testing area of the Laboratory of the Study and Development of Energy Alternatives Group of the Federal University of Pará (GEDAE-UFPA). Besides the experimental evaluation, this work proposes the development of static and dynamic simulation files of the grid developed in the laboratory. To facilitate the understanding of the content developed in this work, this dissertation was divided into four chapters, as follows:

The first chapter presents a theoretical review of DCDM, highlighting the components, topologies and modes of operation found in the literature.

The second chapter describes the installed experimental system and object of study present in this dissertation. It also presents the methodology for dimensioning the grid components installed in the GEDAE laboratory.

The third chapter is dedicated to grid's modeling, presenting the models used and adapted to perform static and dynamic simulations in Matlab/Simulink simulation environment.

The fourth chapter presents grid's operating results under different conditions, comparing with simulation results and evaluating grid's serviceability in different tests. At the end, the conclusions obtained from this dissertation are presented and recommendations for future works associated with this dissertation are listed.

# 1. LOW VOLTAGE DC DISTRIBUTION SYSTEMS

## 1.1 Introduction

The dawn of electrical distribution systems in the late 19$^{th}$ century was marked by a technical and economic dispute that later became known as the 'War of Currents' (HUGHES, 1993). On the one hand, the direct current generation, distribution and loads system designed by Thomas Edison, on the other, the proposal for an alternating current system advocated by George Westinghouse and Nikola Tesla. In fact, the first years of the electric power distribution system in the United States were marked by direct current operation, with loads composed basically by DC resistive lamps and DC brushed motors (HUGHES, 1993).

However, the system proposed by Edison was associated with a series of technical limitations, such as: the impossibility of working efficiently with different voltage levels in the same grid; the high voltage drop in the conductors, so that the generating plants needed to be close to the load centers; the reduced efficiency of the DC motors, among other factors. In this sense, the proposal of a system operating in alternating current has suppressed several of the problems associated with the DC distribution: the use of transformers enabling the efficient step-up and step-down of voltage in a grid; the transmission of energy in three-phase, high voltage, reducing losses in conductors and making the physical distribution of generating plants more flexible; the higher efficiency of the induction motor proposed by Tesla, operating on alternating current, etc.

At the time, the technical superiority of the AC system was decisive in making it a standard in distribution systems around the world, as it continues to this day. However, more than a century after the beginning of power distribution systems, the scenario of power grids is quite different: loads largely comprised of electronic devices (switched power supplies, controllable motor drivers, LED lighting, etc.); increased participation of distributed generation sources (integrated into the grid using power converters); evolution of power electronics, enabling the development of high-efficiency solid state transformers (high frequency transformers), among others.

In this context, the adoption of DC distribution systems may be more advantageous, both technically and economically (ELSAYED; MOHAMED; MOHAMMED, 2015). Factors such as: overall efficiency gain, increased robustness and reliability, simplification of control and management strategies, among others, are keys to this possible technical and economic advantage.

Thus, it is undisputed that AC power distribution grids will still continue to be the basis of complex distribution systems for a long time, however applications of DC systems, especially in the context of nano, mini and micro distribution grids tends to multiply in the coming years, being isolated from the AC mains or not.

At this point, it is important to make a distinction between minigrids, microgrids and nanogrids. Although they are fundamentally the same entity for the electrical system, it is usual to distinguish nanogrids, microgrids and minigrids according to the load power and/or generation capacity, being the nanogrids assigned to the small, lower-power systems. However, there is still no consensus in the literature regarding the power limit between nano, micro and minigrids. In this work, the term nanogrid is considered, given the reduced power of the experimental system, therefore the acronym DCDN corresponds to the direct current distribution nanogrid, being used to specifically address the grid implemented in the laboratory. Also, it is used in this work the acronym DCDM that corresponds to direct current minigrids, being used in a generic sense.

For a good understanding of this work, it is important to present a theoretical background on DC distribution systems. In the next sections of this chapter, a review of the main topologies, modes of operation and control, and the main devices that form an DCDM are shown.

**1.2 General Scheme of a DCDM**

The general scheme of an DCDM and its main constituent elements are presented in Figure 1.1.

**Figure 1.1 – General scheme of a DCDM**

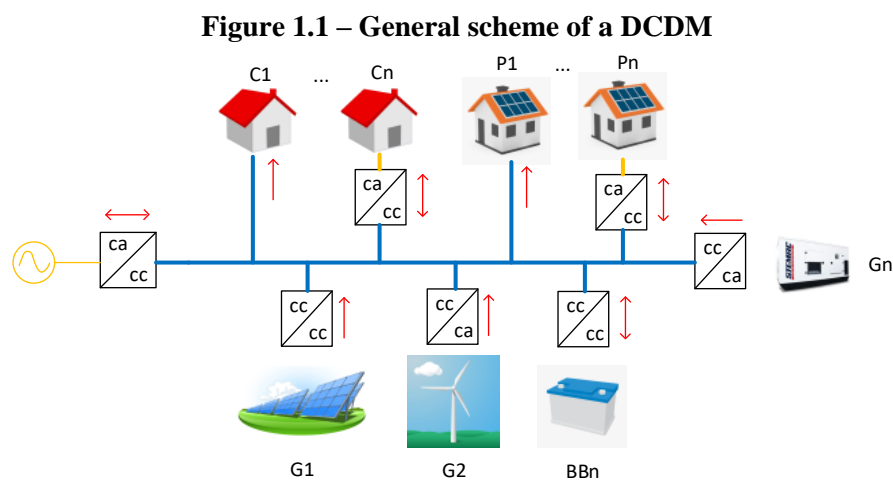

Houses $C_{1\ldots n}$ represent the consumers of this minigrid, which can be consumers served directly by DC ($C_1$) or consumers using AC loads ($C_n$). As well as in an AC grid, the main loads in an DCDM can be of constant power or constant impedance type. The houses $P_{1\ldots n}$ represent the prosumers, this a class of consumers that has a generation system installed, for example, photovoltaic (PV), and can inject power into the DCDM, working as a distributed generation (DG) in the DCDM. Like consumers, prosumers can have DC ($P_1$) or AC ($P_n$) service.

The $DG_{1,2 \ldots n}$ distributed generators and $BB_n$ battery bank are the DCDM's main power sources. In an DCDM it is possible to have different generators operating in parallel, as is the case with $DG_1$ and $DG_2$ generators. The battery bank $BB_n$ operates as a load to the system under conditions of excess of generation from other sources, charging the batteries, and operates as a source when the generation of other sources is insufficient to meet the loads. The diesel generator, identified as $DG_n$, has backup function, and should operate when the other sources are insufficient to meet the loads, in case of an isolated DCDM. Alternatively, the DCDM may have one or more AC mains connection points, so that power may be exchanged between the DCMD and the mains grid in both directions and under certain conditions, such as when no local source of power is available.

## 1.3 Topologies

Two basic characteristics must be taken into consideration when classifying the topology of a DCDM: polarity and architecture, as presented in the following items.

### 1.3.1. Polarity

The polarity is related to the DCMD main bus voltage levels. The most common configurations are unipolar and bipolar, as illustrated in Figure 1.2.

**Figure 1.2 – DCMD (a) unipolar and (b) bipolar.**

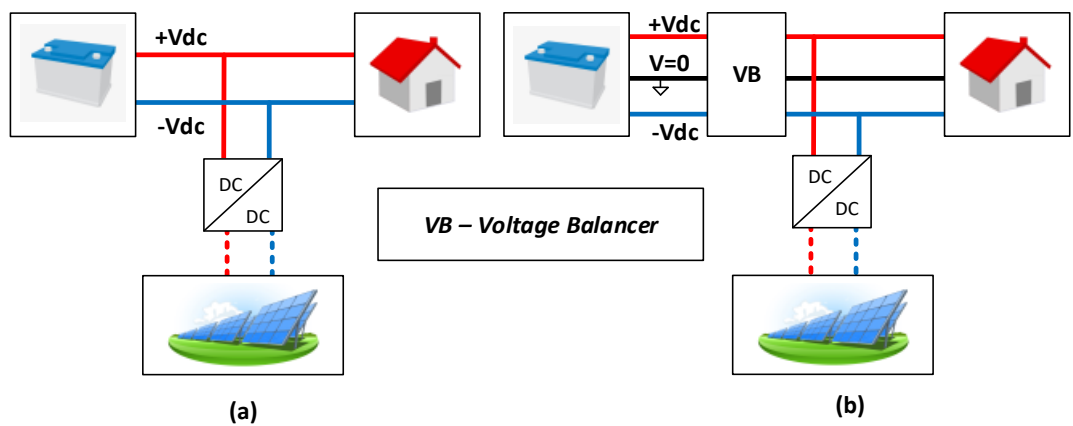

(a)         (b)

In the unipolar configuration, sources and loads are connected to the positive and negative poles of the DC bus, as shown in Figure 1.2 (a). In this configuration, the system operates at a single rated voltage level on the main bus. This configuration is more common in small systems, because to transmit higher power one must increase the voltage in the bus, and this will result in a larger number of power converters to connect loads operating at lower voltages.

The unipolar system has the advantage of simpler implementation and not having the asymmetry problem that can occur in the bipolar system. However, this system has no redundancy and if a line is lost the whole system will be affected. In addition, the power transmission capacity is quite limited compared to the bipolar configuration. The system installed in the GEDAE-UFPA laboratory has unipolar configuration.

The bipolar configuration, shown in Figure 1.2 (b), is the most widely used in larger systems. In this configuration, the system operates at three voltage levels: + Vdc, -Vdc and 2 x Vdc, and the transmitted power is distributed at these three levels. This configuration gives the grid operator greater flexibility in choosing the rated operating voltage, $V_{dc}$, as heavier loads can be connected to the higher voltage terminals and the lighter ones can be distributed between + Vdc and -Vdc. In addition, this configuration has natural redundancy, since if one of the transmission lines fails, there is still voltage between two other lines that can be used to meet critical loads.

In the bipolar configuration there may be asymmetry between positive and negative $V_{dc}$ voltages, either under normal operating conditions, for example: higher power injection at the positive pole, or higher loading at the negative pole; or under fault conditions, such as the loss of a generation subsystem. In order to prevent this asymmetric condition from arising, the bipolar DCDM must have a voltage balancer, VB, or a control system for converters coupled to the generation bars that can act to ensure the symmetry of the voltages.

A third, less common configuration considers a variable nominal voltage on the main bus. This configuration is used, for example, for direct coupling of photovoltaic generation on the DC bus as shown in Figure 1.3. In this configuration, the voltage on the DCDM main bus is controlled by a bidirectional converter AC/DC that interfaces with a AC grid (Figure 1.3 (a)); In an isolated system, the voltage on the DC bus can be controlled by the DC/DC converter connected to the energy storage system (Figure 1.3 (b)). This voltage is set to operate as close as possible to the maximum power voltage, $V_{MP}$, of the PV generator.

**Figure 1.3 – DCDM with variable voltage: (a) connected to the AC mains and (b) isolated.**

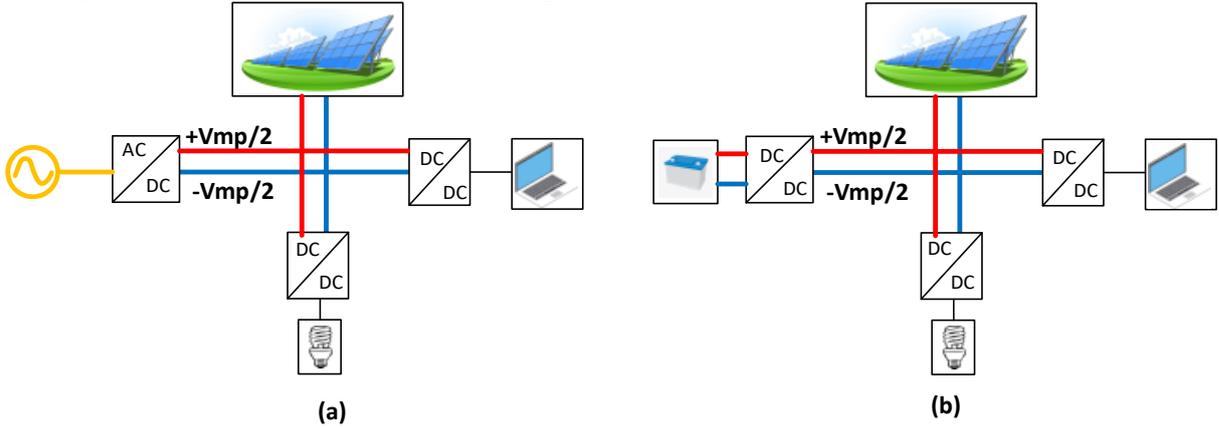

This topology eliminates the need for DC/DC converters to connect the PV generator to the DCMD, and, on the other hand, requires the load-coupling DC/DC converters to operate with variable input voltage.

The main limitation of this topology is that different PV generators, with different configurations, operating temperature, tilt angles and spatial orientations, have different maximum power voltage, and parallel connection on the same bus makes it impossible achieve the maximum power for all the generators.

This setting is recommended only in small grids where all generators operate at close maximum power points. This is the case presented in Fregosi et al (2015), where the PV generators are all on the same roof of an industrial warehouse. In this particular system, it was observed an increase of up to 8% in energy efficiency of the PV generation, compared to an AC grid-connected system.

### 1.3.2. Architecture

The architecture of an DCDM is defined by how loads and sources are connected to the main DC bus and the eventual connection of this bus to the AC grid. In general, an DCDM can be configured in radial or ring architecture, as shown in Figure 1.4.

**Figure 1.4 – DCDM architecture (a) radial, (b) ring and (c) ring under fault.**

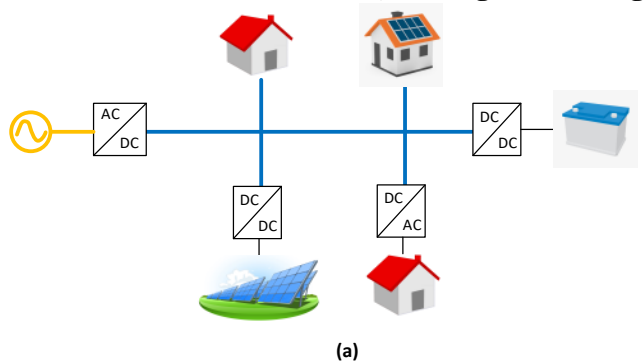

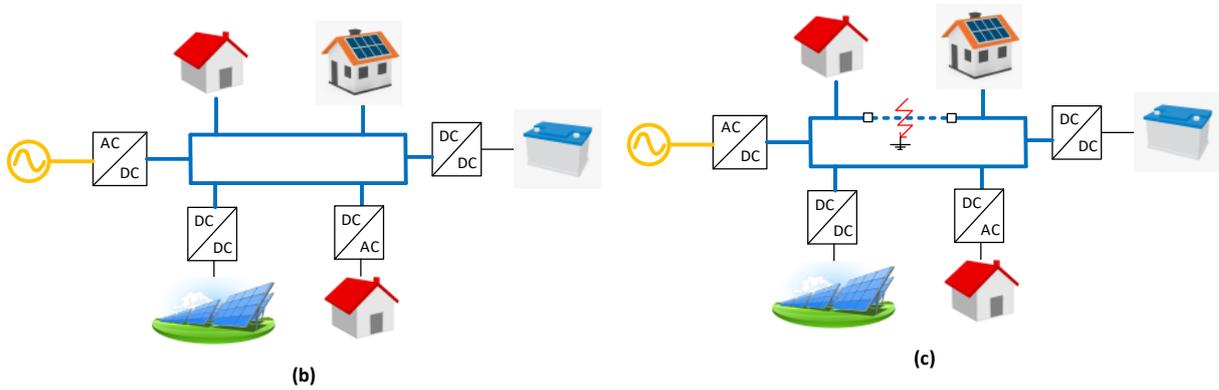

(b)  (c)

In radial architecture, power is transmitted by a single main path and is distributed in branches to meet the loads. The simplicity of the radial architecture allows, in most cases, greater proximity between source and load, reduced distribution losses and reduced grid's implementation costs. However, the low flexibility of this configuration under fault conditions is a problem in networks that require greater reliability, since the loss of a portion of the grid may result in interruption of access to various system loads.

The ring architecture, on the other hand, makes the system more flexible in fault situations, as power can be transmitted from source to load in two or more paths. In Figure 1.4 (c), a ground fault in a portion of the grid is isolated and all system loads continue to be serviced, even though by an alternative and less efficient path. This configuration, while more complex and costly to implement, is recommended for urban and industrial grids that require greater flexibility. The system installed in the GEDAE-UFPA lab has a ring architecture.

In the configurations presented above, there is only one connection point between the DCDM and the AC mains. In the event of a fault in the part of the grid that makes this connection or in the event of an AC/DC converter problem, the DCDM is required to operate isolated and power limitation problems may occur. To work around this problem, DC grids with multiple points of connection to the AC mains can be implemented, known as interconnected grids, which may be of the meshed (also known as multiterminal) or zonal type, as illustrated in Figure 1.5.

In the meshed architecture (Figure 1.5 (a)), multiple AC connection points are distributed over the grid. This type of architecture is found in high-voltage direct current transmission systems (HVDC), such as offshore wind farms, and in underground sub-transmission and distribution systems in urban networks (KUMAR; ZARE; GHOSH, 2017).

**Figure 1.5 – Interconnected DCDM architectures (a) meshed e (b) zonal.**

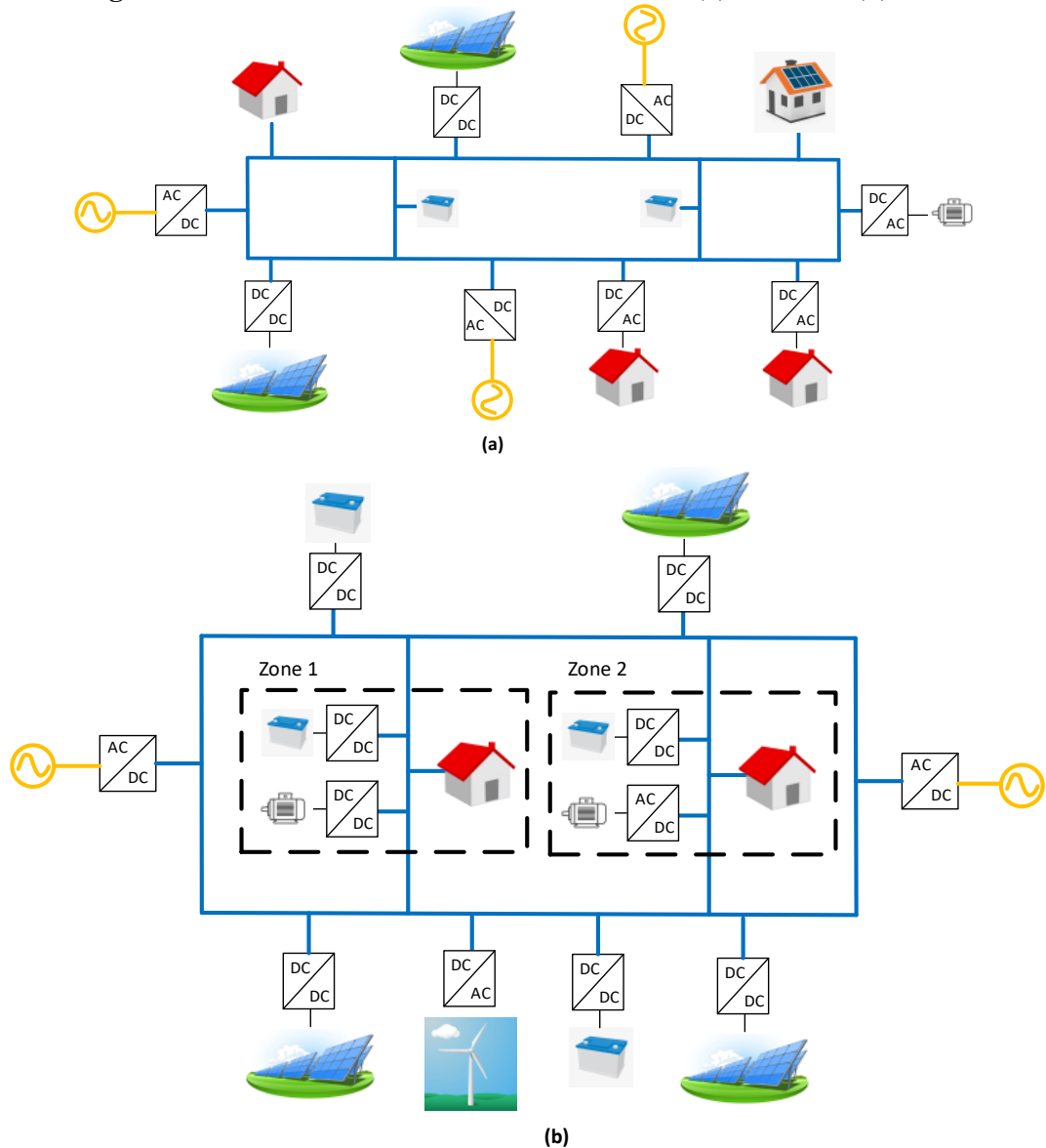

(a)

(b)

The zonal architecture can be understood as a set of cascading DCDM with a symmetrical configuration: the distribution system is subdivided into zones and each zone is served by two redundant DC buses (Figure 1.5 (b)) (KUMAR; ZARE; GHOSH, 2017).

**1.4 Grounding and protection**

Protection technology in DC distribution systems is still an open area that needs more focus and study to reach an advanced stage of maturity comparable to that of AC distribution systems (MOHAMMADI; AJAEI; STEVENS, 2018).

**1.4.1. Grounding**

The grounding scheme in an DCDM, as in any power distribution system, is of great importance to ensure the proper functioning of the protection and to maintain safe transient voltage and current levels. As for grounding, a DC system can be classified into the following categories: insulated from earth (ungrounded), grounded with resistance and direct grounding (FARHADI; MOHAMMED, 2017). Also, on DC systems connected to the AC mains, the type of grounding in the AC grid should be considered when assessing the grounding configuration at the DCDM (KUMAR; ZARE; GHOSH, 2017).

**1.4.1.1.    Ungrounded**

This type of configuration has advantages such as: greater continuity of supply, since in case of contact of a pole, positive or negative, with the ground, the system is able to maintain the operation; reduction of leakage current and consequently reduction of corrosive effects; and greater simplicity in installation. The system installed in the GEDAE-UFPA laboratory is of ungrounded.

However, in an ungrounded system, problems may occur in the fault identification by protective devices. Although the system can operate if a pole is in contact with earth, identifying this fault is difficult given the low fault current. The situation is even more serious if both poles contact the earth, causing a short between poles, which can cause serious damage to the system.

Another problem associated with an ungrounded system is the emergence of noise that can compromise measurement systems and the operation of sensitive equipment. The absence of grounding also requires the equipment to have an insulation level compatible with the pole-to-ground voltage, so that the increase of voltage implies higher expenses with equipment and conductor's insulation.

Considering the problems associated with this configuration, the ungrounded system is not recommended for larger grids, as the increase in the DC bus voltage potentiates the negative effects on fault occurrence.

The American Standard NEC 2011 (NATIONAL FIRE PROTECTION ASSOCIATION, 2011) deals in the article 250, items 160-169, specifically about the protection and grounding in direct current systems. This standard restricts ungrounded configuration to unipolar networks with voltage limit of 50 V. In Europe, the low voltage directive LVD 65/2006 (EUROPEAN COMMISSION, 2016) provides that systems from 75 V must necessarily be grounded. In Brazil, although there is no specific standard for DC distribution networks, the NBR 5410 provides for the use of separated extra-low voltage systems (SELV) with maximum DC voltage of 120 V.

### 1.4.1.2. Resistance grounding

An alternative to improve DCDM fault detection is the use of a resistive indirect ground. In this configuration, one of the poles or, in case of bipolar topology, the neutral conductor, is grounded by means of a grounding resistor, $R_G$. When possible, the neutral must be used for this type of grounding to reduce the dissipated power in $R_G$. In addition, grounding at the negative pole may potentiate the effects of corrosion.

Similarly to the ungrounded system, in the event of a ground fault the system can maintain continuity of service as the fault current is limited by $R_G$.

In this configuration, the grounding resistance must be dimensioned taking into consideration the following compromise relationship: large $R_G$ will reduce the dissipated power, however compromising the operation during transient faults, which may cause unbalance in the pole voltages; too small $R_G$ will improve protection performance in the event of a fault, yet increases the dissipated power continuously in normal operation.

Another factor that should be taken into consideration when using this configuration is that, under a pole-to-ground fault condition, equipment that is connected to the active pole must have insulation capable of withstanding twice the pole voltage (Figure 4). 1.6 (a)).

### 1.4.1.3. Direct grounding

In direct grounding, one of the poles, or neutral conductor in case of bipolar grid, must be directly grounded. This configuration has the lowest level of continuity of supply, since a pole-to-ground fault is equivalent to a pole-to-pole or pole-to-ground short circuit. Also, transient currents and voltages in the event of fault are more severe in this configuration.

On the other hand, since the fault current is high, it is easier to design and actuate the protection, as it improves the selectivity of the protection devices, which should guarantee the safety of the equipment and personnel in case of fault. In addition, in a bipolar system with neutral conductor grounding, the system can be kept operating with the healthy pole in case of ground fault at the other pole, by actuation of circuit breakers and transfer switches. Another advantage of this configuration is that the equipment insulation in the event of a fault should be designed to withstand only half of the rated value of the operating voltage on the DC bus, as shown in Figure 1.6 (b).

In this configuration, negative pole grounding may potentiate the effects of corrosion more intensely than in the resistive grounding configuration. To minimize these effects, diodes can be used to block leakage currents (Figure 1.6c).

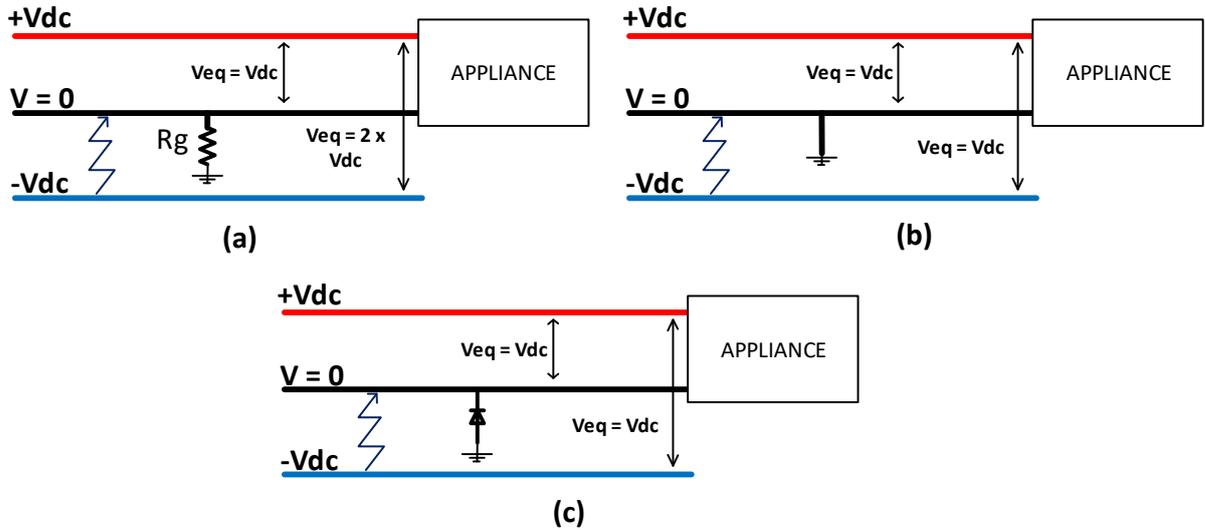

**Figure 1.6 – Voltage at the load terminals under pole-to-neutral fault with (a) resistive grounding, (b) direct grounding (c) direct grounding with diode.**

### 1.4.2. Protection devices in a DCDM

The main difference between DC and AC protection devices is that the first must be capable of extinguishing the arcing formed at the time of interruption of a charged circuit. Arcing is a naturally occurring phenomenon, characterized by the disruption of a dielectric medium, usually the ionization of air, maintaining the flow of electrons in a stochastic manner. This phenomenon is dangerous in electrical installations because its stochastic nature provides the appearance of sparks and is associated with the beginning of fire in installations. A common example is in grid-connected photovoltaic systems where the abrupt interruption in the operating DC bus may cause arcing and fire

In AC the arc extinction occurs naturally, at the moment of the voltage passing through zero. In DC, since the voltage is ideally constant, the arc extinction should be induced by the protection device. Therefore, in a grid-connected photovoltaic system, the inverter de-energization procedure first provides for the disconnection of the AC bus and then the DC bus

In an DCDM, arcing is critical not only in fault situations, where protective devices such as relays and circuit breakers must act, but also in normal system operation at the time of load switching. In addition, generally the inductance in an DCDM is low (FARHADI; MOHAMMED, 2017), contributing to a high fault current growth rate. Therefore, protection devices in an DCDM must act fast to minimize damage from fault currents.

### 1.5 Power quality problems

An DCDM is subject to several types of power quality problems that can affect the operation of loads. These power quality problems can be classified into short term (transient) or long term (steady state) problems and may be voltage and current related in an DCDM. The main phenomena of power quality problems are presented in the following items.

### 1.5.1. Voltage transients

Voltage transients in an AC grid connected to a DCDM can cause overvoltage problems and even change of operating point on the DC bus (KUMAR; ZARE; GHOSH, 2017). A DCDM may be vulnerable to capacitor bank switching, load shifting and generation variations on the AC bus in case of improper operation of the AC/DC coupling converter.

Therefore, the standardization of the acceptable voltage level limits and in transients after disturbances is important for adequacy and dimensioning of the various equipment present in the grid.

### 1.5.2. Harmonics (Spurious Components)

Although, by definition, in a DC system there are no harmonic components frequencies multiple to the fundamental (since the fundamental frequency is 0 Hz), harmonic components in DC grids are considered as any oscillating voltage and current in the system. Power converters are the main source of harmonics in an DCDM.

The DC/AC converters, used in variable frequency motor drivers and for coupling the DCDM to the AC mains, are sources of low frequency harmonics on the DC bus. These low frequency harmonics can generate extra losses in various system components (LANA, 2014), reducing equipment life and decreasing energy efficiency.

DC/DC converters, on the other hand, are commonly associated with high frequency harmonic currents, and may be responsible for severe overvoltage conditions in resonance situations (VAN DEN BROECK; STUYTS; DRIESEN, 2018). An DCDM may have multiple resonant frequencies due to the large number of filter capacitors at the input of the various power converters present in the grid, associated with the impedance of the DC bus cable. Stability issues may occur on the DC bus if one of the resonant frequencies is tuned to the same harmonic range generated by the power converter.

Another problem associated with high frequency harmonics is the interference that can arise in power line communication systems (PLC). This type of communication may be critical to the operation of control and power management systems at DCDM, and excessive noise from high frequency harmonics may make impossible the use of PLC.

One of the simplest ways to filter harmonics in the grid is to use filter capacitors on the DC bus. However, the sizing of these capacitors should consider factors such as changes in mains impedance and in resonant frequency and increase in short circuit current. In Lana,

Kaipia & Partanen (2011) a simulation study of a DCDM connected to the AC mains was conducted, where harmonic currents and voltages were evaluated in various operating scenarios with and without capacitive filters.

*1.5.3. Inrush currents*

When connecting a de-energized power converter into a DC grid, the converter input capacitor (associated with the EMI filter) will charge with a high starting current, limited only by the supply capacity of the other parallel connected converters, capacitances on the DC bus, and batteries in case of direct coupling on the bus. Figure 1.7 illustrates the emergence of this phenomenon. This current, known as the inrush current, may cause damage to the connection and even lead to the emergence of undervoltage that may affect the operation of other components in the grid.

This type of problem can be minimized in two ways: preloading the converter before connecting it to the grid or adopting a soft-start methodology similar to that used in AC grids to avoid inrush current in transformers and motors.

**Figure 1.7 – Emergence of inrush current at the input of a DC converter in a DCDM.**

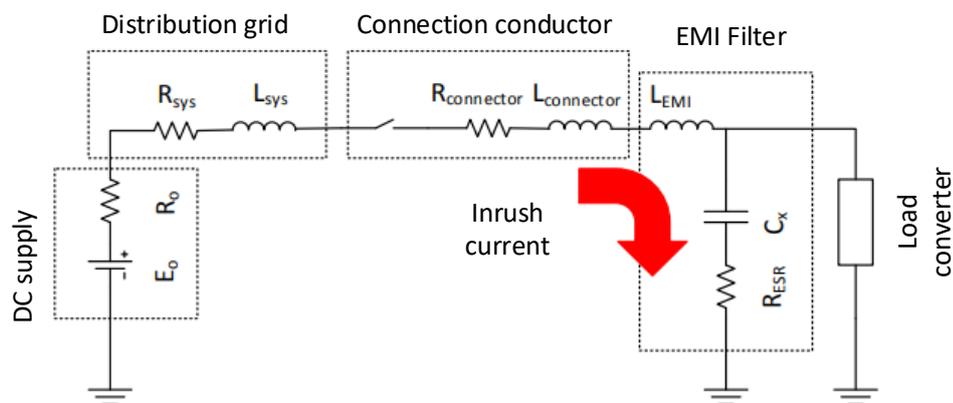

Source: (WHAITE; GRAINGER; KWASINSKI, 2015).

1.5.4. DC Bus faults

Similar to what occurs in an AC system, the short-circuit implications in an MDCC depend on a number of factors such as: type of short, i.e. whether it is a direct fault (direct conductor contact) or a fault with impedance; grid impedance; short circuit current capacity of the various components in the grid (capacitance, batteries).

In a weak grid with high impedance and low short circuit current supply capacity, potentially hazardous overvoltage may occur at various points in the grid as a result of the fault current flowing. In addition, in a weak grid, protective devices can interpret the short circuit as a heavy load on the system, making it difficult to selectively actuate protection.

The short circuit study must be done in a DCDM for the correct sizing and adjustment of protective devices. If the DCDM is connected to the AC mains, the short circuit study should also consider the various faults that may occur on the AC bus, in addition to the DC bus faults that may occur between live conductors and between live conductor and neutral conductor (in case of bipolar grid).

### 1.5.5. Voltage asymmetry in bipolar system

In bipolar systems, asymmetry of positive and negative busbar voltages may arise with respect to the common point. This asymmetry occurs due to unbalanced loading and unequal generation on the positive and negative buses, and in the case of DCDM connected to the mains grid, the unbalance between AC phases can also lead to asymmetry in the DCDM. This asymmetry may compromise the operation of equipment and converters connected to the grid, and the implementation of power conditioners to ensure symmetry should be considered (voltage balancers).

### 1.5.6. Circulating currents

The parallel connection of power converters on the same DC bus may cause circulating currents to arise between the converters if there is a voltage difference between the converter connection points. Figure 1.8 illustrates the emergence of circulating current between two DC/DC converters connected in parallel on an DCDM.

**Figure 1.8 – (a) circulating currents between two DC/DC converters (b) steady state equivalent circuit**

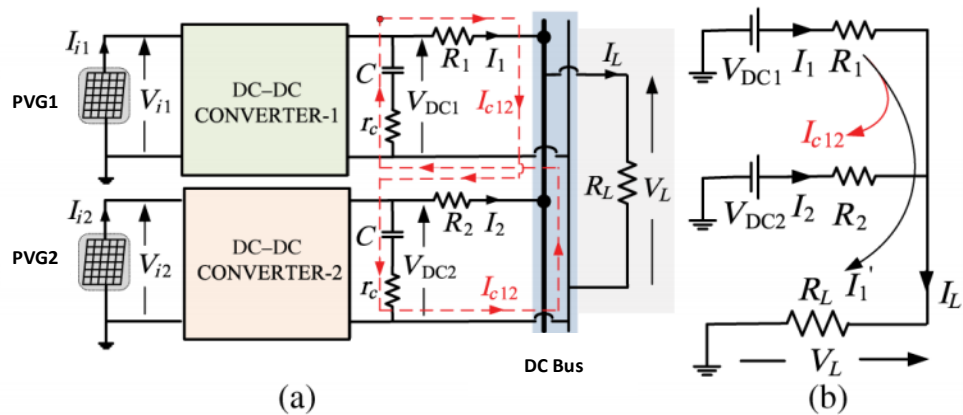

**Source: (AUGUSTINE; MISHRA; LAKSHMINARASAMMA, 2015).**

In the situation shown in Figure 1.8, the output voltage of converter 1, $V_{DC1}$, is greater than the output voltage of controller 2, $V_{DC2}$, so that the output current of converter 1, $I_1$, is given by Equation (1.1):

$$I_1 = -I_2 + I_1' \qquad (1.1)$$

Where $I_1'$ is the load current at R$_L$ and $-I_2$ is the current at the converter 2, such that:

$$-I_2 = I_{c12} \qquad (1.2)$$

Where $I_{c12}$ is the circulating current between converters 1 and 2.

This current is undesirable as it is a source of losses in the system, since power is dissipated at grid's resistors and converter's impedance, even though no load is operating. In addition, this current is more likely to occur in systems with large numbers of converters in parallel. Control strategies are proposed to mitigate the emergence of circulating currents (AUGUSTINE; MISHRA; LAKSHMINARASAMMA, 2015).

Circulating currents can also occur if power converters are associated with energy storage devices such as battery banks. Connecting the converters in parallel can cause circulating currents from one battery bank to the other until the equilibrium between the parallel bank voltages is achieved. This phenomenon occurs in the DCDN installed in the GEDAE-UFPA laboratory; the measurement results associated with this effect are presented in Chapter 4.

**1.6 Control strategies**

In general, the main control objectives in an DCDM are: DC bus voltage regulation, power flow management, and harmonic and reactive power compensation at the AC mains connection point (LAUDANI, 2017). According to Meng et al (2017), the control and management of a minigrid is associated with multiple objectives that are related to various technical areas, time scales and physical levels. The proposal for a multilevel hierarchical control scheme is widely accepted as a standard solution to optimally and efficiently achieve the various objectives associated with the operation of a minigrid (MENG et al., 2017). The same hierarchical control scheme can be applied to a DCDM control as shown in Figure 1.9.

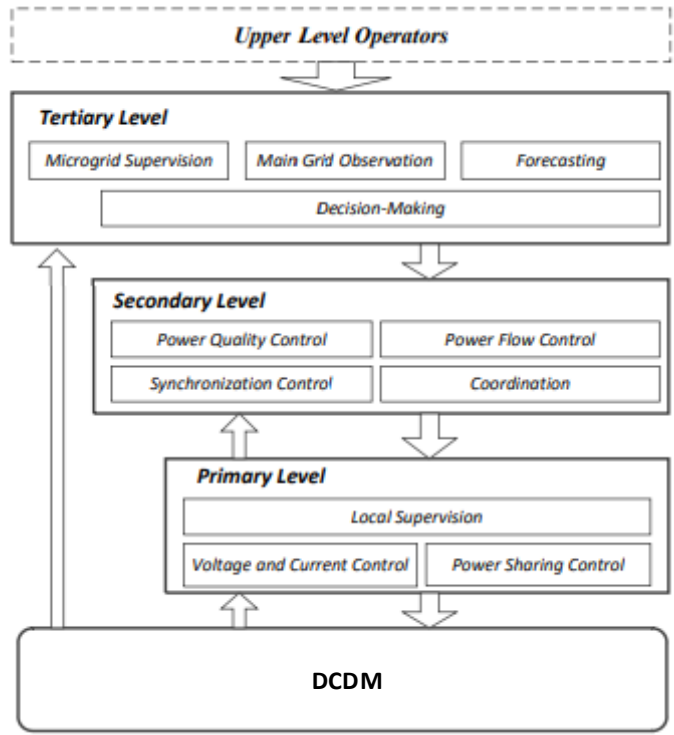

**Figure 1.9 – Hierarchical control scheme in a DCDM.**

Source: (MENG et al., 2017).

The primary level is associated with local voltage and current control, i.e. it is implemented individually in each distributed generation converter connected to the system. This control has the purpose of regulating voltage and/or current of the converters according to the operating point provided by the higher levels.

The secondary level is responsible for system level operation such as power quality control, AC mains connection, connection to the other DCDM, power flow control and generation coordination between distributed sources.

The tertiary level is related to overall system optimization, management and regulation, for example, use of weather forecast data to calculate optimal dispatch between different system sources considering hourly or daily intervals.

Also according to Meng et al. (2017), considering the three-level hierarchical system in an DCDM, this control can be implemented through four different structures: centralized, decentralized, distributed and hierarchical. Figure 1.10 illustrates these four control structures in the context of an DCDM. The choice of which structure will be used depends on factors such as: type of application (residential, university campus, commercial etc.), location, size, type of property, among others.

**Figure 1.10 – Basic control structures: (a) centralized; (b) decentralized; (c) distributed and (d) hierarchical.**

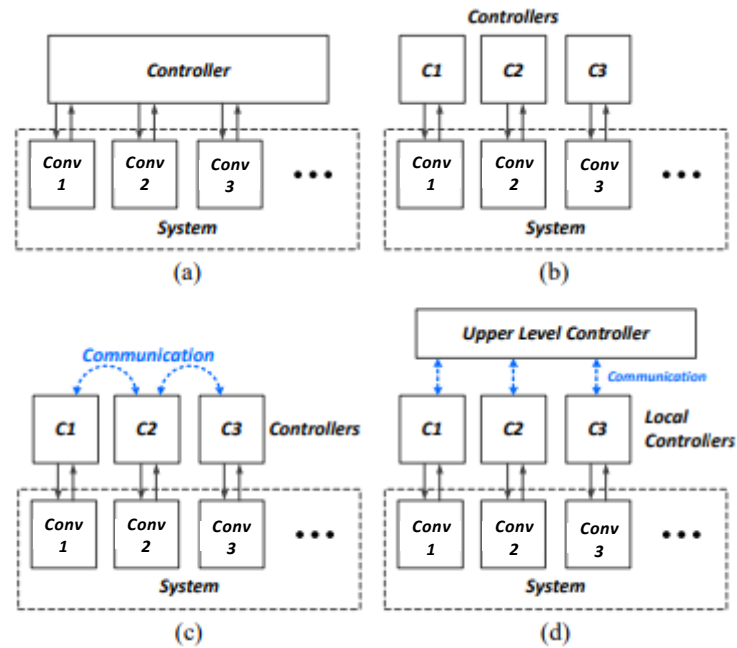

**Source: (MENG et al., 2017).**

In centralized control, the data processing of all DCDM components is performed in a single central controller that receives data of all relevant variables, having high observability and controllability of the system, and its implementation is simpler. However, if the central control unit fails, the entire system collapses, causing the loss of essential functions. In addition, this type of structure has little flexibility and little expandability: for example, the inclusion of a new distributed generation source requires reconfiguration of the central controller, which can lead to disruptions in DCDM operation. This type of configuration is more common in small systems with a small number of control variables and with physically close sources.

In a decentralized method of control, the controller from each system source operates using only locally obtained data, with no need to communicate with other controllers. Thus, system observability and controllability are reduced, limiting the range of more elaborate system-level control actions. However, this configuration has greater flexibility in the geographical distribution of the grid and less need for computational power, as well as no need of high bandwidth communication infrastructure. A typical example of decentralized methodology is the use of droop curve control, which can set the power delivered by each DC to an DCDM.

The distributed and hierarchical structures are based on communication between the various converter's controllers in a system. The distributed structure provides for the sharing of variables of interest among the controllers, so that the control actions on each converter can be based on local and system data as a whole, facilitating coordinated action between the controllers. In addition, unlike the centralized structure, the loss of a controller has a minor impact on the DCDM as a whole.

Despite the advantages associated with the distributed method, more complex systems that require greater intelligence and computational power may be unviable without the presence of a central controller. In this sense, the hierarchical method provides for communication of local controllers with a central controller, so that primary level functions (e.g. voltage regulation) are performed locally, and secondary and tertiary level functions are performed by the central controller. This type of method has become standard in minigrid control.

The DCDN installed in GEDAE does not have any active voltage and power control strategy on the main DC bus, and the bus voltage is a function of the voltages in the battery banks and the currents demanded by the system loads. The simplicity of the implemented system limits the adoption of control strategies and energy management in the system but ensures load servicing.

**1.7 Energy efficiency comparison between DC and AC distribution.**

In Dastgeer et al. (2019), it is presented an extensive survey of several studies related to the comparison of energy efficiency between DC and AC distribution systems. Although these surveys have been conducted for over a decade, it is not yet possible to reach a definitive conclusion on which one is most efficient and appropriate under certain conditions and scenario. The authors conclude that the lack of studies that may have comparable methodologies is a major factor for this absence of definition on the subject, and propose a methodology for future studies on the comparison of efficiency between DC and AC grids.

A numerical example can be used to verify the large variation to which an energy efficiency study is subject, as set out in Dastgeer et al. (2019). Four hypothetical scenarios, A, B, C, and D, of residential consumers are considered, with monthly AC consumption as shown in Figure 1.11. The consumptions adopted follow the references presented in ANEEL (2019).

**Figure 1.11 – Scenarios of monthly energy consumption of a residence.**

| SCENARIO A | | |
|---|---|---|
| Equipment | Un. | E_AC(kWh) |
| TV 29" | 1 | 15.3 |
| PC(CPU+Monitor) | 1 | 27.0 |
| Phone | 1 | 5.4 |
| AC 10000 BTU | 1 | 186 |
| LED bulb 8W | 4 | 7.68 |
| Refrigerator + Freezer 350L | 1 | 53.1 |
| | TOTAL= | 294.48 |

| SCENARIO B | | |
|---|---|---|
| Equipment | Un. | E_AC(kWh) |
| TV 29" | 1 | 15.3 |
| PC(CPU+Monitor) | 1 | 27.0 |
| Phone | 2 | 10.8 |
| AC 10000 BTU | 2 | 372 |
| LED bulb 8W | 4 | 7.68 |
| Refrigerator + Freezer 350L | 1 | 53.1 |
| | TOTAL= | 485.88 |

| SCENARIO C | | |
|---|---|---|
| Equipment | Un. | E_AC(kWh) |
| TV 29" | 2 | 30.6 |
| PC(CPU+Monitor) | 2 | 54.0 |
| Phone | 3 | 16.2 |
| AC Inverter 10000 BTU | 2 | 260.4 |
| LED bulb 8W | 4 | 7.68 |
| Refrigerator + Freezer 350L | 1 | 53.1 |
| | TOTAL= | 421.98 |

| SCENARIO D | | |
|---|---|---|
| Equipment | Un. | E_AC(kWh) |
| TV 29" | 1 | 15.3 |
| PC(CPU+Monitor) | 1 | 27.0 |
| Phone | 1 | 5.4 |
| AC Inverter 10000 BTU | 1 | 130.2 |
| LED bulb 8W | 4 | 7.68 |
| Refrigerator + Freezer 350L | 1 | 53.1 |
| | TOTAL= | 238.68 |

To evaluate the equivalent consumption for the same equipment in a DC distribution grid, the involved DC/AC, AC/DC and DC/DC conversion efficiencies should be verified. Electronic devices (TV, computer, mobile phone charger and LED lamp) operate on direct current, requiring an AC/DC converter to operate on an AC network. These converters consist of two stages: a rectifier and a DC voltage regulator (usually step down). Equipment such as refrigerators and air conditioners operate directly connected to the AC bus, except for the inverter type air conditioner, which operates at variable speed and is connected to the AC mains via an AC/DC/AC converter. This converter is also composed of two stages: a rectifier and an inverter, which will supply the motor of the air compressor with an adjustable frequency according to the cooling needs of the environment.

The maximum efficiency of an ideal full bridge rectifier (disregarding the diode voltage drop - or MOSFET - and the semiconductor internal resistance) is $\eta_{AC/DC} = 81.2\%$. In the stages for DC voltage regulation and inverter, equipment with about $\eta_{DC/DC}= \eta_{DC/AC} = 98\%$ efficiency can be found on the market (KADAVELUGU et al., 2017).

Therefore, for electronic equipment, the equivalent DC consumption, $E_{DC,e}$, is given by Equation (1.3). For refrigerator and air conditioning equipment, except inverter, there is an equivalent consumption $E_{DC,m}$ given by Equation (1.4). For inverter-type air conditioning, there is an equivalent consumption $E_{DC,i}$ given by Equation (1.5).

$$E_{c.c.,e} = \eta_{c.c-c.c.} \times \eta_{c.a-c.c.} \times E_{c.a.,e} \qquad (1.3)$$

$$E_{c.c.,m} = E_{c.a.,m}/\eta_{c.c-c.a.} \qquad (1.4)$$

$$E_{c.c.,i} = \eta_{c.c-c.c.} \times \eta_{c.a-c.c.} \times E_{c.a.,i} \qquad (1.5)$$

Where E$_{AC,e}$ and E$_{AC,m}$, and E$_{AC,i}$, correspond to the energy consumption of electronic equipment, equipment such as refrigerators and air conditioners, and inverter type air conditioner equipment, respectively. The graph in Figure 1.12 shows the comparison between monthly consumption for households in the four scenarios presented in Figure 1.11, considering consumption with AC power and equivalent consumption in DC.

**Figure 1.12 – Monthly consumption considering DC and AC distribution grids for each scenario.**

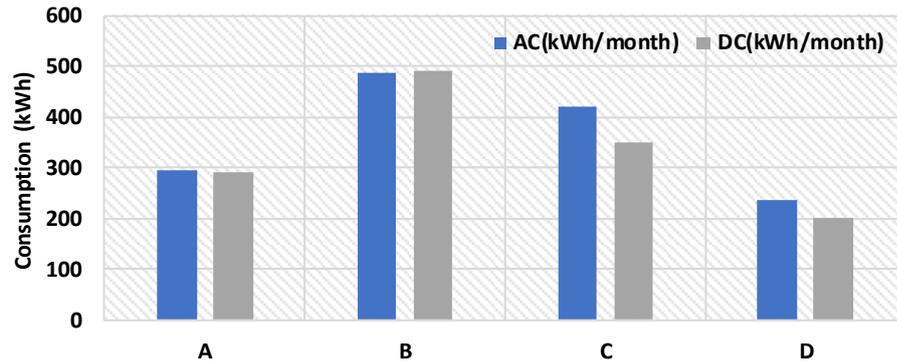

For scenario A, the final energy consumption is practically indifferent when considering DC or AC supply. In consumption scenario B, the DC power supply represents slightly higher consumption, while in scenarios C and D the consumption in DC is lower. For the scenarios in this example, the main factor is air conditioning consumption, so that in the scenarios where inverter equipment is used, there is a significant reduction when using direct coupling to the direct current network.

It is verified that with the few variables involved in this example, it is possible to obtain quite different consumption values, capable of justifying the adoption, or not, of a direct current distribution grid. Increasingly different results can be obtained considering scenarios with more diverse loads, presence of distributed generation and presence of energy storage system.

It is important to stress that other factors besides equipment energy consumption should be considered when justifying the implementation of a direct current system, such as: cost and availability of end-use appliances, cost and availability of power conditioning converters and the power dissipated in the system wiring.

The evaluation developed in this dissertation considered some of these aspects, such as wiring losses and the availability of commercial equipment.

# 2. DESCRIPTION OF THE EVALUATED DC DISTRIBUTION NANOGRID

## 2.1 Introduction

This chapter describes in detail all the DCDN equipment installed in the GEDAE-UFPA laboratory and how this equipment is connected to form the nanogrid.

The DCDN developed and evaluated in this work is formed by three load banks (LB) and three storage and generation systems (GSS). The grid is unipolar, 24 V rated voltage, ring architecture and insulated from the ground. The simplified DCDN single-line diagram is shown in figure 2.1, all voltage and current measurement points have been indicated in red.

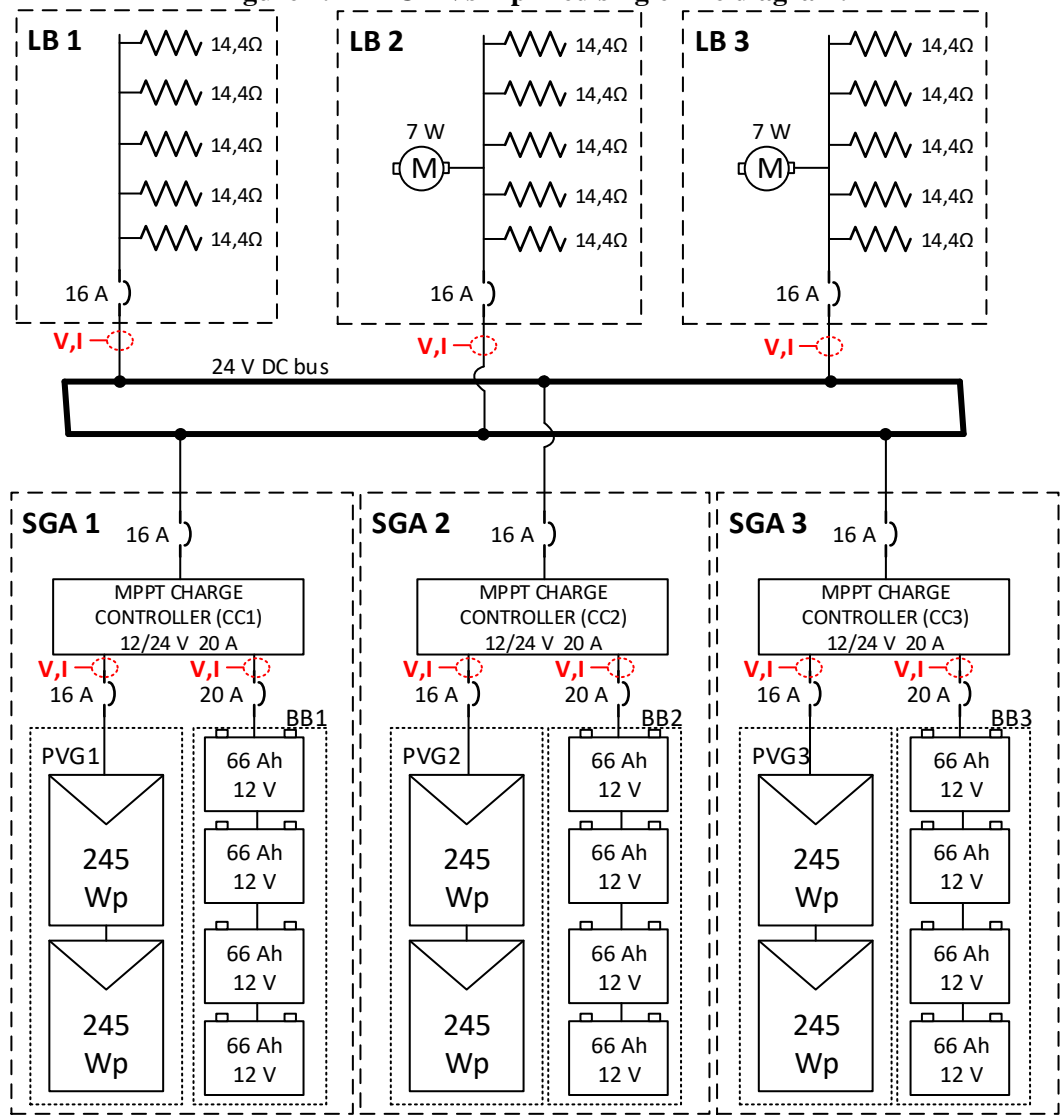

Figure 2.1 – DCDN simplified single-line diagram.

## 2.2 Generation and storage system (GSS)

The three distributed generation and storage systems at the DCDN are the only sources of electricity in the grid and are responsible for serving the load banks. The GSS should be sized to ensure continuity of supply even on days with reduced solar resource availability. Each GSS consists of a photovoltaic generator, a battery bank, a charge controller and protective devices.

### 2.2.1. System sizing

The sizing of the photovoltaic generator and the battery bank in each GSS was performed following the design methodology for standalone photovoltaic systems. For this, it is necessary to access the loads to be met by the DCDN, as well as the operating period of each load.

In this sense, it was adopted as consumption reference a small residential consumer with the following loads, all in direct current: a freezer, lamps, mobile phone charger, laptop charger, and fan. The reference load curve was chosen based on a common equipment hourly usage profile and is depicted in Figure 2.2. For this load curve, there is a daily consumption of $L_{DC}$ = 1.63 kWh and a maximum demand of approximately 200 W. Therefore, for the reference curve, the maximum current demanded by the system will be approximately 25 A, considering the three LB operation (approximately 8.33 A per LB in a balanced load distribution condition).

**Figure 2.2 – Reference load curve per load bank for system sizing.**

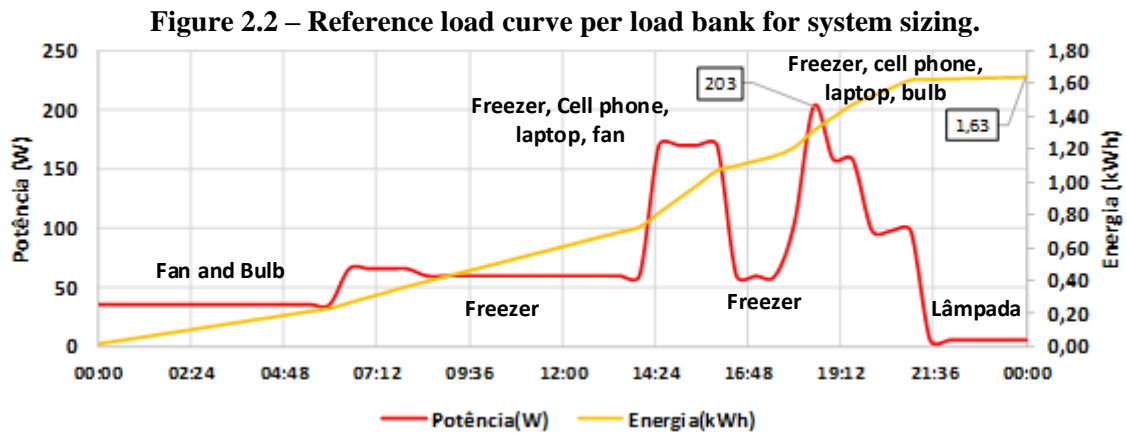

From the daily consumption of DC loads, the corrected consumption shall be calculated considering the battery bank charge and discharge efficiency, $\eta_{cdc}$. The calculation of corrected daily consumption, $L_{DC,cor}$, is expressed by Equation (2.1).

$$L_{DC,cor} = L_{DC}/\eta_{cdc} \qquad (2.1)$$

The typical value used for charge and discharge efficiency of the lead acid battery is $\eta_{cdc}$ = 0.86 (PINHO; GALDINO, 2014). Therefore, there is a corrected daily consumption of $L_{DC,cor}$ = 1.92 kWh for each load bank.

Based on corrected daily consumption, the battery bank should be sized considering the bank's autonomy time, usually expressed in number of days, $N_D$. Bank autonomy can be chosen arbitrarily or through equations that consider the solar resource available throughout the year and other requirements for continuity of load supply (e.g. servicing critical hospital loads). For the DCDN sizing, Equation (2.2) was used to calculate the number of autonomy days, which was obtained empirically by Messenger and Ventre (2003), to meet non-critical loads and considers solar resource availability in the most critical month for the system.

$$N_D = -0.48 \times SLH_{min} + 4.58 \qquad (2.2)$$

$SLH_{min}$ corresponds to the equivalent number of sunlight hours with an irradiance of 1000 W/m$^2$, in other words, it is equivalent to the energy, given in kWh/m$^2$, supplied by the Sun on the average day of the month with the lowest solar resource availability. In the month of worst solar resource availability at the DCDN location, $SLH_{min}$ = 4.2 kWh/m$^2$ (PEREIRA et al., 2017). Therefore, $N_D$ = 2.56.

The battery bank capacity, $C^*_B$, in Wh, is obtained from Equation (2.3), where $P_{D,MAX}$ is the maximum bank depth of discharge.

$$C^*_B = L_{DC,cor} \times N_D / P_{D,MAX} \qquad (2.3)$$

The maximum depth of discharge of the battery bank is specified for each battery type. For deep-cycle batteries used in standalone PV applications, it is considered $P_{D,MAX}$ = 80 %. This value corresponds to the battery bank voltage level at which the charge controller will disconnect the loads, preventing deep discharge of the batteries. Therefore, $C^*_B$ = 6.14 kWh. The bank capacity value in Ampere-hours is obtained by dividing the result in kWh by the rated bank voltage (in the case of the DCDN, 24 V). Therefore, the bank capacity, in Ah, must be $C_{rated}$ = 256 Ah. This capacity shall be obtained in a discharge rate ranging from 10 h to 20 h, compatible with the charge and discharge cycle of an autonomous photovoltaic system.

The power of the photovoltaic generator should be sized considering the corrected daily consumption and solar resource in the period of the year with the lowest availability. This type of sizing ensures that the system will generate enough energy throughout the year, however, in months of higher irradiance, some of the available solar energy will not be harnessed. The rated power of the PV generator is calculated by Equation (2.4), where $k_{ADJ}$ = 1.25 is a safety factor that considers the effects of temperature losses, dispersion losses, dirt accumulation and cabling losses. Therefore, the PV generator must have a rated power of 570 Wp.

$$P_{NOM}(Wp) = k_{AJS} * L_{DC,cor}/HSP_{min} \qquad (2.4)$$

The charge controller shall be compatible with the rated power and voltage of the PV generator, the battery bank voltage and the maximum current demanded by the loads. For example, for this dimensioning, a charge controller for PV generator up to 600 Wp, battery bank voltage at 24 V and maximum load current of 25 A would be enough (considering a contingency condition where only one GSS should meet all the loads).

### 2.2.2. Photovoltaic Generator (PVG)

The photovoltaic generator present in each GSS is formed by the series combination of two PV modules from manufacturer *Yingli Solar*, model YL245P-29b of 245 Wp rated power (YINGLI, 2014). The main electrical characteristics under the standard test conditions (STC: PV cell temperature, $T_{c,STC}$ = 25 °C, global irradiance on the PV generator plane, $G_{i,STC}$ = 1000 W/m² and air mass, AM = 1.5) and thermal coefficients of the modules used are presented in Table 2.1. The values given in this table were obtained for a single module test from a solar simulator of PASAN MEASUREMENT SYSTEMS model HighLight 3c, with accuracy class A+A+A+ (the complete report of this module curve IV test can be seen in Annex I).

It is important to highlight that this module model was chosen due to its availability in the laboratory, although its nominal power is about 15% lower than the calculated power to supply the energy requirement of the system, as verified in the sizing

Table 2.1 - Technical characteristics under the STC of the PV module used.

| | | | |
|---|---|---|---|
| **Rated power (Wp)** | 238.25 | **Nominal operating cell temperature – NOCT (°C)** | 46 |
| **Maximum power voltage ($V_{MP}$)** | 29.22 | **Power thermal coefficient (%/°C)** | -0.45 |
| **Maximum power current ($I_{MP}$)** | 8.15 | **Short circuit current termal coefficient (%/°C)** | 0.06 |
| **Open circuit voltage ($V_{OC}$)** | 37.21 | **Open circuit voltage thermal coefficient (%/°C)** | -0.33 |
| **Short circuit current ($I_{SC}$)** | 8.76 | **Maximum power voltage thermal coefficient (%/°C)** | -0.45 |

The three generators are positioned on the test roof located outside the GEDAE's laboratory. The orientation of all generators is to magnetic north (azimuth deviation equal to 13° northeast) and with tilt angle of 11° to the horizontal plane. Figure 2.3 shows the positioning of each PV generator on the roof.

**Figure 2.3 – Placement of PV modules on the test roof outside the laboratory.**

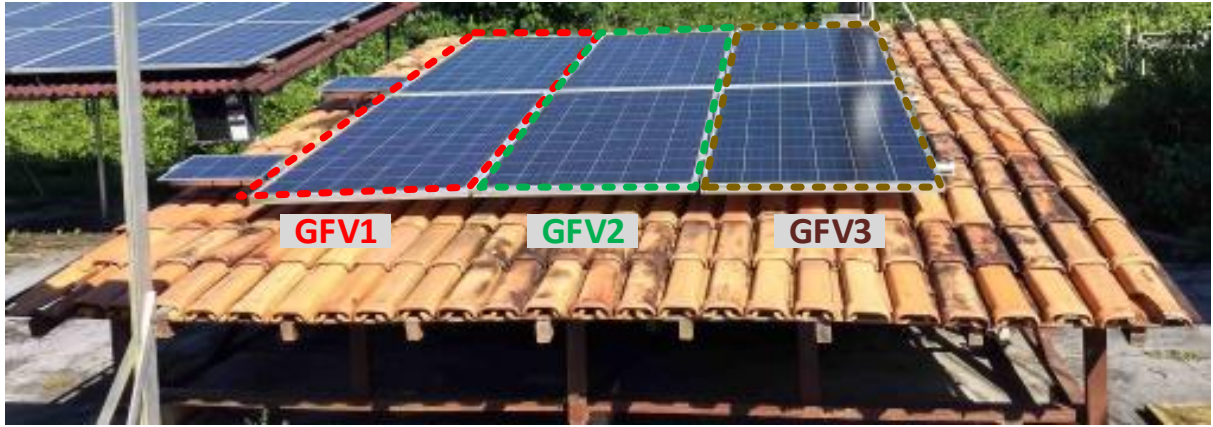

Considering the values presented in Table 2.1, it can be inferred that each PVG has electrical characteristics according to Table 2.2.

**Table 2.2- Electrical characteristics under the STC of each PVG.**

| | |
|---|---|
| **Rated power (Wp)** | 476.5 |
| **Maximum power voltage ($V_{MP}$)** | 58.44 |
| **Maximum power current ($I_{MP}$)** | 8.15 |
| **Open circuit voltage ($V_{OC}$)** | 74.42 |
| **Short circuit current ($I_{SC}$)** | 8.76 |

### 2.2.3. Battery bank (BB)

Each GSS has a sealed lead-acid battery bank formed by the series-parallel association of four batteries. The battery model used to compose the bank is BLUETOP D27M, from the manufacturer Optima Batteries. Each battery has a nominal voltage of 12 V and a nominal capacity of 66 Ah, considering a constant current discharge rating of 3.3 A (C/20 regime). Therefore, each battery bank has 24 V rated voltage and capacity of 132 Ah (C/20).

This battery is designed for use in boats or other marine applications, given its high vibration resistance. According to the manufacturer's catalog, this battery model can be used in two types of applications: electric starters and deep cycle regimes. In relation to starters, the battery is capable of supplying a starting current of 1000 A at a temperature of 0 °C and 800 A at a temperature of -17 °C[1] (OPTIMA BATTERIES, 2014). For battery characterization in deep cycle applications, the manufacturer uses a figure called reserve capacity (RC), which corresponds to the time, usually in minutes, that a fully charged battery takes to reach 10.5 V under a constant 25 A discharge. In this sense, the manufacturer guarantees a nominal RC of

---

[1] In starting applications requiring high current in short time intervals, it is usual to characterize two quantities: CA (Cranking amps) and CCA (Cold cranking amps) which correspond to the starting current supply capacity at 0 ° C and -17 °C, respectively.

140 minutes, which reduces as the number of charge and discharge cycles increases, as shown in Figure 2.4

**Figure 2.4 - Comparison of reserve capacity as a function of the number of charge and discharge cycles between the used battery and a conventional lead acid battery.**

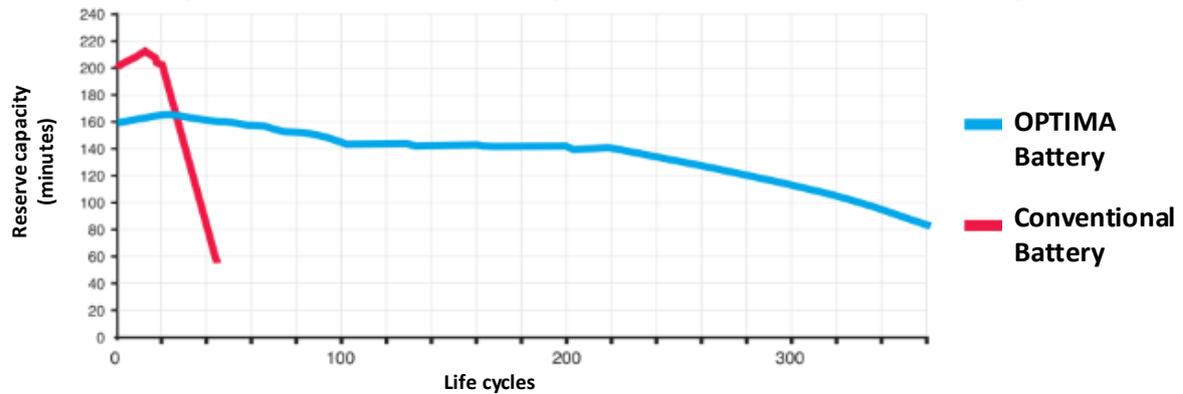

**Source: adapted from (OPTIMA BATTERIES, 2014).**

RC is not the most appropriate parameter for evaluating battery performance on a daily charge and discharge regime, as a constant 25 A discharge is far from reality in the DCDN application. However, it can be considered that in lower current discharge regimes the behavior is equivalent and, by analyzing figure 2.4, it is clear that this battery is able to maintain its capacity in deep cycle regimes for more cycles than other conventional batteries.

To charge this battery in cyclic applications, the manufacturer recommends a 14.7 V constant voltage charging and no current limit as long as the battery temperature remains below 51.7 °C. When a current lower than 1 A is reached, a 2 A constant current charge shall be used for 1 hour, by increasing the charging voltage. To maintain the battery charged after reaching full charge, it's recommended to keep a maximum 1 A charging current indefinitely, by applying a voltage from 13.2 V to 13.8 V. It is important to follow the charging instructions provided by the manufacturer to ensure longer battery life, however this is not always possible, especially in applications where the power supply is intermittent, such as PV.

It is noteworthy that this battery model was chosen to compose the DCDN GSS given its availability in the laboratory and also the excellent RC presented in relation to conventional batteries. However, the batteries used are not new and had been previously used in another application, going through situations of high stress, such as long time without being charged, shelter in excessively humid and wet environment and subjected to deep discharge. These situations compromise battery performance, reduce battery efficiency and charge-retention capacity. In addition, the capacity of each bank is approximately half of the previously sized value (256 Ah).

To evaluate the current state of each battery bank used, charge and discharge tests were performed in the laboratory bench to assess the capacity, in Ah, of each battery bank for different discharge regimes. The instrumentation and configuration used in the tests are presented in Figure 2.5. The procedure adopted in the tests was based on the battery capacity testing methodology proposed in (INMETRO, 2011), and is indicated in the flowchart illustrated in Figure 2.6.

**Figure 2.5 - Diagram of connections and measurements used for testing each battery bank.**

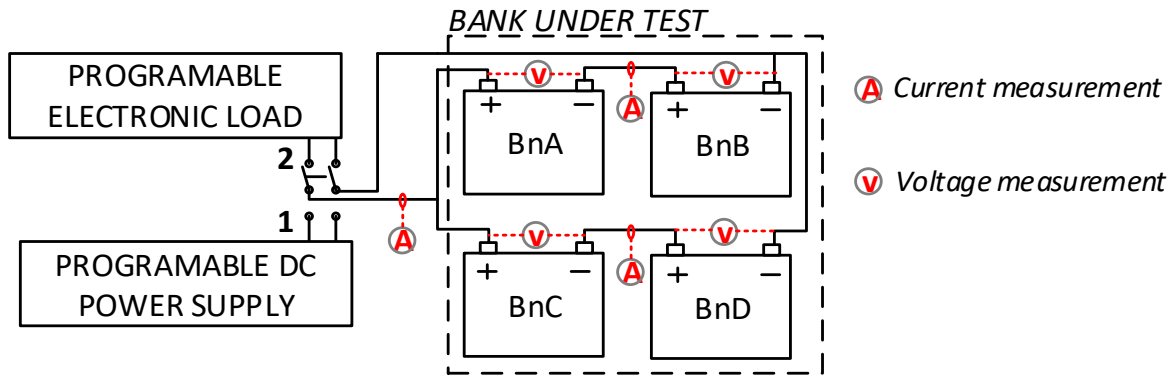

**Figure 2.6 - Procedure for charge and discharge tests.**

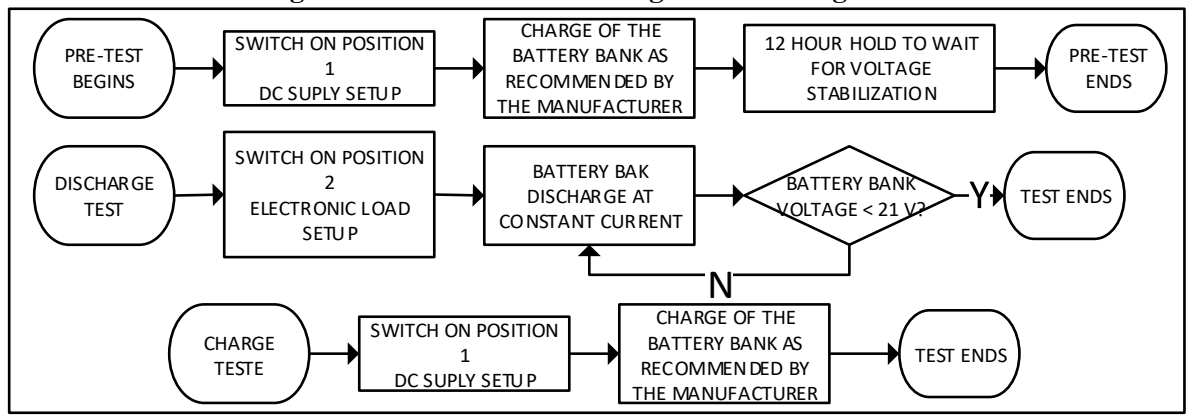

In Figure 2.5, $n = 1, 2$ or $3$ and indicates the battery bank under test. The electronic load used is from *Chroma*, model *63803* up to 3.6 kW and can operate as a DC load with constant current, constant resistance, constant power or constant voltage. The DC source used is from the manufacturer XANTREX, model XDC 60-100, up to 60 V and 100 A, which can operate as a voltage or current source. For measurement, the Chroma Model 66204 digital power meter was used, which has 8 independent input channels, 4 current and 4 voltage, so it is possible to measure the individual voltage of each bank battery and current in each string in parallel.

Before the test, the battery banks were pre-charged to ensure that the discharge test starts with the banks fully charged. All tests were performed in a refrigerated environment to minimize the temperature variation effect on banks performance during the assessment. Figure 2.7 shows the discharge curves for the three battery banks, for tests performed with a 10 A discharge current. $V_A$, $V_B$, $V_C$ e $V_D$ correspond to the voltages of each battery that composes the bank, as shown in Figure 2.5; $I_{a,b}$ and $I_{c,d}$ correspond to the currents in each bank's parallel string. Figure 2.8 presents a comparison of the discharge test results considering 10 A and 6.6 A constant currents. The 6.6 A discharge test sought to evaluate the 132 Ah rated capacity of each battery bank under a discharge of 20 h obtained at a steady current of 6.6 A.

**Figure 2.7 - Battery bank discharge tests for a 10 A constant current (a) BB1, (b) BB2 and (c) BB3.**

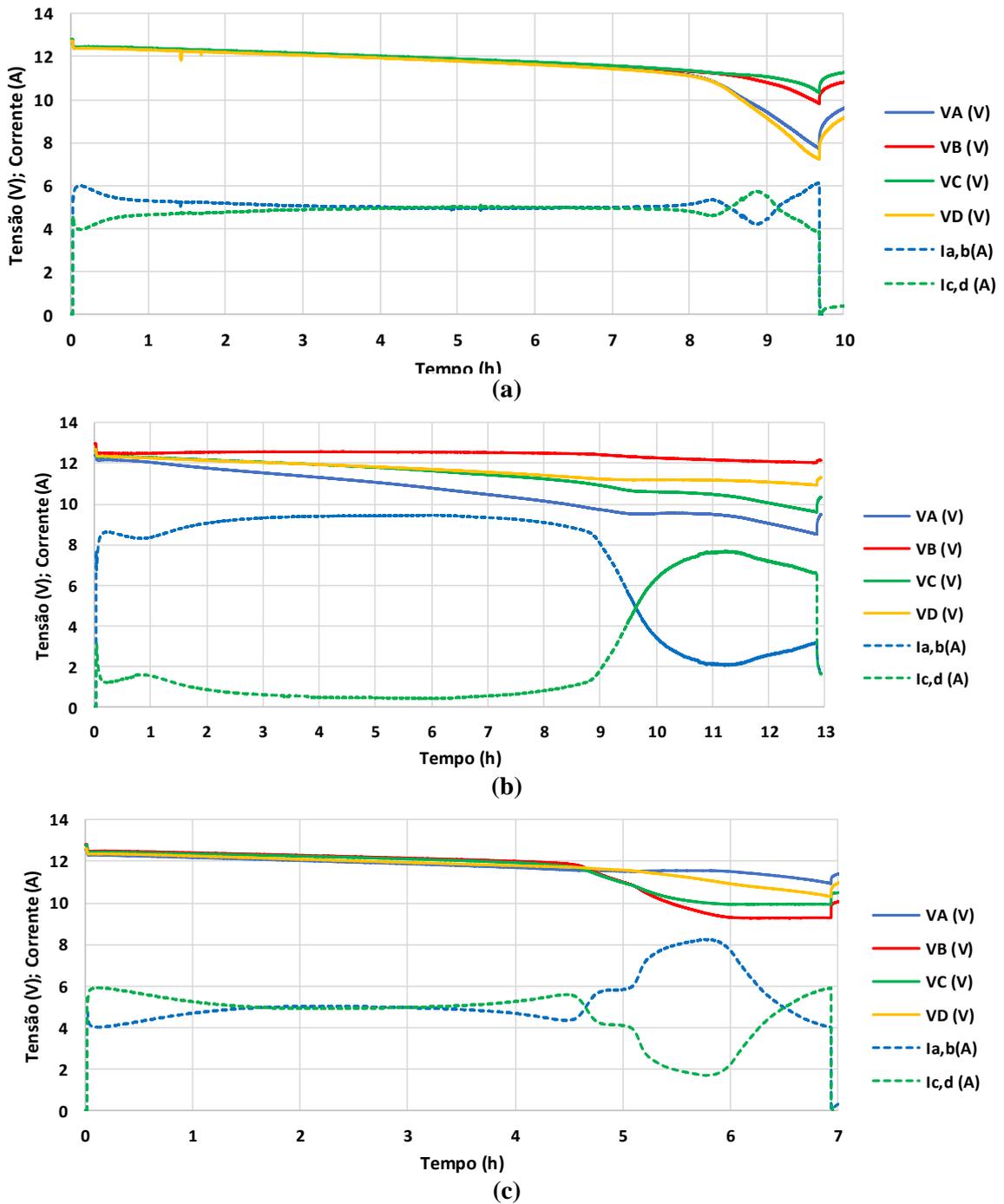

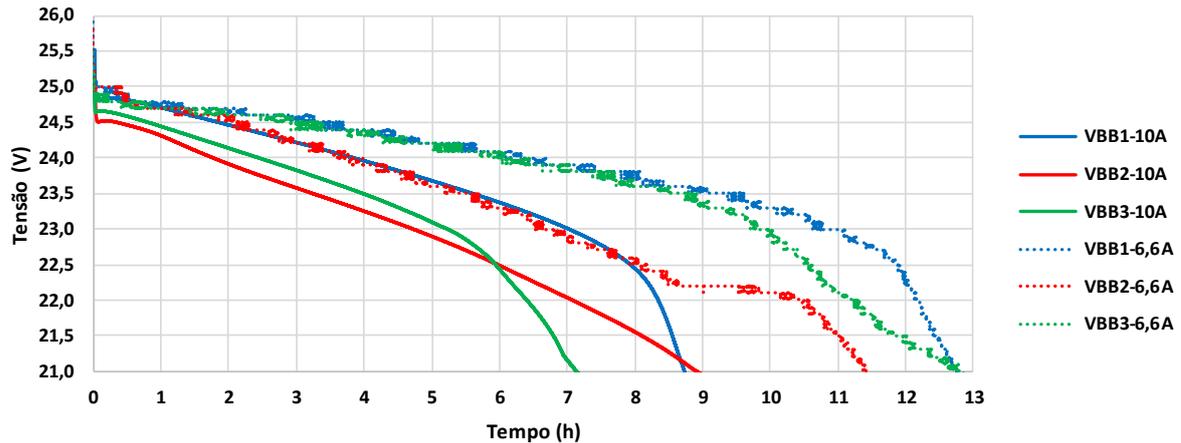

**Figure 2.8 - Battery bank discharge tests for 10 A and 6.6 A constant currents.**

Table 2.3 summarizes the capacities of each bank obtained in the tests.

**Table 2.3 - Capacity (Ah) of each battery bank for 6.6 and 10 A discharge tests.**

| BB | Test 1: I = 6,6 A | Teste : I = 10 A |
|---|---|---|
| | Capacity (Ah) | Capacity (Ah) |
| **BB1** | 84,2 | 87,3 |
| **BB2** | 75,3 | 89,0 |
| **BB3** | 84,9 | 71,3 |

It is noticed that the batteries in the same bank discharge differently, so that one battery reaches the exponential zone before the others. This is an indication that the batteries that composes each bank do not have the same capacity, so the battery of smaller capacity limits the bank as a whole. This is especially evident in Figure 2.7b and Figure 2.7c, where it is possible to see the difference in discharge current between the bank strings. When a battery reaches a high state of charge, the tendency is to increase its resistance to charge reception, so that the string resistance tends to increase and, therefore, reduce the load current in the string.

According to the results presented in Table 2.3, it is clear that for BB1 and BB2 the capacity increases with the discharge current increase, which is an inconsistent behavior for batteries. Again, this uneven performance is explained by the difference in the batteries capacities within a bank, so with the series-parallel association, it cannot be guaranteed that all batteries are reaching full charge or discharging equivalently. In addition, in all tests, banks capacities were much lower than the rated 132 Ah for a 6.6 A test, which proves the batteries deteriorated condition.

With the data obtained from this test, it is possible to group the batteries together to form banks that use only batteries with capacity as close as possible to optimize the use of the available capacity in each battery. In this master's work, it was chosen to keep the distribution of batteries in the GSS's banks according to the tests presented in this section.

The photograph shown in Figure 2.9 presents the battery banks of each GSS that composes the DCDN. The batteries must be sheltered from sun and rain, and in a ventilated environment. For this purpose, it was built a mobile stand to position the benches under the PV modules test roof. Electrical connections should be protected to prevent oxidation, because they are relatively exposed and in a humid environment. To perform this protection, grease lubricant must be applied periodically.

Figure 2.9 – GSS's battery banks.

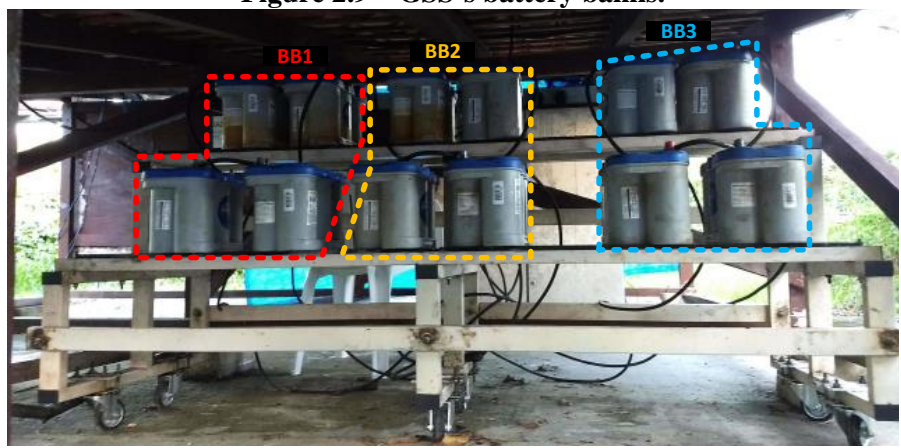

### 2.2.4. Charge controller

Each GSS contains a charge controller to regulate the charge and discharge of battery banks and maximize the use of photovoltaic energy. The charge controller used is SCCM20-100R from Outback Power. This controller operates with a maximum charge and discharge current of 20 A and can work with 12 V or 24 V battery banks. The recommended maximum power of the PV generator is 300 Wp to work with 12 V banks and 600 Wp for 24V banks. The controller features the PV generator's maximum power point tracking (MPPT) function, ensuring that the maximum available power is delivered for a given irradiance and temperature condition. Table 2.4 presents the main technical characteristics of the charge controller used.

Table 2.4 – Technical specifications for the charge controller *SCCM20-100R*.

| Section | Parameter | | 12V | | 24V | |
|---|---|---|---|---|---|---|
| **PV** | Maximum power | | 300 W | | 600 W | |
| | Maximum DC voltage | | 100 V (off); 90 V (operating) | | | |
| | Open circuit voltage | | 20 V – 100 V | | | |
| | Maximum power voltage | | 15 V – 75 V | | 27 V – 75 V | |
| | Maximum short circuit current | | 20 A | | | |
| **Battery** | Input voltage | | 8 V – 15 V | | 16 V – 30 V | |
| | Standby mode current | | < 37 mA | | < 26 mA | |
| | Capacity | | 100 Ah minimal @ C/5 | | | |
| **Charger** | Charging voltage | Stage | Flooded | VRLA | Flooded | VRLA |
| | | Bulk | 14,8 V | 14,6 V | 29,6 V | 29,2 V |
| | | Absorb | 14,8 V | 14,4 V | 29,6 V | 28,8 V |
| | Absorption stage time | | 2 hours | | | |

|              | Floating stage voltage | | 13,2 V | 13,5 V | 26,4 V | 27,0 V |
|---|---|---|---|---|---|---|
|              | Maximum charging current | | 20 A | | | |
|              | Temperature voltage compensation range | | + 30 mV/°C / 12 V de 0 °C a 24 °C | | | |
|              | | | - 30 mV/°C / 12 V de 26 °C a 60 °C | | | |
|              | Battery temperature sensor | | NTC 10K | | | |
| **Equalization** | Note: equalization occurs every 60 days or after a battery low voltage disconnection | | | | | |
|              | Maximum equalization voltage | | 15,5 V | -- | 31,0 V | -- |
|              | Equalization time | | 1 hour | | | |
| **Internal protections** | Internal protection | Electronical | PV and battery polarity reversion | | | |
|              | | Passive | 30 A fuse | | | |
|              | Maximum load output current | | 20 A | | | |
|              | Low-voltage disconnection | | 11,4 V | | 22,8 V | |
|              | Reconnection voltage | | 12,4 V | | 24,8 V | |
|              | Over voltage disconnection | | 15,0 V | | 30,0 V | |
| **Efficiency** | MPPT | | > 99% over 100 W/m² and with a minimal of 10 % PV rated power | | | |
|              | Conversion | | Up to 98,5 % | | | |

This controller also features an RS-485 communication interface that allows the configuration of various parameters such as battery bank's connection and reconnection voltages and monitoring of the following variables: PV generator voltage and current, battery bank voltage and current, load current, internal temperature, battery bank temperature (external NTC sensor), and charging stage. This interface can be used by the DCDN data acquisition system.

The charge controller regulates the charging of the battery bank in 3 stages: constant current (also known as bulk), constant voltage (also known as absorption stage) and fluctuation.

The first stage is constant current and occurs when the battery is under 80 % of rated capacity. At this stage, the battery receive charge more easily, and gradually increase the voltage at its terminals as it charges. In PV source charging applications, the name "constant current charging" is not appropriate as the charging current is a function of solar resource availability and may vary during the process. In addition, this charge controller limits the maximum charging current to 20 A on the low-voltage side. This first stage is responsible for restoring most of the charge of a discharged battery, and its duration varies depending on the availability of the solar resource, the installed PVG power, and the charge profile that the system supplies.

In the second stage, the battery has already recovered most of its charge, presenting greater resistance to charging. At this stage the charge controller regulates the voltage, keeping it constant throughout the stage; The charging current, on the other hand, naturally decreases as charge is absorbed. This stage is also known as the absorption stage, because the charging current is limited so that the battery can absorb the remaining charge and restore the remaining 20% of its capacity.

When the battery bank reaches full charge, the controller enters the floating stage. In this stage the controller also regulates the voltage of the battery bank, provided that the voltage of the PV generator is higher than that of the bank; The small current that is injected into the bank aims to keep the battery charged, reducing the phenomenon of self-discharge in the batteries.

The constant voltage values that the charge controller regulates in the absorption and fluctuation stages can be adjusted through the communication interface of the equipment. In this regard, the values provided by the battery manufacturer should be used, which are: 29.4 V (2 x 14.7 V) for the absorption stage and 26.4 V (2 x 13.2 V) for the floating stage.

Figure 2.10 was obtained from a real operating test at the DCDN to charge the GSS3 battery bank, performed on March 14 and 15, 2019, and illustrates the behavior of the charging voltage and current at each of the stages. It is noted that the initial bank voltage was approximately 22 V, indicative of the low accumulated charge. Beginning the charging by the Bulk stage, it is noticed that the bank voltage increases gradually, being influenced by the charging current, which is a function of solar resource, justifying the stochastic variations over time.

In addition, when analyzing the current waveform in Figure 2.10, one can notice variations that even zero the battery charging current at well-defined intervals that correspond to the PVG operating in the open circuit. This phenomenon occurs due to the controller MPPT algorithm, which periodically sweeps the PVG IxV curve in search for the maximum global power point. This scan is important because it reduces the occurrence of local maximum points, which may arise under shading conditions as depicted in Rodrigues (2017). Through DCDN measurements in operation, it was found that this scan occurs every 30 minutes and lasts about 1 sec.

**Figure 2.10 – Irradiance, PVG3 voltage and GSS3 battery bank voltage and current during charging test.**

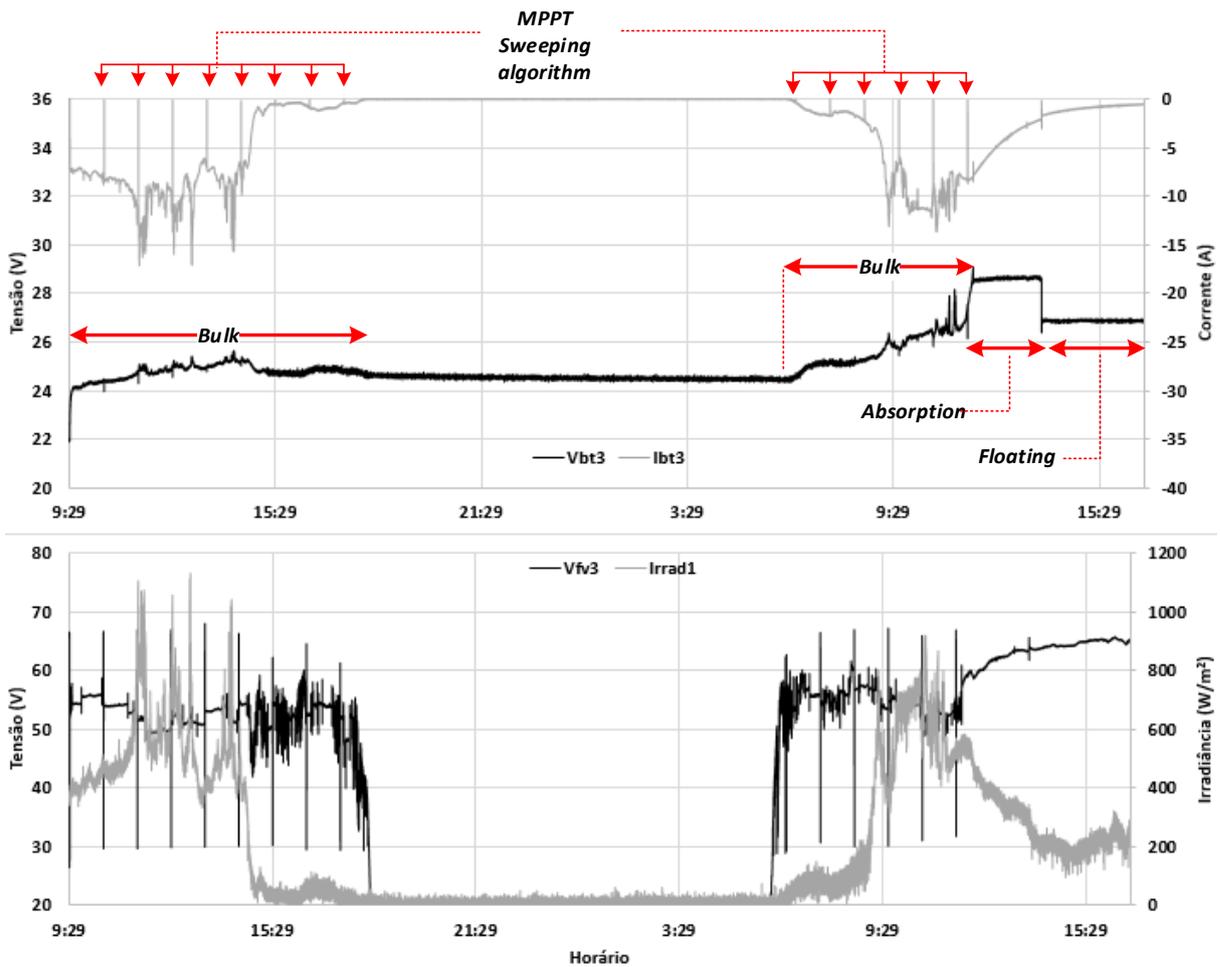

In addition to the 3 charging stages, the charge controller also periodically equalizes the battery bank to reduce the disparity in charge state of each battery that composes the bank. Equalization is performed every 60 days or after disconnection of the charge controller under battery bank undervoltage condition. The procedure adopted by the controller is to set the bank voltage to 31 V (for banks rated at 24 V) for 1 hour.

The charge controller is also responsible for preventing the battery bank from being subjected to a deep discharge as this type of occurrence shortens the battery life and may even lead to an unrecoverable discharge. Thus, the controller is set to cut-out the loads if the battery bank voltage decreases to a preset value, $V_{DESC}$. When this occurs, the loads will only be enabled again when the battery bank voltage is greater than a predetermined reconnect value, $V_{REC}$. Figure 2.11 presents actual DCDN operating data demonstrating the occurrence of load cut-out by the controller.

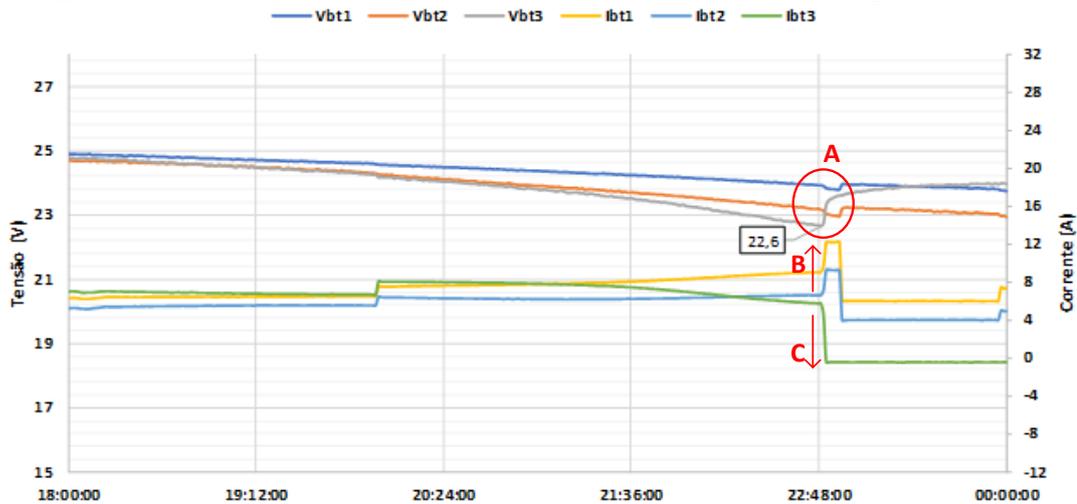

**Figure 2.11 – Voltage and Current at DCDN's battery banks during load cut-out.**

In the situation shown in Figure 2.11, when battery bank 3 reaches the limit value of 22.6 V, the controller associated with this bank cuts-out its load terminal from the grid and, thus, events A, B and C happen. BB 3 voltage rises (event A) since there are no more loading in this bank and its current decreases to zero (event C); the other banks 1 and 2 increase the current (event B) to supply the loads once bank 3 has stopped contributing; As a result, the voltages in banks 1 and 2 fall more sharply (event A). Immediately thereafter, the system load decreases (acting independently of any load cut-out by the controller) and the currents in banks 1 and 2 reduce.

### 2.2.5. Protection devices

Each GSS has two layers of protection: external protection through DC circuit breakers and internal charge controller protection. As shown in Table 2.4, each charge controller has the following protections: against reverse polarity at the PVG input and at the BB input; against overcurrent there are internal 30-amp fuses at the PVG and battery bank terminals, as well as fast overcurrent electronic protection at the load output terminal.

DC circuit breakers are used at the three charge controller connection points' of each GSS (PVG, BB and load output) for overcurrent and short circuit protection. The sizing of the circuit breakers has taken into consideration the maximum allowable current values of the charge controller. The circuit breakers used in GSS1 and GSS2 are from Tongou manufacturer, have type-C breakdown curve and maximum operating voltage up to 1 kV. The circuit breakers used in GSS3 are from manufacturer TOMZN, also have type-C breakdown curve and maximum operating voltage of 440 V. All circuit breakers used are capable of acting in both current directions, for example, they actuate in case of short circuit upstream or downstream of the circuit breaker, provided that the short supply source is in the opposite position (downstream or upstream respectively). It was chosen to select circuit breakers of two different manufacturers for simple performance evaluation only. Table 2.5 provides a summary of the rated characteristics of the circuit breakers used, and Figure 2.12 illustrates the location and type of internal protection in the GSSs.

Table 2.5 – Technical specifications of DC breakers for DCDN's GSS protection.

| GSS | Manufacturer | Localization | Rated current | Number of poles | Insulation voltage | Rupture curve | Maximum breaking current |
|---|---|---|---|---|---|---|---|
| 1 e 2 | TONGOU | PVG | 16 A | 2P | 1000 V | C | 6 kA |
| | | BATTERY BANK | 20 A | | | | |
| | | LOAD OUTPUT | 20 A | | | | |
| 3 | TOMZN | PVG | 16 A | 2P | 440 V | C | 6 kA |
| | | BATTERY BANK | 20 A | | | | |
| | | LOAD OUTPUT | 20 A | | | | |

Figure 2.12 - Illustrative diagram of the protections in each GSS. In red, surge protection devices not yet installed.

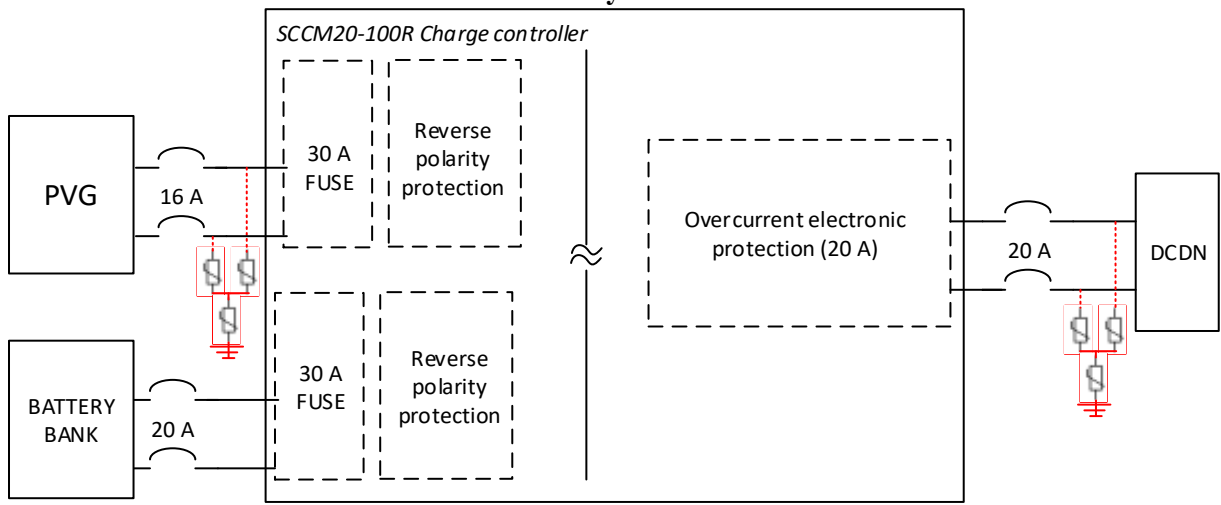

The external devices used are for overcurrent protection only (either overcharge or short circuit). However, it is important to have surge protection devices. Voltage surges may occur mainly due to lightning strikes that may directly strike the DCDN or induce conductor overvoltage. In each GSS, the main device to be protected against voltage surges is the charge controller, which is subject to surges from the distribution grid and the PVG. According to the manufacturer, the maximum working voltage is 100 V. This value can easily be reached during a surge, making the protection of this device critical.

For this type of protection, DC surge protection devices (DC-SPD) should be implemented at the output terminals for the load and at the PVG connection point. The SPD must be sized considering the following factors: the maximum operating voltage of the device to be protected, in this case 100 V; the resistivity voltage of the equipment; the nominal current of the SPD, the SPD operating class and the maximum discharge current. The SPD works by providing a low impedance path to ground in the event of surges. To prevent over voltages at its terminals, it must be connected to ground for the flow of charge.

It is suggested as SPD to be installed the model VCL 75V 12.5 / 60kA Slim from manufacturer CLAMPER or with similar characteristics. This SPD has the maximum continuous operating DC voltage of 100 V, nominal and maximum discharge current of 30 kA and 60 kA respectively and protection level of 1.3 kV.

External surge protection has not yet been installed at NDCC due to its high cost. For comparative purposes, the local market retail value of all external protection devices to be used in a single GSS (three bipolar DC breaker and two DC SPD) is about 70 % of the charge controller cost.[2]

## 2.3 Load bank (LB)

The power consumption at the DCDN is realized by incandescent lamps and fans that are distributed in three independent load banks. The loads were chosen to allow the implementation of load curves with daily consumption equivalent to the reference values adopted by the power distribution utilities in the context of isolated power generation and distribution microsystems (MIGDI), as defined by normative resolution 493/2012 (ANEEL, 2012). In this resolution, it is intended to serve consumer units with daily reference consumption of 435 Wh, 670 Wh, 1000 Wh, 1500 Wh, 2000 Wh and 2650 Wh.

The loads used in each LB are listed in Table 2.6. LB 1 and 2 can demand up to 206 W, in 40 W and 6 W steps. BC3 can demand up to 200 W, with 40 W steps. With this demand, banks 1 and 2 can consume up to 4.94 kWh/day, while the BC3, 4.8 kWh/day.

Table 2.6 – DCDN's loads technical specifications.

| LB | Load | Load type | N. Units | Rated voltage | Rated power |
|---|---|---|---|---|---|
| 1 e | Incandescent bulb | Constant impedance | 5 | 24 V | 40 W |
| 2 | Fan | Constant impedance | 1 | 24 V | 6 W |
| 3 | Incandescent bulb | Constant impedance | 5 | 24 V | 40 W |

All loads from each LB are connected in parallel and can be driven individually by relays and series connected on/off switches. In addition, each LB has a general DC breaker for protection and sectioning. Figure 2.13 illustrates the connections in each LB. In Figure 2.13, each normally open relay contact is represented by Knx, where *n* is the load bank number and *x* is the load connected in series. Snx represents the on/off hand switches in series with each lamp in the LB. The fan connection, represented by the DC motor, M, occurs only in LB2 and LB3. Figure 2.14 shows a photograph of the load bank installation at the DCDN.

---

[2] Costs as accessed on 24/07/2019: 20 A MPPT Charge controller BRL 362.00; C-curve 20 A bipolar DC breaker BRL 45.00; DC SPD BRL 120.00.

**Figure 2.13 – Illustrative diagram of the load bank.**

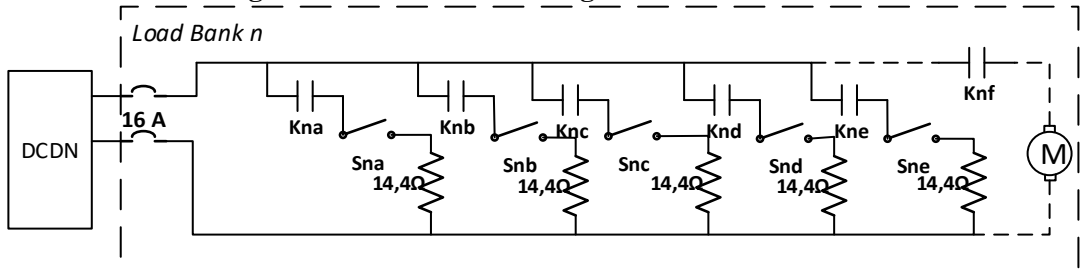

**Figure 2.14 – Loads installation: bulbs and fans.**

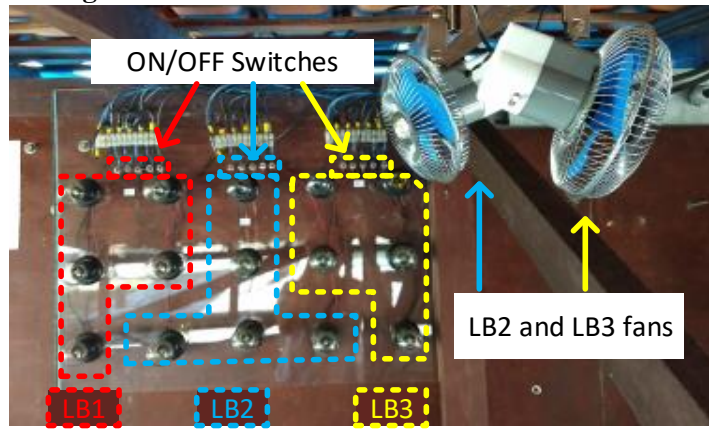

## 2.4 Distribution Grid

The DCDN GSSs and LBs are distributed in an approximately 200 m long grid located outside the GEDAE-UFPA laboratory. The distribution grid is supported by three poles that are shared with the AC grid (ACDM) also installed in the laboratory. The ring-architecture grid is made up of two-conductor aluminum twisted cables (V+ and V-) braided with a third bare cable for mechanical support. The conductors, shown in Figure 2.15, are from the national manufacturer Alubar and have 35 mm$^2$ cross-section with XLPE insulation for 0.6/1 kV voltage class. The technical data of this conductor is given in Table 2.7. This conductor was adopted considering the following factors: its availability in the laboratory, its resistance to UV radiation and its ability to conduct current and meet the loads.

**Table 2.7 – Technical specifications of the conductor.**

| Aluminum cable XLPE 0.6/ 1 kV - Alubar | | | |
|---|---|---|---|
| **Transversal section** | 35 mm$^2$ | **Minimal insulation thickness** | 1.6 mm |
| **String type** | Compacted | **Isolated conductor diameter** | 9.95 mm |
| **Insulation** | XLPE | **Number of wires** | 7 |
| **Temper** | H19 | **DC electrical resistance @ 20 °C** | 0.8037 Ω/km |

| Linear mass | 137.03 kg/km | Thermal resistance coefficient | 0.00403 °C$^{-1}$ |

Source: (ALUBAR, 2015).

**Figure 2.15 – Multiplexed aluminum conductor used in DCDN's distribution grid.**

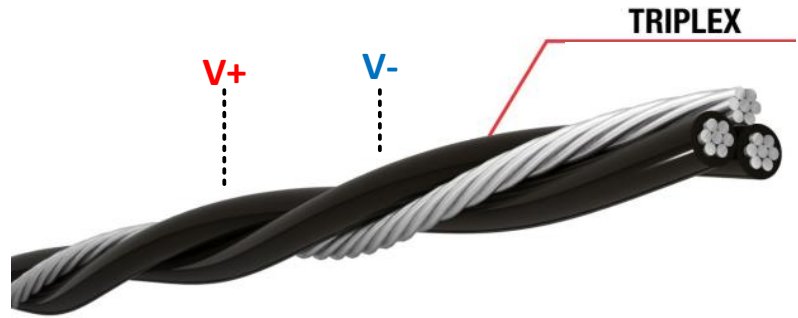

The poles used to support the DCDN are made of reinforced concrete and have a "double t" shape with a height of 7 meters. These poles were already installed in the lab, supporting the AC microgrid. Figure 2.16 illustrates the architecture of the DCDN considering the placement of poles and the topology of the AC grid that was already installed. As illustrated in Figure 2.16 (a), two of the DCDN's distribution sections have poles shared with the ACDM. In order to avoid the emergence of interference between grids, the DCDN installation was performed at a level about 50 cm below the ACDM.

**Figure 2.16 – (a) DCDN and ACDM distribution outside GEDAE laboratory and (b) DCDN's distribution of generation systems and load banks.**

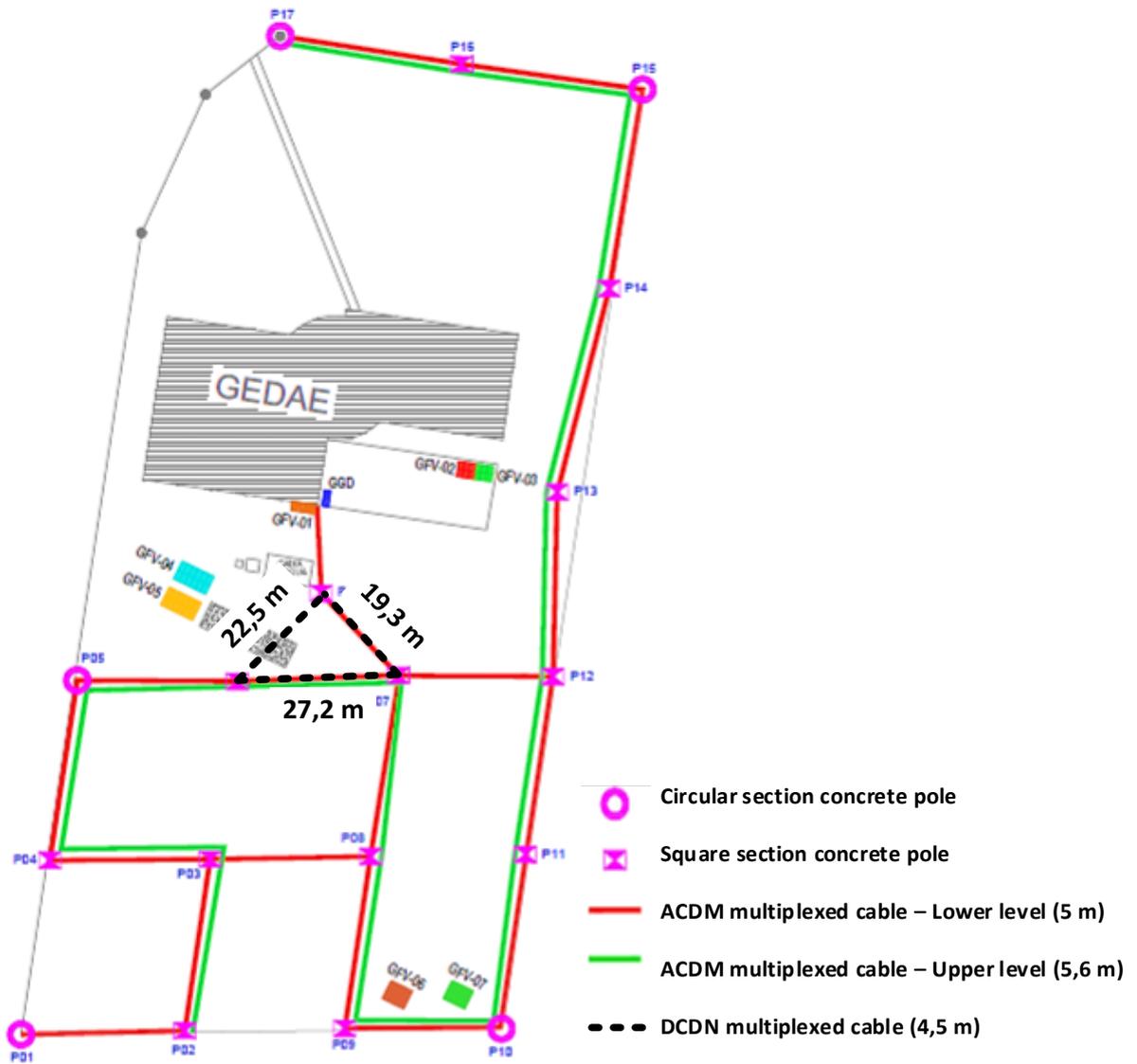

(a)

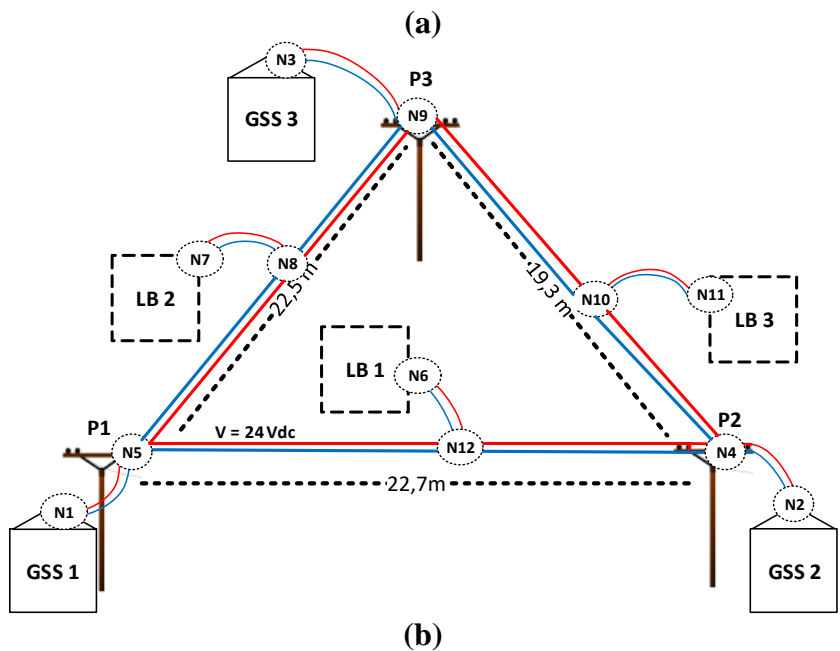

(b)

The location of the GSSs and LBs shown in Figure 2.16 (b) indicates the electrical connection points of each subsystem, so that the GSSs are connected at each pole and the LBs are connected at half the distance between poles. However, despite the electrical connections being distributed throughout the grid, it was decided to house all equipment in a single connection center. This decision was taken by considering the following factors: easier connections to make, the need for a single shelter for all equipment, greater simplicity to implement the load monitoring and driving system, as well as allowing a greater simplicity in changing the position of subsystems in the distribution nanogrid. Figure 2.17 shows a photograph of the DCDN and its connection center.

**Figure 2.17 – DCDN's positioning of poles and connection center.**

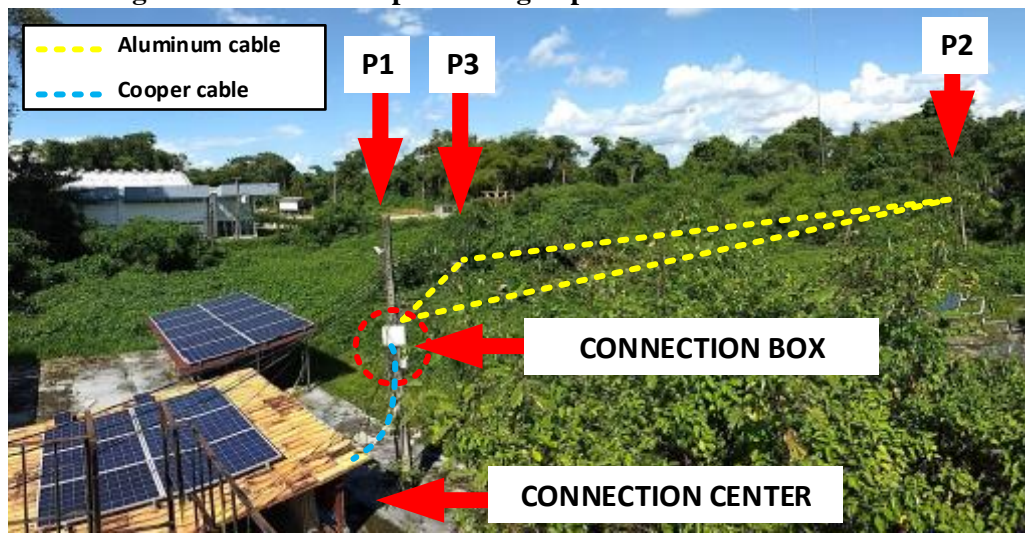

Total distances between all DCDN points N1 through N12 are shown in Table 2.8.

**Table 2.8 – Distances between DCDN's nodes.**

| From | To | Dist. (m) | From | To | Dist. (m) |
|------|-----|-----------|------|-----|-----------|
| N1   | N5  | 6.00      | N4   | N10 | 9.65      |
| N5   | N12 | 11.35     | N8   | N7  | 17.25     |
| N5   | N8  | 11.25     | N8   | N9  | 11.25     |
| N12  | N4  | 11.35     | N10  | N11 | 38.35     |
| N12  | N6  | 17.35     | N9   | N3  | 28.5      |
| N4   | N2  | 28.7      | N9   | N10 | 9.65      |

To make the connections in the DCDN at points N4, N5, N8, N9, N10 and N12, shown in Figure 2.16 (b), it was used piercing connectors for aluminum cables. This type of connector is widely used in distribution networks as it allows the connection to be made with minimal contact of the installer with the grid, since it does not require prior removal of the insulating layer of the cables - making it possible to make connections even with an energized grid.

From the other nodes (N1, N2, N3, N6, N7 and N11) to the general circuit breaker of each subsystem (GSS and LB), a 25 mm$^2$ copper conductor with XLPE insulation for voltage class 0.6/1 kV was used. Therefore, special connector should be used for the connection between an aluminum conductor and a copper conductor in order to suppress the corrosive effect that this connection may trigger. The connector used at these points has special treatment for this type of bimetallic connection, however it requires the prior removal of part of the conductor insulation. The connection of these points was made in the connection box indicated in Figure 2.17, and the copper cables in the output of this box are connected to the general distribution, measurement and control switchboard (QGD). Figure 2.18 illustrates the two types of connectors used in the distribution network.

**Figure 2.18 – (a) Piercing connector for aluminum cables and (b) bimetallic shunt connector for copper-aluminum connection.**

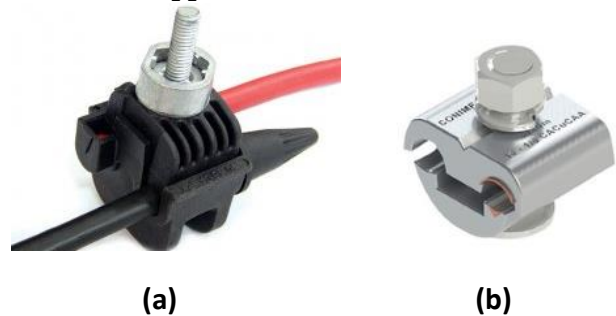

(a)          (b)

## 2.5 Main switchboard for distribution, measurement and command (QGD)

All charge controllers and protection devices are housed within the general distribution, measurement and command switchboard. The QGD also houses the measuring and control circuits developed for this application, as the main measurement points of the DCDN pass through this board, as well as the circuits for the individual loads. The QGD is located within the connection center shown in Figure 2.17, sheltered and easily accessible for maintenance, alterations and punctual measurements. Figure 2.19 shows an internal view of the QGD, specifying the main components housed on it.

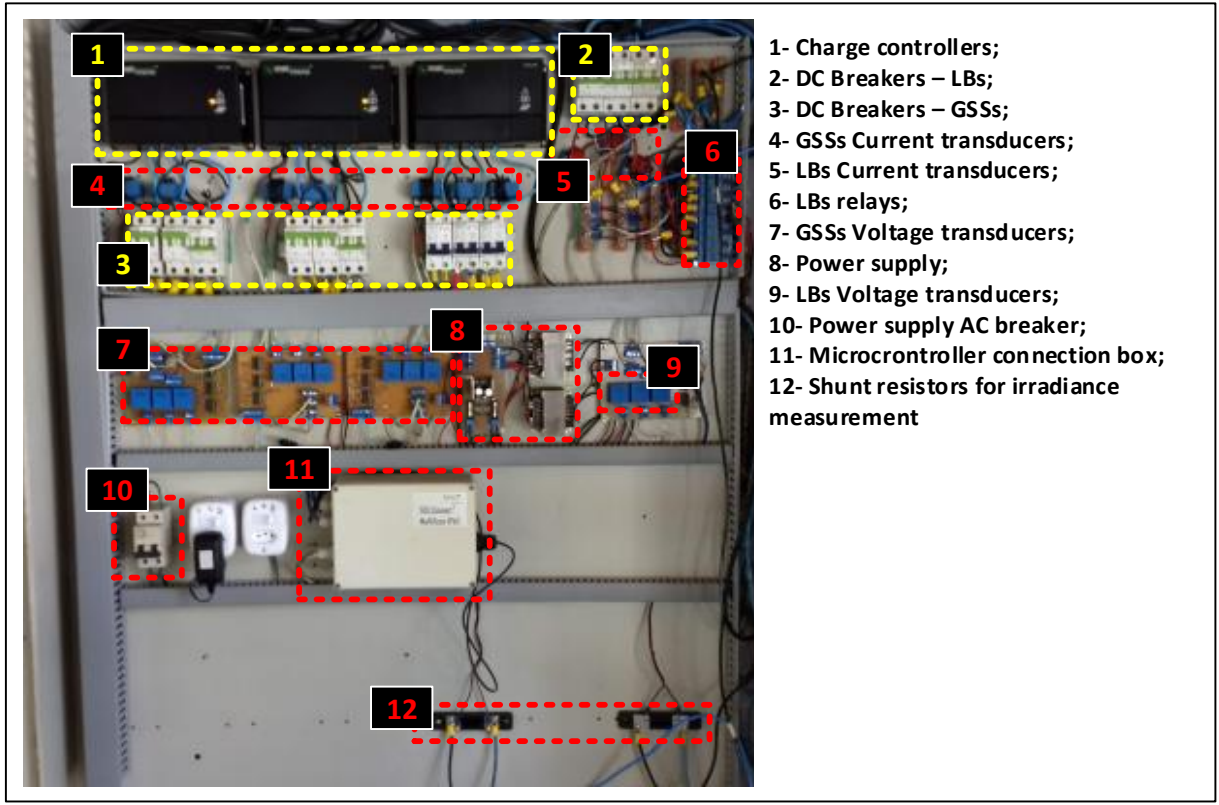

Figure 2.19 – DCDN's QGD internal view.

1- Charge controllers;
2- DC Breakers – LBs;
3- DC Breakers – GSSs;
4- GSSs Current transducers;
5- LBs Current transducers;
6- LBs relays;
7- GSSs Voltage transducers;
8- Power supply;
9- LBs Voltage transducers;
10- Power supply AC breaker;
11- Microcrontroller connection box;
12- Shunt resistors for irradiance measurement

## 2.5.1. System monitoring

The monitoring system developed for the DCDN allows data acquisition and real-time visualization of the following variables: voltage and current of each PVG, voltage and current of each BB, voltage and current in each LB and irradiance in the plane of the PVGs, as shown in Figure 2.1. In addition, the ambient temperature is also monitored by an environmental monitoring station installed near the DCDN. All monitored values are sampled and transmitted every 1 s, in addition, the averaged values are stored locally on a memory card at 1-minute intervals, except for the ambient temperature which is stored at 5-minute intervals. Figure 2.20 illustrates the data acquisition process implemented.

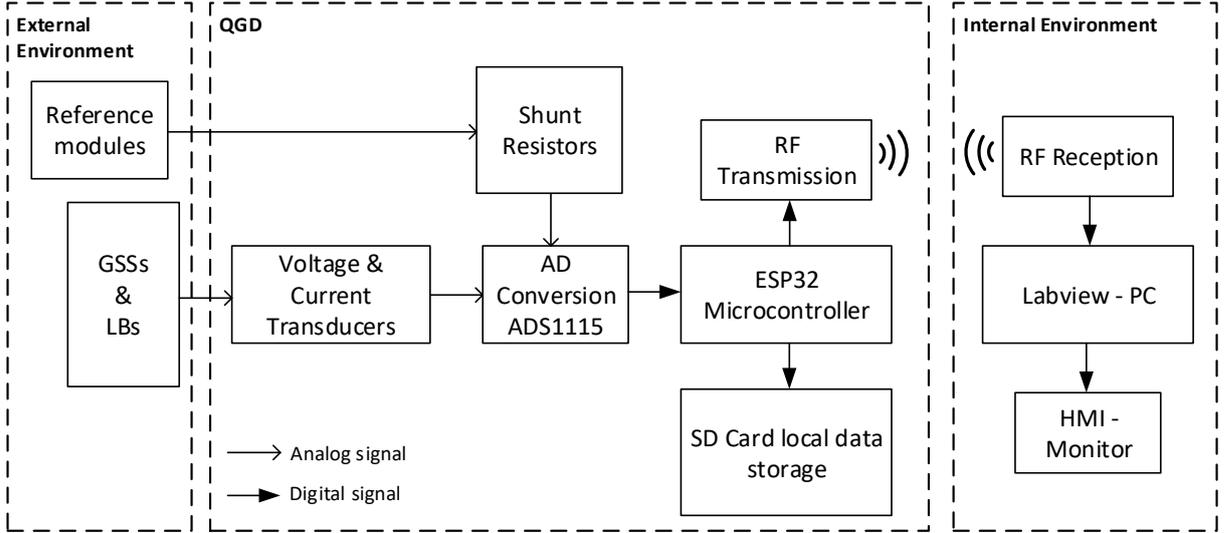

**Figure 2.20 – DCDN Monitoring System Overview.**

Voltage and current signals from PVGs, BBs, and LBs are conditioned in a laboratory-developed circuit consisting of isolated current and voltage transducers, active low-pass analog filters, and voltage attenuators. Transducers reproduce the waveform of the monitored signal and provide an isolated, adjustable magnitude voltage or current signal. The voltage transducers used are of model LV 20-P, manufactured by LEM (LEM, 2012) and current transducers are of model HAS 50-S, also manufactured by LEM (LEM, 2015).

For irradiance measurement, two standard 6-cell PV modules are used. The short circuit current of a module is proportional to the incident radiance on its plane, so, for a given reference condition, the overall incident radiance on the PVG plane, $G_i$, can be obtained by measuring its short circuit current, according to Equation (2.5). The reference short circuit current, $I_{SC,STC}$ was obtained from the GEDAE-UFPA solar simulator test (test report in Annex II) for the standard test irradiance. Short circuit current conditioning is done with 75 mV / 50 A shunt resistors, providing an equivalent voltage signal.

$$G_i = \frac{G_{i,STC} \times I_{SC}}{I_{SC,STC}} \quad (2.5)$$

All conditioned signals are sampled at a rate of 1 Hz using the IC ADS1115, which is an analog to digital converter (ADC). This ADC has 16-bit resolution and adjustable reference voltage range, which allows for high accuracy analog signal sampling. Each ADS1115 IC can measure up to 4 channels in common mode and 2 channels in differential mode. All measurements are common, except for irradiance, which is read in differential mode.

The result of the ADC conversion is sent, via SPI protocol, to the ESP32 microcontroller from the manufacturer Espressif Systems. This microcontroller was chosen for its low cost, high availability in the local market, extensive documentation and compatibility with various peripherals (memory card, external ADC, real time clock, radio frequency transmitter), as well as already being integrated into a prototyping board that facilitated its implementation in the system.

The microcontroller is responsible for processing the AD conversion result of each variable, applying the calibration curves previously obtained with calibrated laboratory instrumentation, and saving the data locally to the memory card. Connected to this microcontroller is the APC-220 radio frequency transmitter, which sends all measured data in real time to a computer located within the GEDAE laboratory.

A program developed on the Labview 2016 platform runs on the computer that receives measurement data from the DCDN and graphically displays the monitored variables, and processes these variables to calculate power and energy generated/consumed by the DCDN sources and loads and save the data on computer in text file.

### 2.5.2. Load driving

The activation of each load in the system is individualized and can be programmed through microcontrolled relays. For this, two relay modules of 8 channels each were implemented, totaling 16 circuits that can be operated independently. This type of microcontrolled driving enables the easy implementation of different load curves, such as the one shown in Figure 2.2, so it is possible to evaluate the system behavior under different charging conditions and energy consumption.

The load curve is implemented by means of a text file that is saved to the memory card and read by the microcontroller. From this file, a 1440 x 16 matrix is generated, where each row equals 1 minute of the day and each column equals one relay. For each minute of the day, if the value in a matrix position is null, the controller interprets it as a load that should be turned off, otherwise the load should be turned on.

# 3. CHARACTERIZATION, MODELING AND VALIDATION OF RESULTS

## 3.1 Introduction

Modeling the various components of a system and their interconnection to form a nanogrid allows the evaluation of the system behavior under different operating conditions, as well as making possible studies of generation and load expansion, identification of operational constraints and inefficient operating conditions, estimation of equipment lifecycle and need for replacement and maintenance, among others. The mathematical model and the computational simulation tool should be selected according to the simulation objectives. For example, if the objective is to assess the capacity to service new loads after expansion of the distribution grid, the priority should be to verify the voltage levels in the new buses added to the system, under different conditions of generation, storage and loading. This type of evaluation is done in a load flow study, which requires static modeling of the grid's components. This type of study should be solved with the aid of computer programs that work with numerical calculation tools.

On the other hand, if the objective of the simulation is to evaluate the system stability in case of short circuit faults, the voltage and current transients should be checked at different points of the grid during the fault occurrence. For this, dynamic models of grid-connected power converters, energy storage systems and loads should be used, as well as simulation tools dedicated to this type of study, for example, specialized software.

The simulation objectives proposed in this dissertation are as follows:

1- DCDN static evaluation, allowing the verification of power flow, voltage levels and distribution line loading in different operational scenarios and;
2- DCDN dynamic evaluation, in order to observe the behavior of the grid over a day of operation considering actual data of solar resource and ambient temperature.

The DCDN static modeling was implemented in Matlab R2017b environment from Mathworks, given the ease of implementing lines of code for engineering applications and extensive documentation available, being a widely used platform in many power system modeling studies. Dynamic modeling was implemented in Simulink environment, which is a tool for modeling, simulation and analysis of dynamic systems, integrated with Matlab software. Simulink has the advantage of facilitating the implementation of algorithms by using blocks and configuring graphical interfaces rather than lines of code.

## 3.2 Characterization and modeling of the PV generators

A PV module is formed by the serial and/or parallel association of encapsulated PV cells to form a rigid and robust structure for use in practical engineering applications. Therefore, the modeling of the PV cell is fundamental to obtain the model of a PV module and generator.

### 3.2.1 Single diode circuit model (standard 5-parameter model)

The most common way to model the behavior of a PV cell is to consider fundamental components of electrical circuits, although other methods such as artificial neural networks are also used (JORDEHI, 2016). In this sense, the standard 5-parameter model is widely used for crystalline silicon cells, given its relative simplicity, ease of computational processing and good accuracy in practical applications of power generation. The circuit used by this model is shown in Figure 3.1.

**Figure 3.1 – Circuit used in the 5-parameter model of a PV cell.**

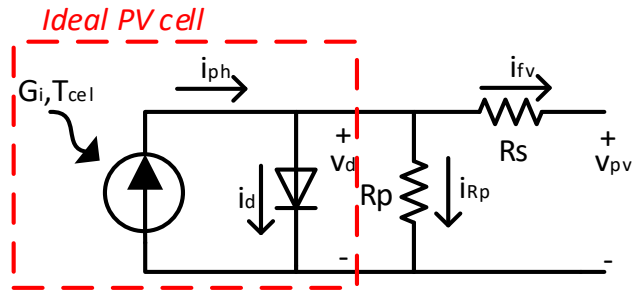

According to Figure 3.1, an ideal PV cell is formed by a current source controlled by the incident irradiance ($G_i$) and cell temperature ($T_c$), associated in parallel with a diode. Series, $R_S$, and parallel, $R_P$ resistances represent constructive non-idealities of the cell: resistance between silicon and electrodes, resistance of electrodes, leakage current at p-n junction, etc. Considering the circuit of this model, the current of a PV cell is given according to Equation (3.1).

$$i_{fv} = i_{ph} - i_d - {v_d}/{R_p} \qquad (3.1)$$

$i_{ph}$ is the current photogenerated by the PV cell. This current is a direct function of cell's incident irradiance and temperature, according to Equation (3.2). $α_{sc}$ (A/°C) is the cell's short circuit current temperature coefficient. $i_{sc,STC}$ is the photogenerated current under standard test conditions and is considered equal to the PV cell short circuit current obtained under standard test conditions.[3]

---

[3] In this calculation of photogenerated current, the parallel current $i_{Rp}$ is disregarded for practicality in the development of the model.

$$i_{ph} = \frac{G_i}{G_{i,STC}} i_{sc,STC} + \alpha_{sc}(T_c - T_{c,STC}) \tag{3.2}$$

The diode voltage is given by:

$$v_d = v_{fv} + i_{fv} R_s \tag{3.3}$$

And the diode current is obtained by (SHOCKLEY, 1949):

$$i_d = i_s \left[ e^{\left(\frac{qv_d}{kT_c A_n}\right)} - 1 \right] \tag{3.4}$$

Where $k = 1{,}38 \times 10^{-23}$ J/K is the Boltzmann constant, $q = 1{,}6 \times 10^{-19}$ C is the elementary charge and $T_{c,k}$ (K) is the absolute junction temperature. $A_n$ is the diode's ideality factor and corresponds to which extent the modeled diode compares to an ideal one ($A_n = 1$), this value usually ranges from 1 to 2. The ideality factor is rarely given by the PV module's manufacturer, and a standard value of $A_n \approx 1{,}3$ can be considered for polycrystalline silicon (TSAI; TU; SU, 2008). $i_s$ the diode's reverse saturation current, given by:

$$i_s = i_{s,STC} \left(\frac{T_c}{T_{c,STC}}\right)^3 e^{\left(\frac{E_g}{k}\left(\frac{1}{T_{c,STC}} - \frac{1}{T}\right)\right)} \tag{3.5}$$

Where $E_g \approx 1.11$ eV is the bandgap energy of the semiconductor (mono and polycrystalline silicon), and $i_{s,STC}$ is the reverse saturation current obtained at standard test conditions. Considering the open circuit, $v_d = v_{OC}$, and disregarding the current in the parallel resistance, $i_{s,STC}$ is given by:

$$i_{s,STC} = i_{sc,STC} \bigg/ \left[ e^{\left(\frac{qv_{oc,STC}}{kA_n T_{c,STC,k}}\right)} - 1 \right] \tag{3.6}$$

Substituting equations (3.2), (3.3) and (3.4) in Equation (3.1):

$$i_{fv} = \frac{G_i}{G_{i,STC}} i_{sc,STC} + \alpha_{sc}(T_c - T_{c,STC}) - i_s \left[ e^{\left(\frac{q(v_{fv} + i_{fv} R_s)}{kT_{c,k} A_n}\right)} - 1 \right] - \frac{(v_{fv} + i_{fv} R_s)}{R_p} \tag{3.7}$$

From Equation (3.7) that corresponds to the model of a single PV silicon cell, one can obtain the equation for a PV module composed by the series association of $N_S$ cells, as expressed in Equation (3.8), where $V_{FV}$ and $I_{FV}$ are the voltage and current of the PV module, respectively.

$$I_{FV} = \frac{G_i}{G_{i,STC}} I_{sc,STC} + \alpha_{sc}(T_c - T_{c,STC}) - I_s \left[ e^{\left(\frac{q(V_{FV} + I_{FV} R_s)}{N_S kT_{c,k} A_n}\right)} - 1 \right] - \frac{(V_{FV} + I_{FV} R_s)}{R_p} \tag{3.8}$$

### 3.2.2 Obtaining the series $R_s$ and parallel $R_p$ resistances

Series and parallel resistance values are rarely provided by manufacturers in PV module datasheets. One way to obtain these parameters is to consider a known point of the PV module operating curve: the maximum power point. This point is provided in the PV module datasheet and corresponds to the pair $V_{FV} = V_{MP}$ and $I_{FV} = I_{MP}$ such that the power delivered by the module is maximum under the standard test condition ($P_{FV} = P_{MP}$).

$$P_{MP} = \max(I_{FV} \times V_{FV}) \tag{3.9}$$

Equations (3.10) and (3.11) are used iteratively to obtain the series and parallel resistances of the PV module. $P_{MP,m}$ and $P_{MP,e}$ corresponds respectively to the maximum power values obtained with the model and the experimental data supplied by the manufacturer or obtained under test. The $R_S$ and $R_P$ values are initially estimated and incremented with each iteration until the difference between the model and experimental powers is less than a stopping criterion. The algorithm implemented in Matlab used to obtain the resistances is described in Ferreira (2018).

$$P_{MP,m} = V_{MP}\left\{I_{ph} - I_s\left[e^{\left(\frac{q(V_{MP}+I_{MP}R_s)}{N_SkT_{c,k}A_n}\right)} - 1\right] - \frac{(V_{MP}+I_{MP}R_s)}{R_p}\right\} \tag{3.10}$$

$$R_P = \frac{V_{MP}(V_{MP}+I_{MP}R_s)}{\left\{V_{MP}I_{MP} - V_{MP}I_s e^{\left[\frac{q(V_{MP}+I_{MP}R_s)}{N_SA_nkT_{c,k}}\right]} - V_{MP}I_s - P_{MP,e}\right\}} \tag{3.11}$$

### 3.2.3 Obtaining the PV cell temperature

The operating temperature of the PV cell is one of the input parameters of the model used, however, this data is not being monitored on the implemented DCDN. Only the ambient temperature is monitored, so a mathematical model should be used to obtain the operating temperature of the PV cells. One of the standard procedures used to obtain the PV cell operating temperature as a function of ambient temperature involves a parameter known as the nominal operating cell temperature (NOCT), which is defined as the temperature of a PV cell operating under the following reference conditions: irradiance $G_{i,NOCT} = 800$ W/m², room temperature $T_{a,NOCT} = 20$ °C, average wind speed of 1 m/s. According to the manufacturer's data for the PV module used, NOCT = 46 °C. Equation (3.12) is used to obtain the operating temperature of the PV cell, as a function of the incident global irradiance and ambient temperature, $T_a$, (SKOPLAKI; PALYVOS, 2009).

$$T_c = T_a + \left(\frac{G_i}{G_{i,TNOC}}\right)\left(\frac{U_{L,TNOC}}{U_L}\right)(TNOC - T_{a,TNOC})\left[1 - \left(\frac{\eta_c}{\xi\alpha_f}\right)\right] \quad (3.12)$$

Where $U_L$ and $U_{L,NOCT}$ correspond to the thermal loss coefficients at ambient temperature and at nominal operating temperature, respectively. $\eta_c$ is the electrical efficiency of the PV module, $\xi$ is the thermal transmittance coefficient of the glass and $\alpha_f$ corresponds to the thermal absorption coefficient of the PV cell. Assuming that the coefficient of thermal loss is constant for the temperature range considered, and approximating $\eta_c/\xi\alpha_f \approx 0$, Equation (3.12) can be reduced to:

$$T_c = T_a + \left(\frac{G_i}{G_{TNOC}}\right)(TNOC - T_{a,TNOC}) \quad (3.13)$$

### 3.2.4 Model evaluation

To verify if the model adequately represents the behavior of the PV modules installed in the DCDN, two experimental I-V curve tests were performed and compared to the simulated results on the Simulink platform. Figure 3.2 presents the block developed to perform model validation tests.

**Figure 3.2 – Simulink block developed for the model evaluation.**

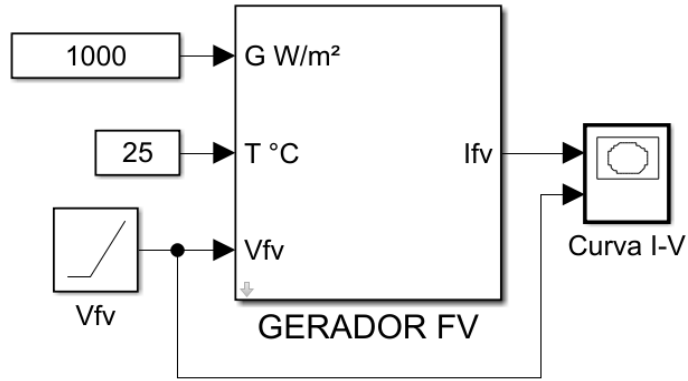

The first test was performed under standard test conditions in a solar simulator for PV module testing installed in the GEDAE laboratory. In this test, the I-V curve of a single PV module from the same manufacturer and model of the modules installed at the DCDN was tested. Figure 3.3 shows the tested and the modeled I-V curves.

**Figure 3.3 – Comparison between the I-V curves obtained from experimental test and mathematical model, for operation under standard conditions.**

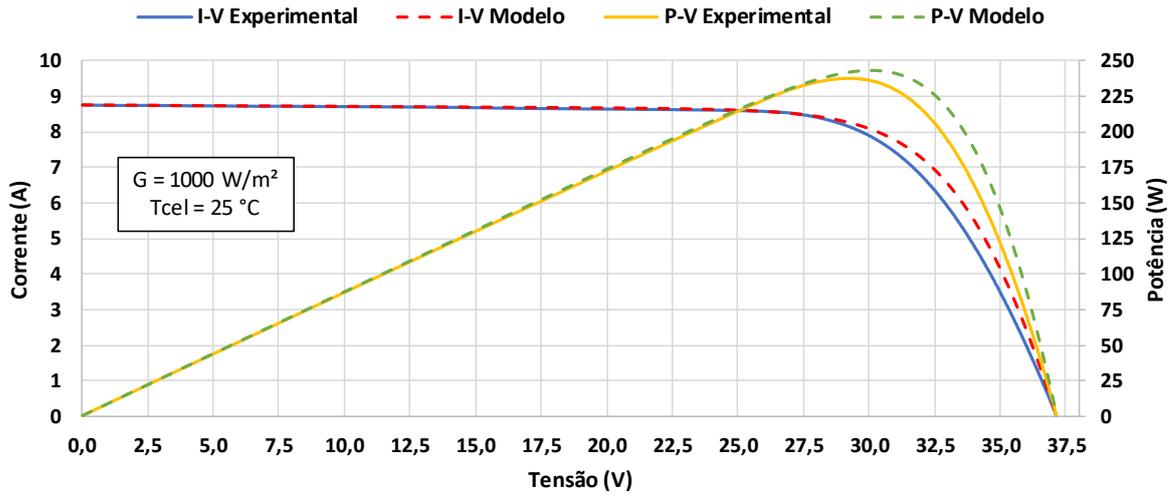

To evaluate the performance of the model, it is calculated the deviation obtained between the measured and modeled values in different points of interest: open circuit, short circuit and maximum power. Equations (3.14), (3.15) and (3.16) are used to calculate deviations in open circuit, $D_{OC}$, short circuit, $D_{SC}$ and maximum power, $D_{MP}$, respectively (XIAO, 2017).

$$D_{OC} = \left| \frac{V_{OC,m}}{V_{OC,e}} - 1 \right| \qquad (3.14)$$

$$D_{SC} = \left| \frac{I_{SC,m}}{I_{SC,e}} - 1 \right| \qquad (3.15)$$

$$D_{MP} = \sqrt{\left(\frac{P_{MP,m}}{P_{MP,e}} - 1\right)^2 + \left(\frac{V_{MP,m}}{V_{MP,e}} - 1\right)^2} \qquad (3.16)$$

$V_{OC,m}$ and $V_{OC,e}$ correspond to the open circuit voltage obtained by the model and by the experimental test, respectively; $I_{SC,m}$ and $I_{SC,e}$ correspond to the short circuit current obtained by the model and by the experimental test, respectively; $P_{MP,m}$ and $P_{MP,e}$ correspond to the maximum power obtained by the model and by the experimental test, respectively, and $V_{MP,m}$ and $V_{MP,e}$ are the maximum power point voltage obtained in the model and by the experimental test, respectively.

It can be seen from the curve shown in Figure 3.3 that the model follows well the experimental values, especially in the region of lower voltages. From the zone known as the "knee" of the I-V curve, the model tends to present slightly higher current values than the measured data. This error is mainly due to the difference between the actual diode ideality factor and that adopted in the model, as well as the difference between the parallel resistance adopted in the model (which was obtained through the iterative parameter extraction process) and the actual value.

From the deviation's equations, $D_{OC} = 0.1367\%$, $D_{SC} = 0.0274\%$ and $D_{MP} = 4.0596\%$. The model is considered adequate for the DCDN simulation purpose, since the deviations obtained do not represent significant errors in terms of grid's performance, regarding power and energy produced. In addition, it is important to consider the simplicity and reduced computational effort required to simulate this model.

The other test performed to validate the model of the PV module was to obtain the I-V curve for the PV generators already installed in the DCDN. This field test is also important as partial commissioning the installation, as the field-obtained I-V curve can assist in the identification of shadings, defective PV cells, incorrect connections, and other problems that may occur in system installation.

The I-V curve of PV generators installed in the DCDN was obtained using a capacitive load developed in the GEDAE laboratory, together with an oscilloscope. The oscilloscope used was model 190-204, made by Fluke, which has four independent channels that can be used to measure current and voltage signals with bandwidth up to 20 MHz. The methodology used in the capacitive load tests followed the procedure described by Brito (2018), considering the 2-wire measurement configuration and without the use of a reference PV module for verification of irradiance and cell temperature.

The irradiance measurement at the time of the tests was verified using a reference calibrated PV cell, model Spektron 210, which provides a voltage that varies linearly with the incident irradiance. During the tests, this cell was positioned at the same inclination and orientation as the DCDN PVGs. The temperature measurement on the back of the module was verified using the MINIPA MT-350 thermal sensor, which has a reading accuracy of ±2 °C. Figure 3.4 shows the results obtained for real sun testing of the three DCDN's PVGs, compared with the I-V curves resulting from the model simulation.

**Figure 3.4 - Comparison between I-V curves obtained from experimental test and mathematical model, considering real-sun operation. (a) PVG1, (b) PVG2 and (c) PVG3.**

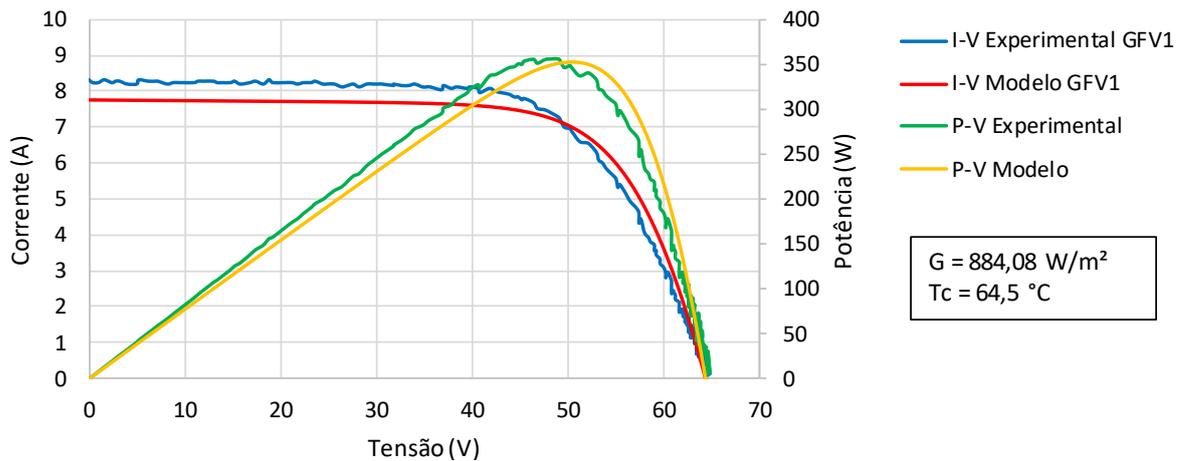

(a)

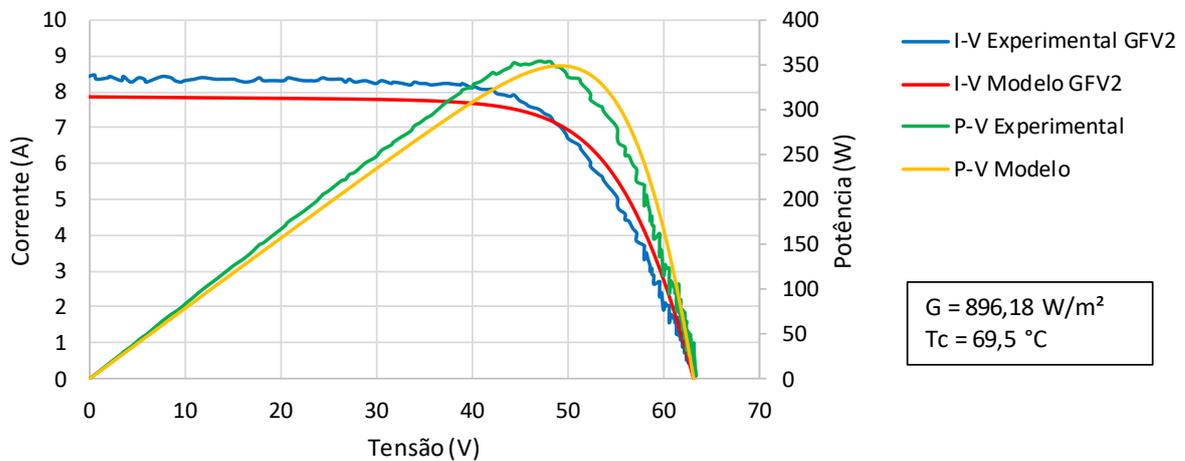

(b)

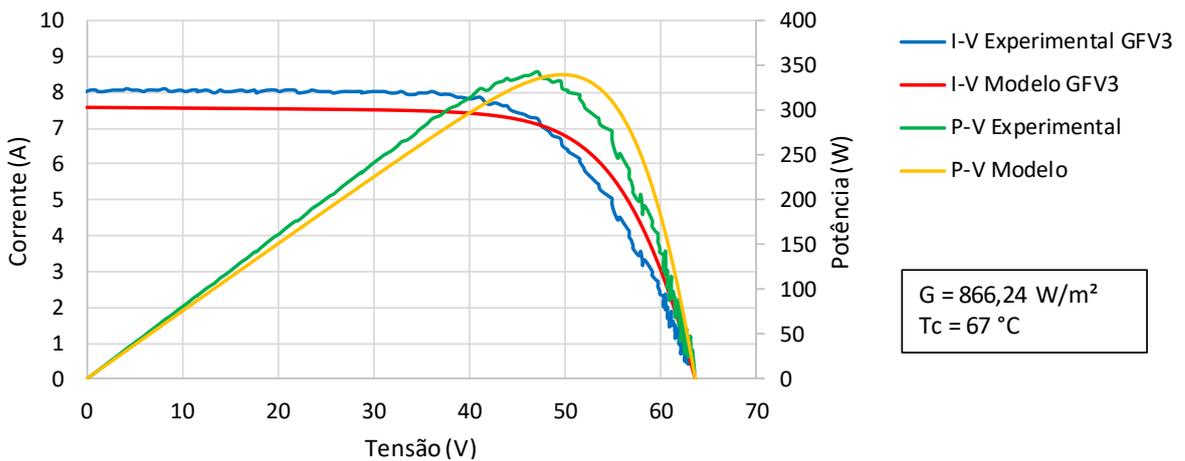

(c)

Regarding the model, it is noticed that the calculated maximum power is close to the maximum power measured in all cases. However, the measured value of short circuit current differs from the simulated value and is an indicative of the difference between the actual series resistance and the simulated series resistance. For the real-sun tests, the percentage deviation values of the points of interest are presented in Table 3.1. The obtained values are satisfactory for the application proposed in the DCDN simulation, however, results closer to the experimental one can be achieved if the values of series and parallel resistance and ideality factor are obtained based on several measurements under real sun.

Table 3.1 – Deviations between simulated value and experimental value for real sun testing of the DCDN generators.

| Generator | $D_{OC}$ (%) | $D_{SC}$ (%) | $D_{MP}$ (%) |
|---|---|---|---|
| PVG1 | 0,6173 | 6,7539 | 3,0819 |
| PVG2 | 0,6289 | 6,7952 | 4,9214 |
| PVG3 | 0,4687 | 5,4332 | 5,3723 |

## 3.3 Characterization and modeling of the battery bank

The search for the development of mathematical models that accurately express the behavior of lead-acid batteries is old in the literature. However, characteristics such as high nonlinearity, interdependence between parameters and independent reaction phenomena make the description of these elements quite complex and require extensive computational effort to simulate the system in detail (BADEDA et al., 2017).

In this sense, there is a great variety of models for lead-acid batteries presenting significant differences in the degree of complexity, being classified into: analytical models, electrical models and electrochemical models (GARCHE et al., 2013). Analytical models are those that require the least computational effort, since they are based on analytical equations capable of calculating battery capacity and voltage as a function of parameters such as charge and discharge current, temperature and other additional parameters.

Electrical models use fundamental circuit elements such as voltage sources, resistors and capacitors to build the equivalent circuit of a battery. This type of model can be more intuitive in obtaining parameters, since they are related to physical quantities such as capacitance and electrical resistances. Electrochemical models are the most complete and describe in detail the electrochemical reactions in a battery, considering the kinetic and thermodynamic effects of chemical reactions. These models are more complex and require a large number of parameters that are difficult to obtain in many practical applications.

Battery modeling in photovoltaic applications is even more complex when considering the characteristic dynamic behavior of solar resource and load variations, since many models are defined for constant charge or discharge currents. In addition, voltage discontinuity in models in the transition from charge to discharge and vice-versa may lead to simulation errors.

In this sense, the model used to simulate the DCDN battery banks is based on the analysis proposed by Guasch and Silvestre (2003), which is based on the model of Copetti, Lorenzo and Chenlo (1993), but considering the dynamic characteristics present in a PV system in addition to assessing the aging effects of the batteries. This model considers electrical circuit components but adopts analytical equations in its development.

### 3.3.1 Description of the model

Figure 3.5 represents the basic circuit used in the modeling, consisting of a voltage source $V_{int}$ in series with an internal resistance $R_{bat}$. The voltage source is a function of batterie's state of charge (SoC), and the internal resistance is a function of the operating temperature, $T_{bat}$, battery current, $I_{bat}$, and SoC. Equation (3.17) is used to calculate the voltage at the battery terminals, $V_{bat}$.

**Figure 3.5 – Fundamental circuit of the battery model**

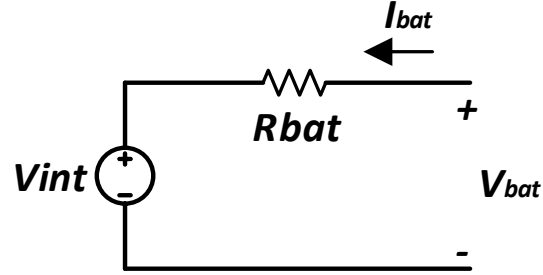

$$V_{bat} = V_{int} + I_{bat}R_g \begin{cases} V_{int} = f(SoC) \\ R_g = f(SoC, T_{bat}, I_{bat}) \end{cases} \quad (3.17)$$

The SoC indicates how much electrical charge is stored in the battery, and ranges from 0 <SoC <1. Equations (3.18), (3.19) and (3.20) characterize the SoC.

$$SoC(t) = \frac{1}{C(t)} \int_{-\infty}^{t} \eta_{ch}(t) I_{bat}(t) dt \quad (3.18)$$

$$C(t) = \frac{C_{nominal} C_{t,coef}}{1 + Acap \left(\frac{|I_{bat}(t)|}{I_{nominal}}\right)^{Bcap}} (1 + \alpha_c \Delta T_b(t) + \beta_c \Delta T_b(t)^2) \quad (3.19)$$

$$I_{nominal} = \frac{C_{nominal}}{n} \quad (3.20)$$

Where: *C(t)* is the battery capacity (given in Ah), $\eta_{ch}$ is the charging efficiency, $C_{nominal}$ is the nominal battery capacity (obtained in *n* hours), $C_{t,coef}$, $A_{cap}$ and $B_{cap}$ are constants of the model, $\Delta T_b$ is the temperature difference from the reference value of 25 °C, and $\alpha_c$ and $\beta_c$ are temperature coefficients.

Another parameter used to characterize the available energy capacity in a battery is the so-called level of energy, *LoE*, which is defined as:

$$LoE(t) = \frac{1}{C_n} \int_{-\infty}^{t} \eta_{ch}(t) I_{bat}(t) dt \qquad (3.21)$$

$$C_n = \max(C) \Big|_{\substack{I_{bat}=0 \\ T_b=[-T_1,T_2]}} \qquad (3.22)$$

$C_n$ is the maximum battery capacity, given by equation (3.19) for a battery in open circuit and considering the range of temperature in which the battery would operate. The *LoE* represents directly the energy stored in the battery, independently of the charge or discharge current. For example, a battery can have at the same time *SoC = 1* and *LoE ≠ 1*, indicating that the battery is saturated but only at a percentage of the nominal capacity.

Guasch and Silvestre (2003) define five regions of battery operation: saturation, overcharge, charge, discharge, over discharge and exhaustion. Figure 3.6 presents the variation of the battery voltage (considering a 2 V element) in the different operating regions, as the current flow varies over time. In the first 16 hours shown in Figure 3.6 current is flowing towards the battery (positive direction according to the reference shown in Figure 3.5), evolving from the charging region to overcharge until it reaches saturation. From 16h to 27h, the current is negative, discharging the battery, passing through a transition region between charging and discharging, discharging, over discharging and then exhaustion. Table 3.2 shows the notation used for battery voltage in each of the operating regions, including the voltage and current conditions for the battery to operate in each region.

**Figure 3.6 – Operating regions of a battery.**

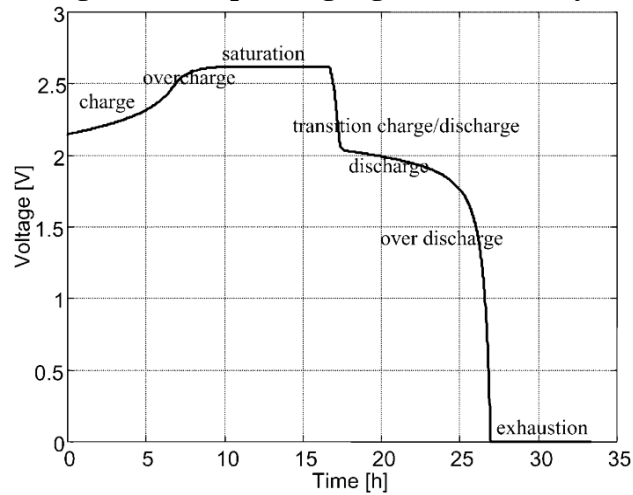

**Source: (GUASCH; SILVESTRE, 2003).**

Table 3.2 – Operating conditions and notation used in the battery model.

| Battery voltage | Operating region | Operating conditions | | |
|---|---|---|---|---|
| $V_{sc}$ | Saturation | $I_{bat} > 0$ | $V_{bat} = V_{ec}$[4] | $\eta_{ch} \approx 0$ |
| | Overcharge | | $V_{ec} \geq V_{bat} \geq V_g$[5] | $0 < \eta_{ch} < 1$ |
| $V_c$ | Charge | | $V_{bat} < V_g$ | |
| $V_{cdc}$ | Charge/discharge transition | $I_{bat} \approx 0$ | $V_c \geq V_{bat} \geq V_{dc}$ | |
| $V_{dc}$ | Discharge | $I_{bat} < 0$ | $V_{bat} > 0,9\ V_n$[6] | $\eta_{ch} \approx 1$ |
| | Over discharge | | $0,9\ V_n \geq V_{bat} \geq 0,7\ V_n$ | |
| | Exhaustion | | $V_{bat} < 0,7\ V_n$ | |

Equation (3.23) was proposed by Copetti, Lorenzo and Chenlo (1993) based on experimental values obtained through various tests on commercial batteries for PV applications. This equation describes the dynamic behavior of battery voltage in the discharge to exhaustion regions as a function of temperature, state of charge and discharge current.

$$V_{dc}(t) = [V_{bodc} - K_{bodc}(1 - SoC(t))] - \frac{|I_{bat}(t)|}{C_{10}}\left(\frac{P_{1dc}}{1+|I_{bat}(t)|^{P2dc}} + \frac{P_{3dc}}{SoC(t)^{P4dc}} + P_{5dc}\right)(1 - \alpha_{rdc}\Delta T_b(t)) \quad (3.23)$$

Where $V_{bodc}$ and $K_{bodc}$ model constants related to constructive characteristics of the battery, for example the density of the electrolyte, and directly influence the batterie's open circuit voltage, $V_{oc,bat}$. $P_{1dc}$, $P_{2dc}$, $P_{3dc}$, $P_{4dc}$ and $P_{5dc}$ are model constants related to resistive losses and directly influence on the value of the internal resistance, $R_{bat}$. $\alpha_{rdc}$ is the thermal coefficient, which also influences $R_{bat}$. $C_{10}$ corresponds to the rated battery capacity at a discharge regime of 10 h.

On the charge region, the equation that defines the behavior of the battery voltage is expressed in (3.24) (COPETTI; LORENZO; CHENLO, 1993).

$$V_c(t) = [V_{boc} + K_{boc}(SoC(t))] + \frac{|I_{bat}(t)|}{C_{10}}\left(\frac{P_{1c}}{1+|I_{bat}(t)|^{P2c}} + \frac{P_{3c}}{(1-SoC(t))^{P4c}} + P_{5c}\right)(1 - \alpha_{rc}\Delta T_b(t)) \quad (3.24)$$

On the charging region, the charging efficiency is given by (3.25), hence, only a percentage of the charging current is actually converted into charge for the battery.

$$\eta_c(t) = 1 - e^{\left[\frac{a_{cmt}}{\left(\frac{I_{bat}(t)}{I_{10}}+b_{cmt}\right)}\right](SoC(t)-1)} \quad (3.25)$$

Where $V_{boc}$ and $K_{boc}$ are model constants related to the constructive characteristics of the battery, for example the density of the electrolyte, and directly influences the open circuit voltage of the battery, $V_{oc,bat}$. $P_{1c}$, $P_{2c}$, $P_{3c}$, $P_{4c}$ and $P_{5c}$ are model constants related to the resistive losses and directly influence the value of batterie's internal resistance, $R_{bat}$. $\alpha_{rc}$ is the thermal constant, which also influences $R_{bat}$. $a_{cmt}$ and $b_{cmt}$ are coefficients related to the time constant of the efficiency reduction in battery charging.

---

[4] $V_{ec}$ corresponds to batteries saturation voltage and is given by Equation (3.28).
[5] $V_g$ corresponds batteries gasification voltage and is given by Equation (3.26).
[6] $V_n$ corresponds to batteries rated voltage.

When excess energy is stored in the battery, it begins to show a reduction in charge efficiency, as part of the energy is used in the electrolyte gasification reaction - thus entering the overcharge region. Considering the concept of level of energy, it should be understood that when the battery reaches the overcharge region, not necessarily the battery has too much energy stored, it can only be said that the battery is nearing its maximum capacity for a given charging regime. By the proposed model, the battery reaches this region when its voltage is higher than the gasification voltage $V_g$, given by:

$$V_g(t) = \left[A_{gas} + B_{gas} ln\left(1 + \frac{I_{bat}(t)}{C_{10}}\right)\right](1 - \alpha_{gas}\Delta T_b(t)) \quad (3.26)$$

Battery voltage in the regions of overcharge and saturation is given by equation (3.27).

$$V_{sc}(t) = V_g(t) + \left(V_{ec}(t) - V_g(t)\right)\left[1 - e^{\left(\frac{LoE(t)C_n - SoC_{vg}(t)C(t)}{I_{bat}(t)\tau(t)}\right)}\right] \quad (3.27)$$

Where:

$$V_{ec}(t) = \left[A_{fonsc} + B_{fonsc} ln\left(1 + \frac{I_{bat}(t)}{C_{10}}\right)\right](1 - \alpha_{fc}\Delta T_b(t)) \quad (3.28)$$

$$\tau(t) = \frac{A_{\tau sc}}{1 + B_{\tau sc}\left(\frac{I_{bat}(t)}{C_{10}}\right)^{C_{\tau sc}}} \quad (3.29)$$

$$SoC_{vg} = SoC_{|Vc=Vg} \quad (3.30)$$

$A_{gas}$, $B_{gas}$, $A_{fonsc}$, $B_{fonsc}$, $\alpha_{gas}$, $\alpha_{fc}$, $A_{\tau sc}$, $B_{\tau sc}$ and $C_{\tau sc}$ are constants obtained empirically from a series of tests with different battery models and under different conditions of charging current and temperature.

Equation (3.27) shows that $V_{sc}$ ranges from $V_g$ (initial gasification voltage, which represents the beginning of the overcharge region) to $V_{ec}$ (exhaustion voltage). The $\tau(t)$ function represents the time constant that the battery remains in the overcharge region before reaching saturation. In the saturation region the battery will no longer receive charge, and prolonged exposure in this region affects the health of the battery, shortening its life. In a PV system, the charge controller is responsible for limiting PV generation so that the battery does not reach this region.

Considering the battery voltage equations corresponding to the charge (3.24) and discharge (3.23) regions, there is a singularity in the transition between these regions, given that:

$$V_c|I_{bat=0} \neq V_{dc}|I_{bat=0} \tag{3.31}$$

This singularity can lead to simulation problems in the charge-discharge transition and vice versa. The solution proposed by Guasch and Silvestre (2003) considers the inclusion of one more region in the model: the transition region charge/discharge, as occurs between 16h and 17h in the graph of Figure 3.6. Battery voltage in this transition region is assumed to correspond to a linear function between the limits of the charge and discharge functions. This limit is defined when the charging or discharging current reaches a sufficiently small value $|I_\delta|$, thus entering the transition region, whose voltage is defined by Equation (3.32). For the DCDN simulation presented in this work, $I_\delta = 0.05$ A.

$$V_{cdc} = \frac{V_{c|I_\delta} - V_{dc|I_\delta}}{2 I_\delta} I_{bat} + \frac{V_{c|I_\delta} + V_{dc|I_\delta}}{2} \tag{3.32}$$

The empirically obtained constants presented in the works of Copetti, Lorenzo and Chenlo (1993) and Guasch and Silvestre (2003) are listed in Table 3.3.

Table 3.3 – **Parameters used in the battery model.**

| | | | |
|---|---|---|---|
| $C_{t,coef} = 1.67$ | $P_{2dc} = 1.3$ | $P_{1c} = 6$ VAh | $\alpha_{fc} = 0.002$ °C$^{-1}$ |
| $\alpha_c = 0.005$ °C$^{-1}$ | $P_{3dc} = 0.27$ Vh | $P_{2c} = 0.86$ | $\alpha_{rdc} = 0.007$ °C$^{-1}$ |
| $\beta_c = 0$ °C$^{-2}$ | $P_{4dc} = 1.5$ | $P_{3c} = 0.48$ Vh | $\alpha_{rc} = 0.025$ °C$^{-1}$ |
| $A_{cap} = 0.67$ | $P_{5dc} = 0.02$ Vh | $P_{4c} = 1.2$ | $A_{fonsc} = 2.45$ V |
| $B_{cap} = 0.9$ | $a_{cmt} = 20.73$ | $P_{5c} = 0.036$ Vh | $B_{fonsc} = 2.011$ Vh |
| $V_{bodc} = 2.085$ V | $b_{cmt} = 0.55$ | $A_{gas} = 2.24$ V | $A_{\tau sc} = 17.3$ h |
| $K_{bodc} = 0.12$ V | $V_{boc} = 2$ V | $B_{gas} = 1.97$ Vh | $B_{\tau sc} = 852$ h |
| $P_{1dc} = 4$ VAh | $K_{boc} = 0.16$ V | $\alpha_{gas} = 0.002$ °C$^{-1}$ | $C_{\tau sc} = 1.67$ |

The values of the parameters listed in Table 3.3 were obtained based on several experimental tests for different types of lead-acid batteries. Therefore, they are generic parameters and, depending on the battery to be modeled, can lead to significant errors. Ideally, one must extract specific parameters for the equipment under consideration. Several methodologies are proposed in the literature for parameter extraction based on experimental tests: Guasch and Silvestre (2003) proposed the use of Levenberg-Marquardt optimization algorithm for parameter identification, Chacón et al (2018) use evolutionary algorithms, Blaifi et al. (2016) uses genetic algorithms. However, the major disadvantage associated with this type of methodology is the need for a large measurement database for model training. For the study presented in this dissertation, it was chosen to use generic parameter values, given the small availability of real data for value extraction.

The equations and parameters presented were developed considering an energy storage cell whose nominal voltage is 2 V. Considering that each battery is formed by the series association of six cells, and each bank is formed by the series association of two batteries, the bank voltage is rated at 24 V. Therefore, the results of the discharge, charge and overcharge equations must be multiplied by 12 to match the model to the bank's rated operating voltage.

Electrochemical effects inherent in a battery such as corrosion or water loss lead to a degradation in battery performance. Two main effects of this degradation are: reduced battery capacity and the presence of self-discharge current (GUASCH; SILVESTRE, 2003). To evaluate battery performance over long operating periods, the effects of battery degradation can be considered using a state of health indicator (SoH) defined in Equation (3.33).

$$SoH(t_i) = 1 - \int_{-\infty}^{t_i} (\eta_T + \eta_{wz}) dt \tag{3.33}$$

Where $\eta_T$ and $\eta_{wz}$ are health factors of temperature and working zone, respectively. $\eta_T$ is defined by equation (3.34), where $\alpha_T$ and $\beta_T$ are thermal coefficients and $T_{ref,s} = 10\ °C$ is the reference temperature.

$$\eta_T = \alpha_T |T - T_{ref,s}| + \beta_T \tag{3.34}$$

$\eta_{wz}$ is related to the degradation that the battery undergoes due to the operating region allowed in a daily operating cycle, the values of this factor are listed in Table 3.4.

Table 3.4 – $\eta_{wz}$ values as a function of the allowed working zone.

| Admitted operating region | $\eta_{wz}$ |
|---|---|
| Saturation and exhaustion | 5,5 x $10^{-6}$ |
| Over charge and over discharge | 5,5 x $10^{-7}$ |
| Charge and discharge | 2,7 x $10^{-7}$ |

To quantify the influence of SoH on reducing battery capacity, it is considered that its capacity decreases to 25 % of nominal value when the battery is fully damaged (SoH = 0) and that this reduction in capacity exhibits linear behavior. Therefore, a battery capacity reduction coefficient, $\eta_{c10}$, is given by:

$$\eta_{c10} = 0,75 \times SoH + 0,25 \tag{3.35}$$

This coefficient is added to Equation (3.19), resulting in Equation (3.36).

$$C(t) = \frac{C_{nominal} C_{t,coef} \eta_{c10}}{1 + Acap \left(\frac{|I_{bat}(t)|}{I_{nominal}}\right)^{Bcap}} (1 + \alpha_c \Delta T_b(t) + \beta_c \Delta T_b(t)^2) \tag{3.36}$$

The self-discharge current is also a function of battery health and accumulated charge. The equivalent battery circuit, considering the effect of self-discharge, is shown in Figure 3.7. The self-discharge current, $I_{adc}$, is given by Equation (3.37).

**Figure 3.7 – Self discharge effect on the battery equivalent circuit.**

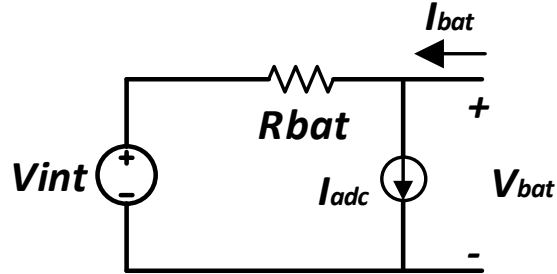

$$I_{adc} = \eta_q \frac{Q(t)}{24h} \Delta t \tag{3.37}$$

Where Q (t) is the charge stored in the battery over a time interval Δt. $\eta_q$ is the self-discharge coefficient, given by:

$$\eta_q = 0{,}01 - 0{,}009 \times SoH \tag{3.38}$$

### 3.3.2 Model's evaluation

A Simulink model was implemented from the equation presented in the previous item, whose input variables are the battery current, $I_{bat}$, and the ambient temperature, $T_{amb}$. The nominal battery bank voltage, initial state of charge, initial level of energy, initial state of health, nominal battery bank discharge capacity and 10 h-discharge capacity values must be initialized to begin the simulation. Figure 3.8 illustrates the block developed in Simulink that represents the battery bank.

**Figure 3.8 - Block developed for battery bank simulation.**

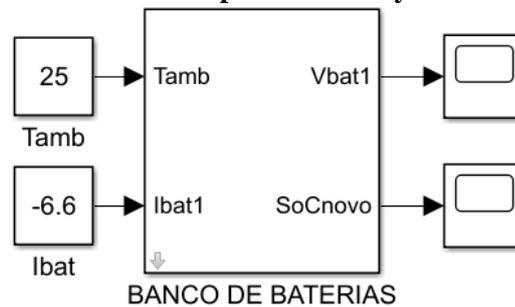

The tests whose results are shown in Figure 2.8 were used to evaluate the model in the discharge region. Figure 3.9 (a) shows the measurement and model results for a constant current discharge of 6.6 A. It was considered in simulations the maximum initial state of charge and initial level of energy (SoC (0) = LoE (0) = 1), varying only the bank's initial state of health (SoH (0) = 1, SoH (0) = 0.7 and SoH (0) = 0.57). Figure 3.9 (b) shows the results for a constant discharge current of 10 A. The temperature considered in all simulations was 25 ° C.

**Figure 3.9 - Comparison between experimental tests and battery discharge model. Constant discharge current of (a) 6.6 A and (b) 10 A.**

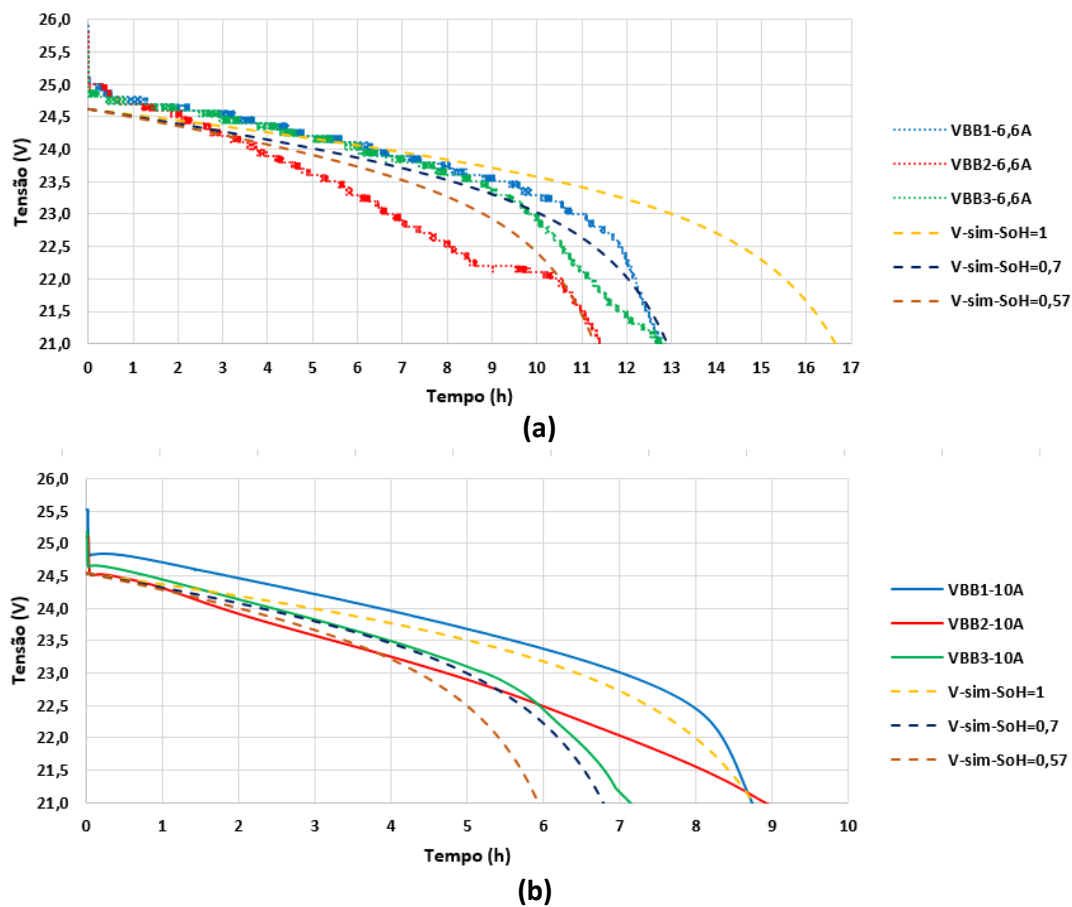

(a)

(b)

For the 6.6 A discharge current test, it is observed that for SoH = 0.7, the model behavior is close to the tests for BB1 and BB3. Also, in terms of capacity, the value of SoH = 0.57 approximates the experimental value for BB2. Nevertheless, a large difference is observed between the experimental and modeled values for low state of charge.

The modeled capacity of a new battery (SoH = 1) is in accordance with the nominal capacity given by the manufacturer, considering that the discharge test presented consumed about 80% of the battery capacity, considering approximately 16.7 h of discharge until the bank reached 21 V.

For the 10 A discharge test, greater discrepancies are observed between the modeled and experimentally obtained values, especially when compared with the results obtained in the 6.6 A tests. For BB3, there is again a good approximation, considering that the bank's SoH is 0.7. However, for BB1, the best approximation occurs for SoH = 1, contradicting the result obtained in the 6.6 A discharge test. In addition, BB2, which had the worst performance in the previous test, showed better correspondence with the model for SoH = 1.

To verify the performance of the model in the charge and overcharge regions, bench tests were performed for BB1 and BB3 battery banks, considering charging with a constant current of 10 A. For simulation of the charging test, it was considered, based on the measured voltage values, that the banks had initial SoC of 5 %, and initial LoE of 5 %, varying only the initial state of health (SoH = 1, SoH = 0.7 and SoH = 0.57). As the test was performed in a refrigerated environment in the laboratory, the temperature of 25 ° C was considered in the simulation. The result of the test and simulations is shown in Figure 3.10.

**Figure 3.10 - Comparison between experimental test and model simulation for 10 A constant current battery charging.**

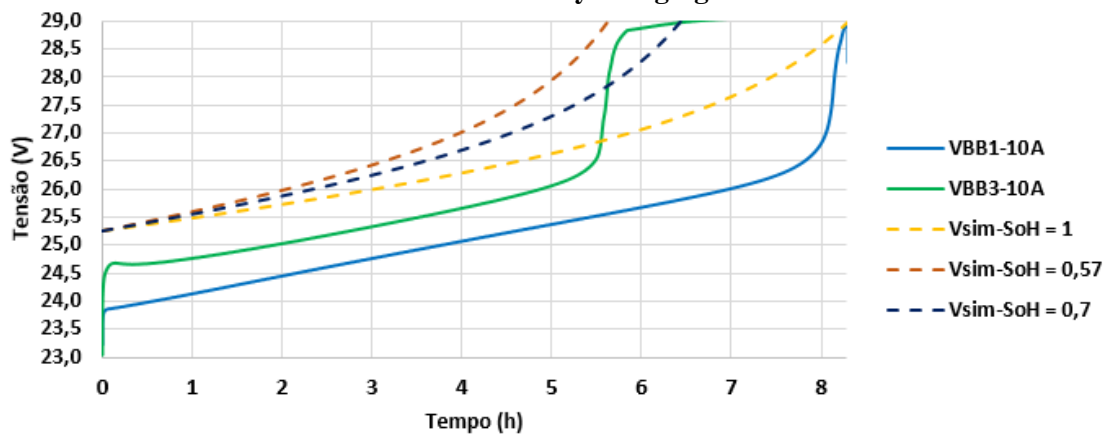

There is a large difference between the experimentally obtained and the simulated voltages. In general, in low SoC regions the simulated voltage is always higher than the measured value. In addition, the voltages evolve differently, reaching the 29 V voltage limit considered in the test at different times, which is indicative of the different capacities of the batteries. Regarding the state of health, it was observed that for BB1, the best result in relation to capacity was obtained for SoH = 1, and for BB3, the best result was for SoH = 0.7, similar to the results of discharge to this bank.

It can be concluded that the model presents poor performance in simulating the actual behavior of the batteries used in charge and discharge tests. The poor performance of the model is mainly explained by two factors:

1- Generic values were used for the various parameters present in the model, thus disregarding peculiarities of the battery used in the DCDN;
2- The series-parallel association of batteries with different states of health compromises the performance of the model, since each battery tends to behave differently and influence the voltage of the battery bank as a whole. This results in a very different experimental results from the simulation.

To improve model performance, one should extract the specific parameters for the battery tested, and individually characterize each battery within the bank. With this, it is possible to consider in the simulation the differences between the capacities of the batteries within a bank, where each block of Figure 3.8 can be used to represent only one battery, being the battery bank formed by the series-parallel association of the four blocks.

## 3.4 Characterization and modelling of the loads

The load banks are formed by the parallel association of resistive lamps and, for BC2 and BC3, fans.

### 3.4.1 Incandescent lamps

Incandescent lamps are formed by a small tungsten filament that emits light and heat as it is passed through the electric current. This filament is characterized as a resistance to current flow. This type of lamp is inefficient since much of the power demanded is converted to heat, only about 8% of the energy is effectively converted to lighting (INEE, 2019). It has been chosen to use this type of load because of its low cost and the power consumed, suitable for the creation of load banks with variable power.

The electrical resistance of each lamp is calculated based on the rated voltage and power values:

$$R_{lamp} = \frac{V_{nom,l}^2}{P_{nom,l}} \tag{3.39}$$

Each lamp demands 40 W at a voltage of 24 V, so the resistance of each one is 14.4 Ω. In the model, each load bank is formed by the parallel association of five 14.4 Ω resistances, which can be activated individually. This type of load is classified as constant impedance load, and the power demanded is a function of the supply voltage.

### 3.4.2 DC Fan

The two fans used are from manufacturer SkyAuto, model SKY2406, usually employed in automotive applications. Given the lack of information provided by the manufacturer on equipment performance, a test was done to characterize this load and the results obtained are illustrated in Figure 3.11. It is noticed that the current demanded by the equipment varies with the supply voltage in a very linear way for the tested voltage range.

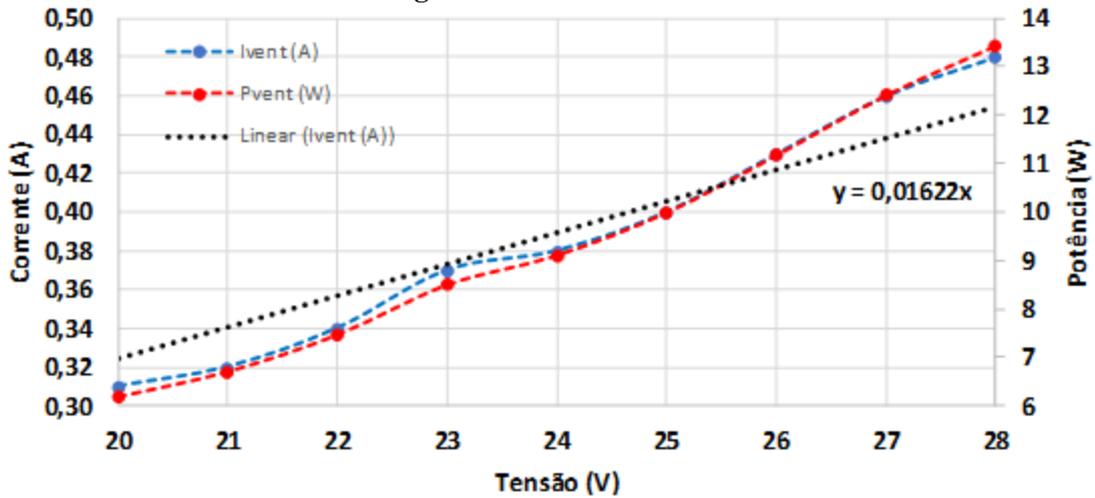
Figure 3.11 – DC fan test.

Given the linear characteristic of the fan voltage-current curve, it has been chosen to model this load as a constant resistance $R_{VENT}$, such that:

$$R_{vent} = \frac{1}{9}\sum_{k=1}^{9}\frac{v(k)}{i(k)} = 62{,}24\ \Omega \tag{3.40}$$

Where *v(k)* and *i(k)* are the measured voltage and current samples, respectively.

The characteristic curve of voltage and current obtained in Figure 3.11 could be best approximated by an affine function with angular and linear coefficients. However, it would not be possible to characterize this load in a simplified way as a resistance, making it difficult to implement the simulation.

**3.5 Characterization e modeling of the charge controller**

The charge controller is a power converter responsible for managing the charge and discharge of the battery, as well as optimizing the use of photovoltaic energy through the MPPT algorithm. As this is a commercial equipment, the schematic of the charge controller used was not provided by the manufacturer. Thus, to identify the converter topology, an inspection of the circuit board connections was done through continuity tests. After this inspection, it was found that the adopted charge controller consists of an uninsulated, step-down, DC/DC converter, known as buck, as shown in Figure 3.12.

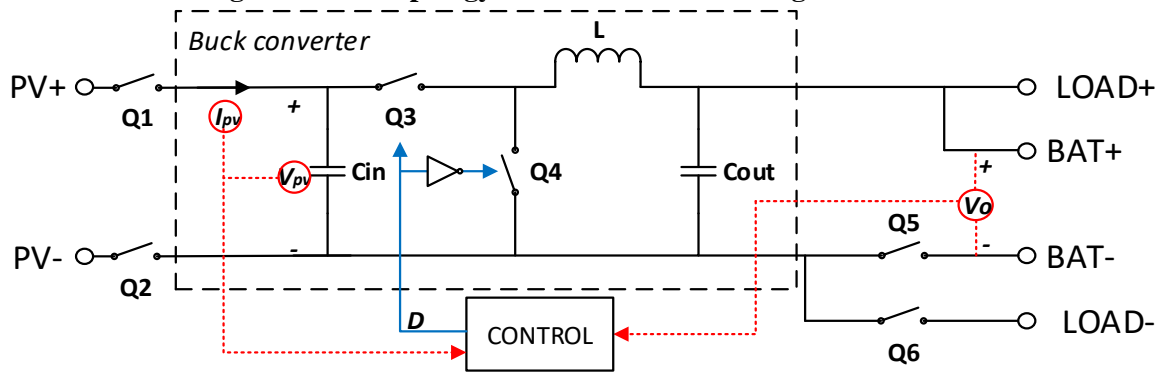

**Figure 3.12 – Topology of the commercial charge controller.**

In the circuit in Figure 3.12, switches Q1 and Q2 are implemented with power MOSFETs, model IRFB4410, with each switch in the diagram consisting in the parallel association of two MOSFETs. This parallel association is a design technique used by the manufacturer to reduce conduction losses on the switches as it reduces the equivalent internal resistance of the switch in the circuit. Switches Q1 and Q2 are used for PVG connection and disconnection and are not applied in high frequency switching. As shown in Table 2.4, this controller operates at a maximum PVG voltage of 90 V, if the voltage exceeds this value, switches Q1 and Q2 are turned off. As indicated by the manufacturer, the maximum input voltage of the PVG is 100 V, and if this value is exceeded it may cause the defect in the switches.

Switches Q5 and Q6 are also power MOSFETs of model IRFB4410, and switch Q5 is used for electronic overcurrent and overvoltage protection (maximum current of 20 A and maximum voltage of 30 V, according to manufacturer data). The Q6 switch is used for load cut-out to prevent deep discharge from the battery bank. This load cut-out is governed by a hysteresis function that prevents uncontrolled connection/reconnection of loads. According to data provided by the manufacturer in Table 2.4, the standard undervoltage disconnect value is 22.8 V and once the battery bank reaches this limit, Q6 is turned off. Once disconnected, Q6 is reconnected only when the voltage exceeds the undervoltage reconnect level, which default value is 24.2 V.

Switches Q3 and Q4 are used in the implementation of the buck converter in synchronous rectifying configuration, where switch Q4 is used in place of a diode, ensuring higher DC/DC conversion efficiency. Switches Q3 and Q4 are implemented with power MOSFETs, model 057N06N, and operate according to pulse width modulated (PWM) signal with duty cycle D, generated by the control loop. In a buck converter, the output voltage $V_o$ is a function of the duty cycle and the input voltage, which in this case is the voltage of the PV generator itself, $V_{FV}$. In an ideal power converter, the input power is the same as the output power, so in an ideal buck:

$$V_o = D \times V_{PV} \tag{3.41}$$

$$I_{o,ideal} = I_{PV}/D \tag{3.42}$$

The control of the buck converter depends on the battery bank charging stage: bulk, absorption or floating. When the controller is operating at the bulk stage (*not to be confused with buck*), the MPPT algorithm is active and is usually implemented by voltage control of the PVG. In this case, the duty cycle can be obtained from the control loop illustrated in Figure 3.13, where $V_{fv}^*$ indicates the reference voltage of the PV generator, obtained from the MPPT algorithm, and $V_{fv}$ is the measured PVG voltage. The manufacturer does not say which MPPT algorithm is used, however, the most commonly implemented in commercial solutions is perturb and observe (P&O). Other algorithms such as incremental conductance, fuzzy logic, among others, can also be used (XIAO, 2017).

The MPPT algorithm has as input the measured power of the PVG and returns a reference voltage to the PVG. This voltage is compared to the measured voltage of the PVG, and the error value obtained is applied to the input of a PID regulator. The control variable of this regulator is the duty cycle, D, which is passed to the PWM generator for Q3 and Q4 switching. Therefore, at the absorption stage, the output voltage of the buck converter is not regulated (being a function of the state of charge, temperature and battery current) and the input voltage is regulated.

**Figure 3.13 - Control loop to obtain the duty cycle in the bulk stage.**

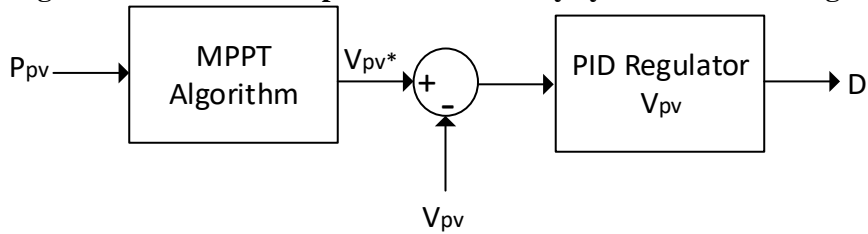

For simulation purposes, it was assumed that the MPPT algorithm implemented by the charge controller can track the maximum power voltage with a constant efficiency of $\eta_{MPPT} = 1$, i.e. it was considered that in the bulk stage the PVG was operating exactly at the maximum power point. The voltage at the maximum power point is calculated using Equation (3.43), where $\beta_{MP}$ corresponds to the thermal coefficient of variation of the maximum power voltage. It has been chosen not to use any practical MPPT algorithm to reduce the computational effort of the simulation, in order to reduce the sampling time and allow the efficient simulation of long periods. The voltage sweeping behavior illustrated in Figure 2.10 was not implemented in simulation, since shading effects are not being considered in the modeling developed in this work. The algorithm implemented in simulation to obtain the PVG voltage at the bulk stage is presented in Figure 3.14.

$$V_{MP} = V_{MP,STC}(1 + \beta_{MP}\Delta T_f) + N_S V_{th} ln\left(\frac{G_i}{G_{i,STC}}\right) \qquad (3.43)$$

$$V_{th} = \frac{A_n k T_{c,k}}{q} N_s \qquad (3.44)$$

**Figure 3.14 - Obtaining the PVG voltage at the bulk charging stage.**

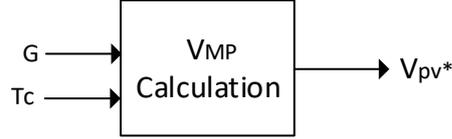

When battery charging reaches the absorption stage, the duty cycle is controlled to regulate the converter output voltage, with the default value of $V_{abs}$ = 28.8 V for a period of two hours. At this stage, the MPPT algorithm stops operating, and the PVG operating point tends to the open circuit, limiting the generator power.

At the end of the 2-hour absorption period, the charge controller assumes that the bank is fully charged and reduces the voltage and current injected into the battery by entering the float region. The duty cycle is still being set to control the converter output voltage, $V_o$, to a standard constant voltage of $V_{flt}$ = 27 V, remaining at this stage until the PV generation is insufficient. The control loop used in the absorption and fluctuation stages is illustrated in Figure 3.15.

**Figure 3.15 - Control loop to obtain the duty cycle in the absorption and fluctuation stages.**

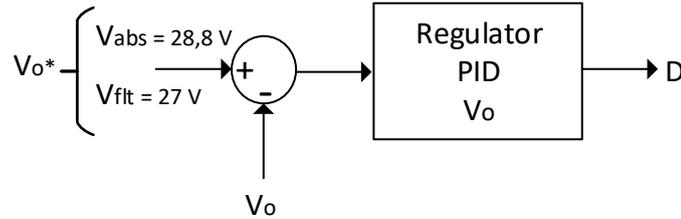

### 3.5.1 Conversion efficiency

The main energy losses associated with the charge controller are related to the DC/DC conversion for charging the battery bank by the PVG, as well as the self-consumption of the equipment. In the developed model, the following sources of power losses were considered: dissipated power in the conductor resistance of the MOSFETs, $P_{Rds\_on}$; switching losses, $P_{SW}$; dead time losses in each switching cycle, $P_{DT}$; dissipated power in the internal resistance of the inductor, $P_{RL}$; and the self-consumption of the equipment, $P_{auto}$. In the model, these losses were implemented as a reduction in the buck converter output current, Io, so that the non-ideal converter output current equation is given by:

$$I_o = \frac{I_{PV}}{D} - \frac{P_{Rds\_on} + P_{SW} + P_{DT} + P_{RL} + P_{auto}}{V_o} \tag{3.45}$$

Conduction loss in the MOSFETs is due to the drain-source conduction resistance. For simulation purposes, it has been assumed that this resistance is fixed and equal to $R_{ds\_on} = 2.5$ mΩ (obtained from the MOSFET manufacturer data sheet). Therefore, the dissipated power in the two MOSFETs is calculated by:

$$P_{Rds\_on} = 2 \times R_{ds\_on} \left(\frac{I_{PV}}{D}\right)^2 \tag{3.46}$$

Switching losses correspond to the rise and fall times in the transition between the on and off states of a MOSFET. This loss is calculated by Equation (3.55), where $t_s$ and $t_d$ are the rise and fall times, and $f_{sw}$ is the converter's switching frequency.

$$P_{SW} = 0{,}5 \times V_{PV} \times \frac{I_{PV}}{D} \times f_{SW} \times (t_s + t_d) \tag{3.47}$$

Dead time losses correspond to the interval at which both MOSFETS Q3 and Q4 remain in the off position at the same time. This event occurs between on-off transitions to avoid situations where the two switches are in the on position, causing a short in the power supply. This loss is calculated by Equation (3.56), where $T_{dead\_time}$ is the time both keys remain off.

$$P_{DT} = 2 \times V_{PV} \times \frac{I_{PV}}{D} \times t_{dead\_time} \tag{3.48}$$

The resistive loss in the inductor is a function of its internal DC resistance, $R_L$. The power dissipated in the inductor is given by:

$$P_{RL} = R_L \left(\frac{I_{PV}}{D}\right)^2 \tag{3.49}$$

Self-consumption loss corresponds to the power required to supply the control circuits of the charge controller and its indication LEDs. This power was obtained in experimental measurement by connecting only the battery to the charge controller and measuring the power consumed. A $P_{auto} = 5$ W experimental value was obtained.

Table 3.5 presents all values considered in the calculation of the dissipated powers relative to the charge controller.

**Table 3.5- Parameters used in charge controller loss simulation.**

| $R_{ds\_on}$ | 2.5 mΩ | $t_{dead\_time}$ | 5 ns | $f_{sw}$ | 50 kHz |
|---|---|---|---|---|---|
| $R_L$ | 3 mΩ | $t_s$ | 10 ns | $t_d$ | 10 ns |

From Equation (3.45) one can calculate the conversion efficiency of the charge controller using Equation (3.50).

$$\eta_{DC/DC} = \frac{V_o I_o}{V_{PV} I_{PV}} \tag{3.50}$$

### 3.5.2 Charge controller parallelism

A commercial charge controller for PV systems is designed to operate isolated, being the PVG and the battery bank that are connected to it the only power sources for the system, and the power flow is unique in the BB/PVG direction to the loads. In the proposed application of the DCDN, however, connecting multiple controllers in parallel to the grid introduces a parallelism effect that confuses the charge controller control system.

Since the load terminal is connected to an energized nanogrid, power flow from the load terminal to the BB may occur, given the topology shown in Figure 3.12, if switches Q5 and Q6 are closed. In other words, the voltage level between different battery banks tends to be the same as a more charged battery bank can charge a less charged bank via the load terminal of the controller.

In addition, in the event of a contingency where a battery bank is out of the system, the PVG corresponding to the GSS whose BB is out can continue to inject power into the grid. This is because the charge controller interprets that the voltage at the load terminal of the DCDN corresponds to the voltage at the BB.

### 3.5.3 Description of the model

The block developed on the Simulink platform has the following variables as inputs: PVG current, BB voltage, load current, incident irradiance and ambient temperature. From these inputs, and considering the logic of the battery charging stages, the block returns as output the PVG voltage and the BB current.

It was decided not to implement a circuit model, referring to Figure 3.12, since switches Q3 and Q4 are switched at high frequency, in the order of tens of kHz, which would make the simulation excessively slow, making it impossible to perform long term simulations. In this sense, the developed model cannot be used to evaluate certain inherent dynamic characteristics of high frequency switching in the power converter. Nevertheless, effects such as power losses due to high frequency switching are considered in the model.

Figure 3.16 presents the summary flowchart for the charge controller simulation.

**Figure 3.16 - Simulation flowchart for obtaining (a) PVG voltage and (b) BB current.**

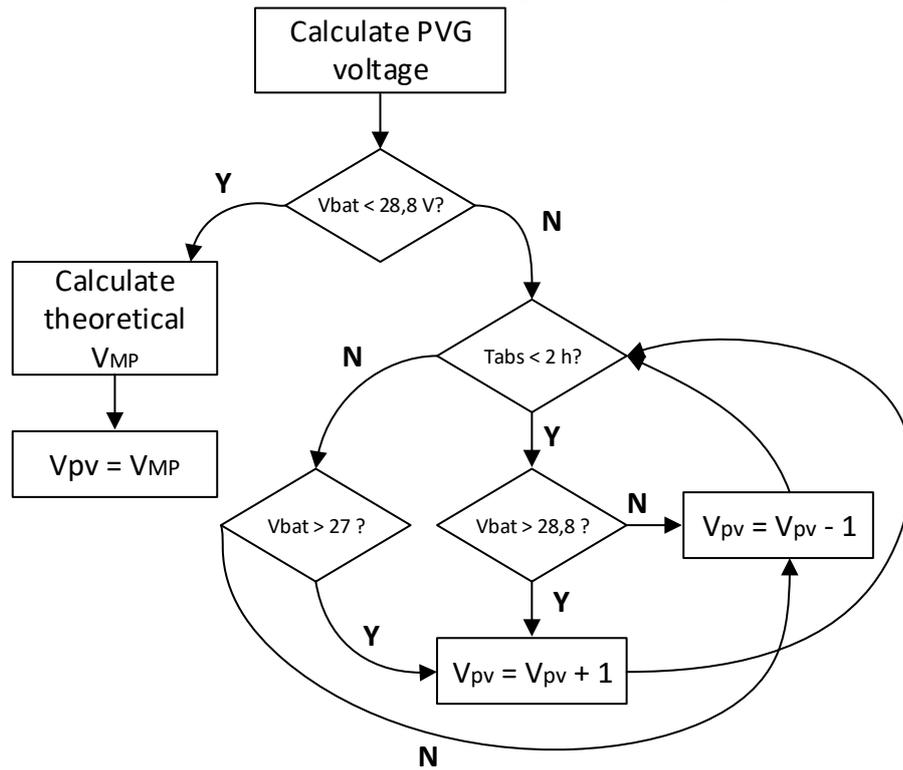

(a)

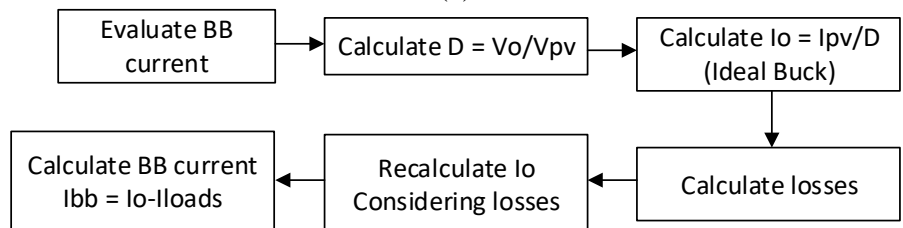

(b)

The battery voltage signal is connected to a Simulink block of voltage-controlled source, and the terminals of this block are connected to the DCDN model. The connection of the PVG, BB and charge controller blocks forms an GSS, implemented on the Simulink platform as illustrated in Figure 3.17.

**Figure 3.17 - GSS implementation on Simulink platform.**

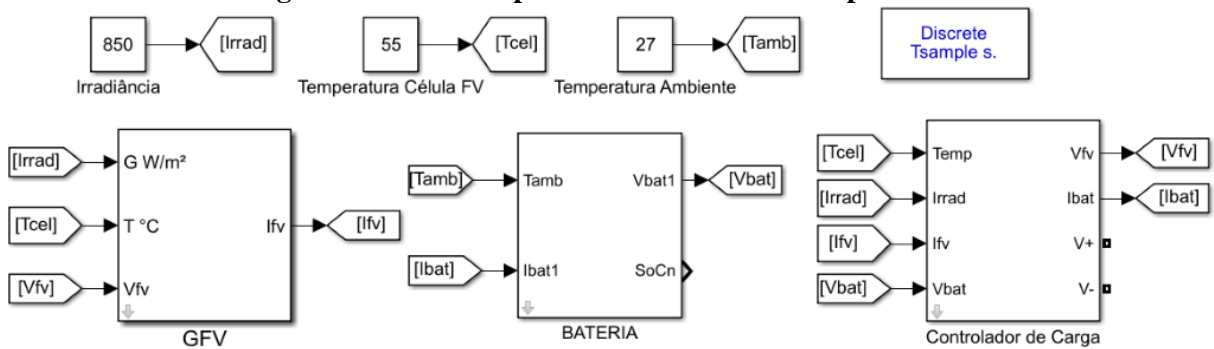

## 3.5.4 Model evaluation

Charge and discharge tests of the BBs in operation in the DCDN were performed with the charge controller to evaluate de adequacy of the proposed model, being the PVG the sole source for battery charging. In the simulation, the GSS was implemented as illustrated in Figure 3.17.

Figure 3.18 presents the results obtained for battery charging test. The presented values are integrated in 1-minute intervals. The total duration of the test was approximately 52 hours, starting at 9h30 am on 11/01/2019 and ending at 12 pm on 13/01/2019. Global irradiance data at the PVGs plane and ambient temperature are shown in Figure 3.19. In this test, the initial SoC of each BB was assumed 20 %.

**Figure 3.18 - Battery voltage and current - comparing experimental and simulated values: (a) GSS1, (b) GSS2 and (c) GSS3.**

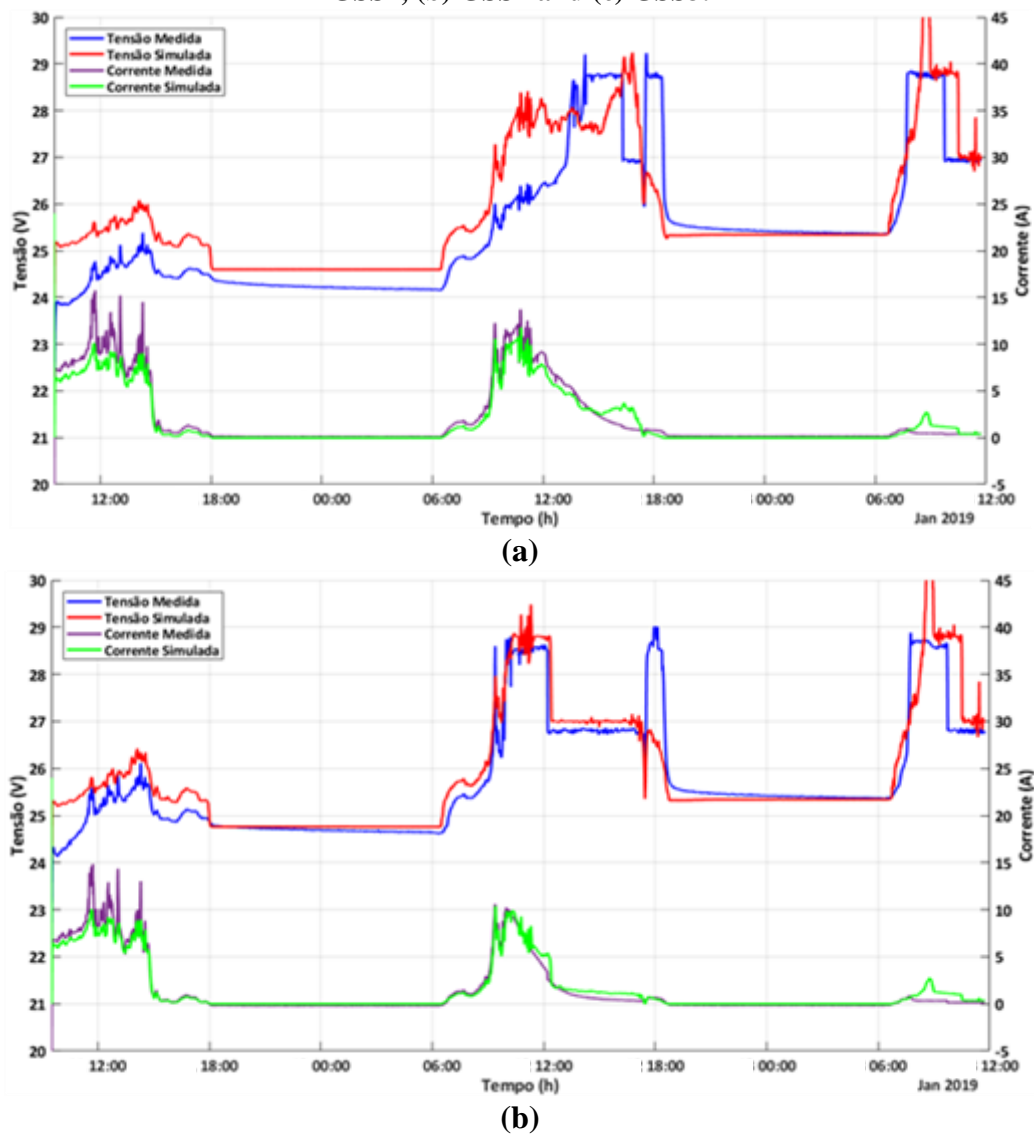

(a)

(b)

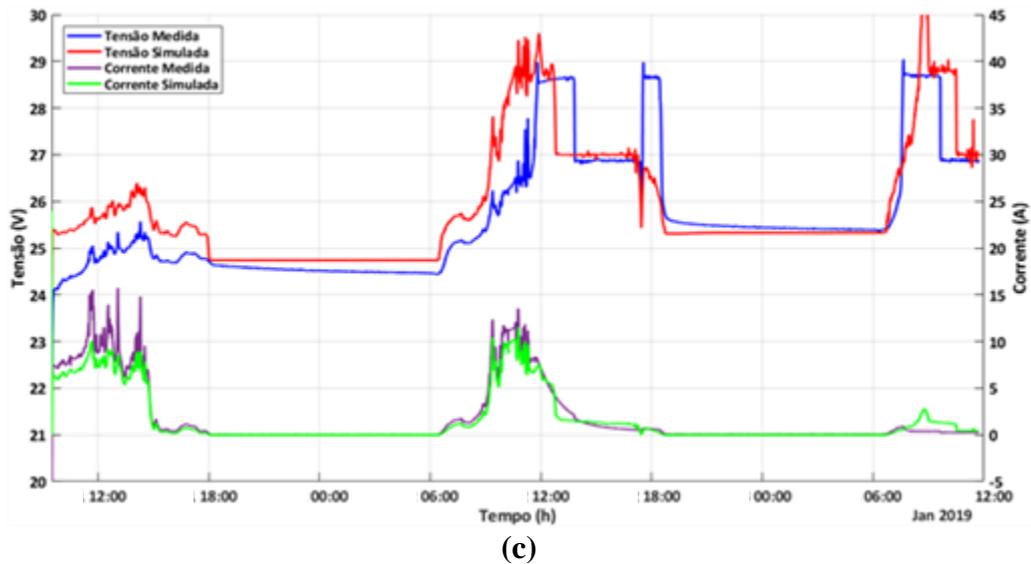

**(c)**

**Figure 3.19 - Environmental data measured and used in the simulation.**

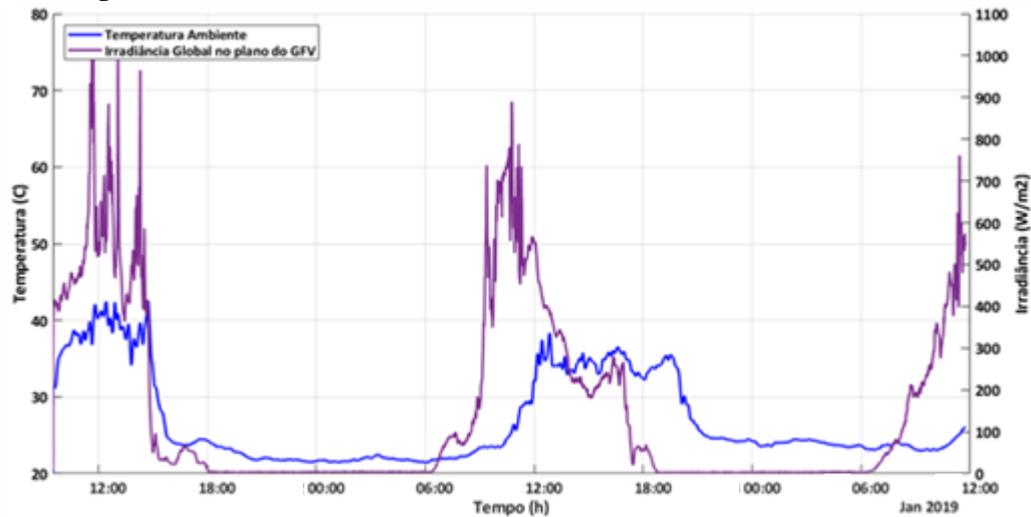

The control actions of the charge controller can be observed in the experimental tests: when the BB reaches 28.8 V, the voltage starts to be regulated and the current gradually decreases over a period of two hours. At the end of the absorption stage, the float stage begins, where the voltage is set to 27 V. Another characteristic behavior of this charge controller occurs at the end of the day if it is in the floating stage: the voltage is raised to the limit value of the absorption stage and then decays when there is no more PV generation.

BBs reached full charge only on the second day, as the test started with the BB fully discharged. On the third day, the voltage variation results from absorption and fluctuation control only, and a minimum current is injected into the BB. It is also noticed that the BB1 took the longest to reach full charge, leaving the absorption stage only in the late afternoon. This is justified as BB1 obtained the highest rated capacity in bench testing.

Despite following the voltage profile presented in experimental data, the simulation results presented discrepancy in relation to the measured data, especially in the region of low state of charge, where higher voltages were always obtained. The simulation for GSS2 was the closest to the experimental values, so that it was possible to reproduce with good fidelity the behavior of the charge controller in the transitions between the bulk, absorption and fluctuation stages. Regarding the charging current, in all cases there was a good performance of the model, and the greatest divergences occurred on the third day, when simulations returned a current higher than the experimental data.

Besides the difficulty of characterizing the battery parameters, another factor that increases model errors is due the sampling time used in the simulation. In order to simulate long periods (in this case, more than two days), a low sampling frequency should be adopted, which compromises the control strategies implemented in the charge controller model. Otherwise, high sample rate simulation makes long time simulations unfeasible.

The other test performed to validate the behavior of the charge controller refers to load cut-out to prevent deep discharge of the BB. The test shown in Figure 3.20 corresponds to the discharge of BB2, performed with DCDN loads on 16/01/2019.

**Figure 3.20 - Charge controller discharge test for BB2.**

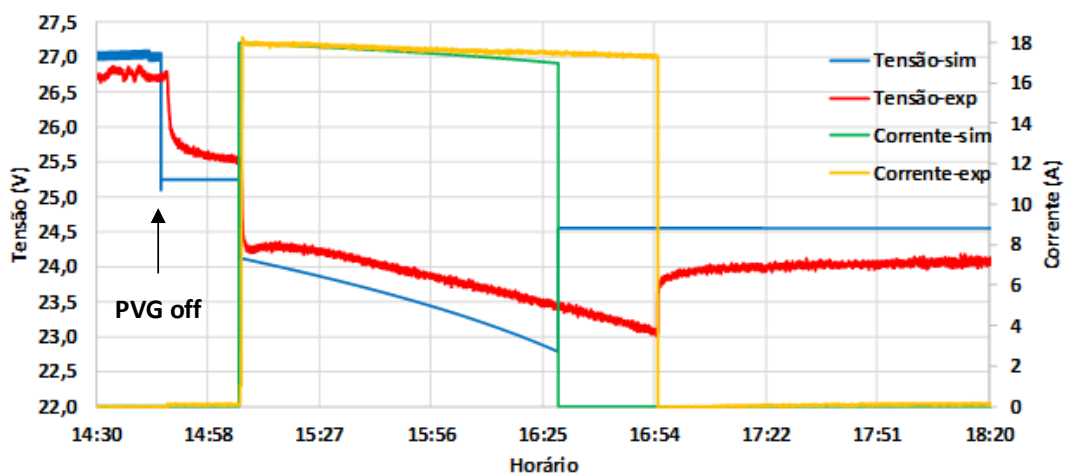

Initially, the BB was fully charged, being regulated by the charge controller to operate in the floating stage (27 V). Around 2h45 pm, the GSS2 PVG was turned off to start the BB2 discharge test. Thus, it is noticed the voltage drop in the BB before the beginning of the discharge, stabilizing at about 25.5 V. Regarding the behavior of the simulation, it is noticed that the battery voltage varies abruptly in the transition between the float stage and the PVG shutdown. This is because the model used does not consider the capacitive effect that occurs in battery transients.

At about 3h10 pm the discharge test begins: four lamps in each LB and the two fans, totaling an initial discharge current of 18 A. As the BB discharges and its voltage decreases, is observed a reduction in discharge current, given the constant impedance characteristic of lamps and fans. When the BB reaches the lower voltage limit, the charge controller disconnects the loads.

As can be seen in Figure 3.20, the charge controller model can reproduce the experimentally observed behavior: voltage regulation in the fluctuation stage and load cut-out to prevent deep discharge of the batteries. The biggest difference observed between experimental and simulated values is due to the battery bank voltage, which is justified by the inaccuracies of the parameters used in the battery modeling. For example, if the internal resistance of the battery is lower than the value obtained from the parameters used in the literature, the simulated voltage shown in Figure 3.20 tends to approximate to the measured voltage.

## 3.6 Conductors characterization and modeling

The distribution network conductors were modeled as purely resistive elements. It is important to highlight that although the distribution is in direct current, capacitive and inductive effects may appear in the conductors due to the presence of high frequency harmonic components originated in the grid-connected power converters, however these effects will not be evaluated in this work.

Considering the conductor used, the resistance between two nodes n and m of the grid is a function of the conductor length in kilometers, $L_{COND}$, and its temperature, $T_{COND}$, according to Equation (3.51). Resistance values per kilometer $r_{DC}$ = 0.8037 Ω/km and the thermal coefficient $\alpha_r$ = 0.00403 Ω/°C were obtained from the manufacturer's catalog.

$$R_{n,m} = 2 \times [L_{COND} \times r_{c.c.} + \alpha_r \times (T_{COND} - 20)] \tag{3.51}$$

Table 3.6 presents the resistance values obtained for the distances of each branch of the grid, as well as considering a conductor temperature scenario of 30 °C.

**Table 3.6 – Conductor resistances for each DCDN branch**

| Branch | Distance (m) | Resistance at 30 °C |
|---|---|---|
| N1 - N5 | 6.00 | 0.0902 |
| N5 - N12 | 11.35 | 0.0988 |
| N5 - N8 | 11.25 | 0.0987 |
| N12 - N4 | 11.35 | 0.0988 |
| N12 - N6 | 17.35 | 0.1085 |
| N4 - N2 | 28.70 | 0.1267 |
| N4 - N10 | 9.65 | 0.0961 |
| N8 - N7 | 17.25 | 0.0987 |
| N8 - N9 | 11.25 | 0.1083 |
| N10 - N11 | 38.35 | 0.1422 |
| N9 - N3 | 28.50 | 0.1264 |
| N9 - N10 | 9.65 | 0.0961 |

In the Simulink environment model, the resistive branch block was used to implement the distribution grid conductors. Considering the DCDN topology indicated in Figure 2.16b and the resistance values in each section indicated in Table 3.6, the distribution grid was implemented in Simulink as illustrated in Figure 3.21.

The implemented resistive branches are part of the SimPowerSystems library, also developed by MathWorks in partnership with Hydro-Quebéc and is embedded in the Simulink platform.

**Figure 3.21 - Distribution grid implemented in Simulink environment.**

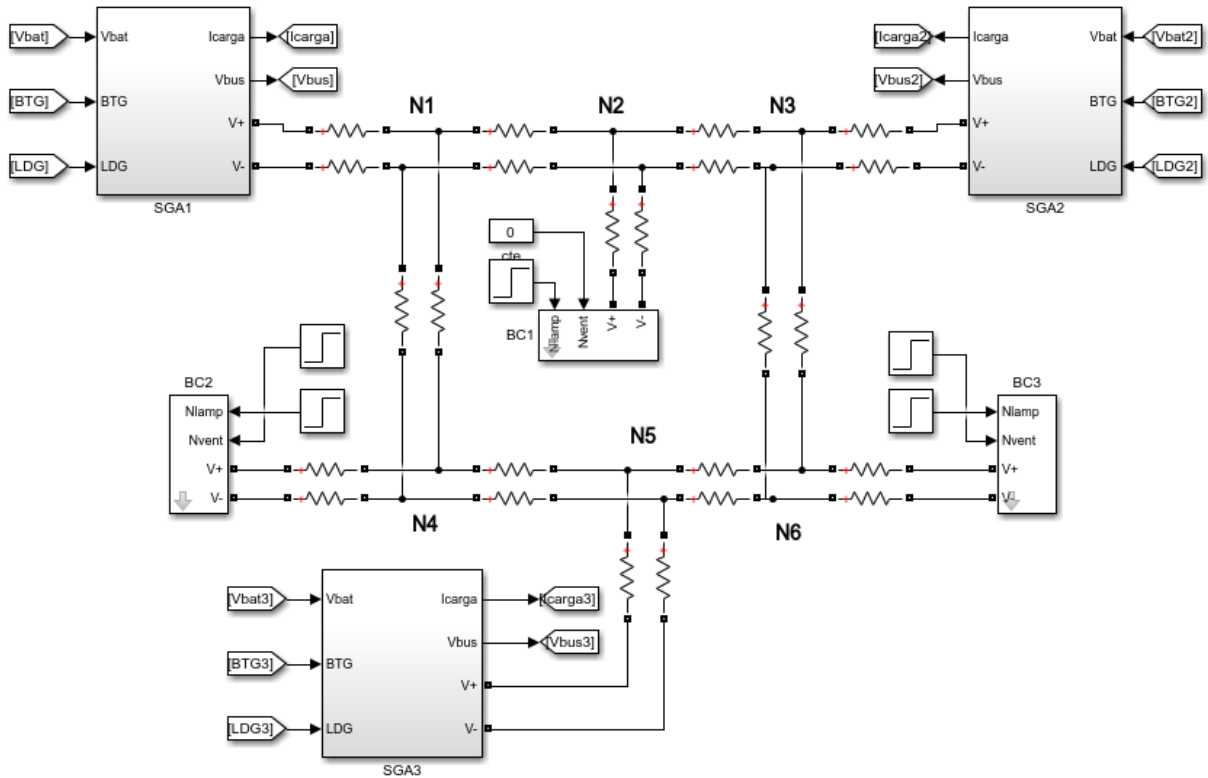

## 3.7 Static Simulation: Newton-Raphson Method for DC Power Flow

The DCDN static analysis corresponds to the load flow (or power flow) in the grid under specific operating conditions. The load flow study allows the grid operating point to be obtained and is the first step towards other more complex studies such as fault analysis and transient and permanent stability studies. Load flow analysis alone is also important for verifying system serviceability in various steady-state scenarios.

Iterative numerical methods are widely used in solving AC load flow problems and can be adapted for application in direct current networks (FLEISCHER; MUNNINGS, 1996) (FAROOQ et al., 2014) (JAYARATHNA et al., 2014). The main methods described in the literature are Gauss-Siedel, Newton-Raphson and Forward-Backward Sweep (the latter for radial grids only). For this dissertation it was chosen to use the Newton-Raphson (NR) method, given its faster convergence and wide use in this type of study.

The standard problematization of DC power flow can be defined as follows: given a distribution grid with defined voltage sources and constant power loads, obtain the voltages in the grid nodes and the currents in the lines. For a better understanding of the problem formulation, the three-bus grid is adopted as an example in Figure 3.22.

**Figure 3.22 - Three-bus grid to illustrate DC power flow.**

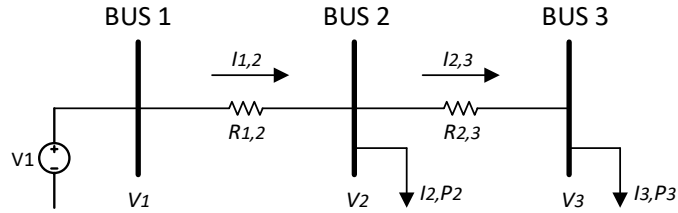

In the example in Figure 3.22, the bus 1 is a generation bus because it has a known constant voltage source, V1. Busses 2 and 3 are load busses, as they are connected to known constant power loads P2 = P2 'and P3 = P3', respectively. In a DC grid, it is considered for steady state analysis that the distribution lines are purely resistive, represented by the resistances $R_{1,2}$ and $R_{2,3}$. The load flow aims to obtain the voltages V2 and V3 in the load busses 2 and 3, respectively and the currents in the lines, $I_{1,2}$ and $I_{2,3}$.

The first step to solve the problem is to obtain the grid's conductance matrix. The process is similar to that used to obtain the admittance matrix of AC grid. Considering that the impedances in the lines are purely resistive, the admittance matrix is reduced to a conductance matrix $G_{N,N}$, where N is the number of busses in the grid. For the example of Figure 3.22, the grid's conductance matrix is given by:

$$G = \begin{bmatrix} G_{11} & G_{12} & G_{13} \\ G_{21} & G_{22} & G_{23} \\ G_{31} & G_{32} & G_{33} \end{bmatrix} = \begin{bmatrix} 1/R_{1,2} & -1/R_{1,2} & 0 \\ -1/R_{1,2} & 1/(R_{1,2}+R_{2,3}) & -1/R_{2,3} \\ 0 & -1/R_{2,3} & 1/R_{2,3} \end{bmatrix} \quad (3.52)$$

The powers in the busses can be expressed as a function of the voltages in the busses, according to Equation (3.53):

$$\begin{bmatrix} P_1 \\ P_2 \\ P_3 \end{bmatrix} = \begin{bmatrix} V_1(G_{11}V_1 + G_{12}V_2 + G_{13}V_3) \\ V_2(G_{21}V_1 + G_{22}V_2 + G_{23}V_3) \\ V_3(G_{31}V_1 + G_{32}V_2 + G_{33}V_3) \end{bmatrix} \quad (3.53)$$

Taking a Taylor series approximation, and considering only the first degree of expansion, small variations in voltage ($\Delta V$) correspond to small variations in power ($\Delta P$):

$$\begin{bmatrix} \Delta P_1 \\ \Delta P_2 \\ \Delta P_3 \end{bmatrix} = \begin{bmatrix} \dfrac{\partial P_1}{\partial V_1} & \dfrac{\partial P_1}{\partial V_2} & \dfrac{\partial P_1}{\partial V_3} \\ \dfrac{\partial P_2}{\partial V_1} & \dfrac{\partial P_2}{\partial V_2} & \dfrac{\partial P_2}{\partial V_3} \\ \dfrac{\partial P_3}{\partial V_1} & \dfrac{\partial P_3}{\partial V_2} & \dfrac{\partial P_3}{\partial V_3} \end{bmatrix} \begin{bmatrix} \Delta V_1 \\ \Delta V_2 \\ \Delta V_3 \end{bmatrix} \quad (3.54)$$

Equation (3.54) can be rewritten symbolically, according to Equation (3.55), where $J$ is known as the Jacobian matrix, containing the partial derivatives of the powers in relation to the bus voltages.

$$[\Delta P] = [J][\Delta V] \quad (3.55)$$

In the formulation of the load flow problem of the example, the voltage V1 is known, so Equation (3.54) can be reduced to:

$$\begin{bmatrix} \Delta P_2 \\ \Delta P_3 \end{bmatrix} = \begin{bmatrix} \dfrac{\partial P_2}{\partial V_2} & \dfrac{\partial P_2}{\partial V_3} \\ \dfrac{\partial P_3}{\partial V_2} & \dfrac{\partial P_3}{\partial V_3} \end{bmatrix} \begin{bmatrix} \Delta V_2 \\ \Delta V_3 \end{bmatrix} \quad (3.56)$$

The Jacobian matrix can be inverted to obtain residual changes in voltage as a function of residual changes in power:

$$[\Delta V] = [J]^{-1}[\Delta P] \quad (3.57)$$

From these equations, the iterative process begins considering the following steps:

1- Assign estimated values for V2 and V3 (initial guess, usually 1 p.u.);
2- Calculate P2 and P3 by Equation (3.53);
3- Calculate the difference between the calculated and the specified powers:

$$\begin{bmatrix} \Delta P_2 \\ \Delta P_3 \end{bmatrix} = \begin{bmatrix} P_2 - P_2' \\ P_3 - P_3' \end{bmatrix} \quad (3.58)$$

4- Calculate the new estimated value for the voltages in each bus, where $i$ indicates the iteration number:

$$\begin{bmatrix} V_2 \\ V_3 \end{bmatrix}_{i+1} = \begin{bmatrix} V_2 \\ V_3 \end{bmatrix}_i - [J]^{-1} \begin{bmatrix} \Delta P_2 \\ \Delta P_3 \end{bmatrix} \quad (3.59)$$

5- Repeat steps 2 to 4 until the difference between the calculated and specified powers is negligible (convergence criterion).

The GEDAE's lab DCDN model is shown in the 12-bar diagram in Figure 3.23. To use the NR methodology presented above, it is considered that the voltages in the battery banks are known and constant (VBB1, VBB2 and VBB3), and that the loads are of constant power type: $P_{LB,1}$, $P_{LB,2}$ and $P_{LB,3}$. Each bus $N_n$ (n = 1, 2, ..., 12) corresponds to a node indicated in Figure 2.16b. The conductance matrix was obtained using the resistances presented in Table 3.6.

**Figure 3.23 - 12-bar grid used for DCDN modeling - considering constant power loads and constant battery bank voltage.**

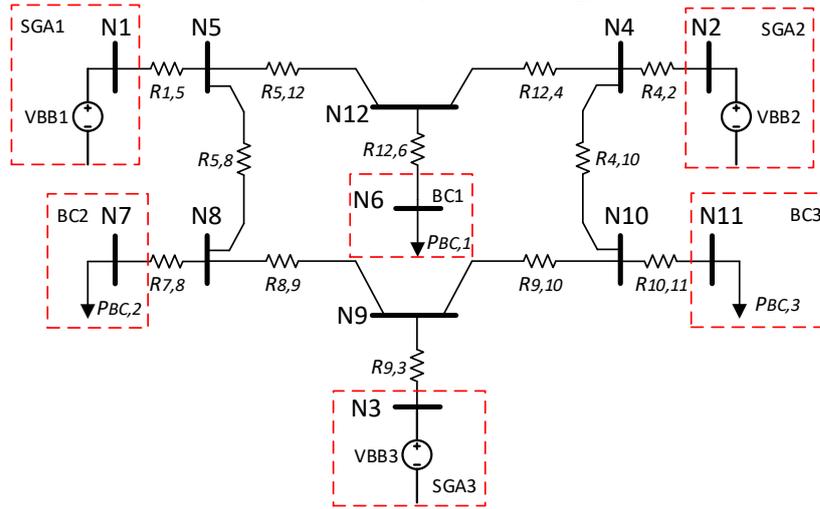

The load flow algorithm presented assumes that the system loads are of constant power type. However, according to the load bank model previously developed, incandescent lamps are purely resistive loads and their power is a function of the operating voltage, the fans also have similar behavior. To consider this characteristic in the load flow model, one must specify the load resistance rather than the power. Considering the example of Figure 3.22, given that resistances $R_{B2}$ and $R_{B3}$ represent the loads in bars 2 and 3, respectively, Equation (3.58) should be replaced by:

$$\begin{bmatrix} \Delta P_2 \\ \Delta P_3 \end{bmatrix} = \begin{bmatrix} P_2 - \dfrac{V_2^2}{R_{B2}} \\ P_3 - \dfrac{V_3^2}{R_{B3}} \end{bmatrix} \quad (3.60)$$

The restriction of this model is linked to the fact that at least one of the system bars must have regulated voltage (generation bar or constant voltage). Since the DCDN is connected directly to the battery banks, the voltage on the generation busses corresponds to the voltage of the BBs. As shown in the model used for the batteries, it is known that the battery voltage is a function of the bank's SoC, temperature and charge or discharge current, and for a more precise assessment of the load flow, these quantities should be taken into consideration.

To consider the battery voltage behavior in the model, the following steps are proposed:

1- Given a SoC and temperature, calculate the voltage of each BB for an initial discharge current (initial guess) using Equation (3.23);
2- The voltage for each BB obtained in the previous step will be used as constant voltage in the generation buses. Run the load flow as previously described;
3- Identify the currents obtained in the generation buses resulting from the load flow and compare with the currents of the previous iteration (in the first iteration, consider the initial guess);
4- Recalculate the voltage of each BB using the currents obtained in the load flow and repeat steps 2 and 3 until the difference in currents is insignificant (convergence criterion used of 0.1%[7]).

The main limitation of the proposed load flow is the non-inclusion of PV generation in the grid. Because PVG is inserted as a constant power bus and is very close to the constant voltage bus (BB), the algorithm has difficulty converging. In addition, in situations where the PVG is charging the battery, the Equation to be used to determine BB voltage is different from the equation used in case of battery discharge, which makes it even more difficult for the Newton-Raphson algorithm to converge. This constraint should be further investigated to identify a methodology for including PV generation in the simulation without compromising simulation convergence.

Despite this limitation, it is possible to consider the influence of PVG on specific operating conditions: BB voltage in absorption and floating stages. As the voltage of the BB in these stages is regulated, the load flow can be realized considering that the BB are with constant voltage equal to the absorption or floating voltages, obtaining the voltages in the loads and currents in the lines for these scenarios. Still, the most energy and power quality critical scenarios for load servicing occur when there is no PVG, as the buses voltages are lower and may make it impossible to meet the loads.

### 3.7.1 Power flow model evaluation

To verify the behavior of the developed load flow, three tests were performed at DCDN with the PVGs disconnected, and the load banks operating with the following loads:

Test 1: BC1, 4 lamps; BC2, 4 lamps and 1 fan; BC3, 3 lamps

Test 2: BC1, 5 lamps; BC2, 4 lamps and 1 fan; BC3, 3 lamps

---

[7] This convergence criterion was chosen after verifying the convergence capacity of the algorithm for different percentage values; thus an appropriate value was identified for the purpose of the simulation and the computational effort required.

Test 3: BC1, 3 lamps; BC2, 4 lamps; BC3, 2 lamps

The results of the measured and simulated load flow values are presented in Figure 3.24. It is noticed that the load flow presents voltage values close to those measured for the three DCDN loading conditions. Adopting the experimentally obtained voltage values as reference, there is a percentage error of less than 4% in all tests. What is noticeable is that the biggest errors occur in the region of highest grid loading.

The good model approximation for voltage measurements does not occur for current values, where large variations between measured and simulated data are observed (deviations above 10% occur in approximately one out of four measurements). However, one should take into account the current measurement uncertainty that is being taken as a reference.

Figure 3.24 - Comparison between measured and simulated values for tests (a) test 1, (b) test 2 and (c) test 3.

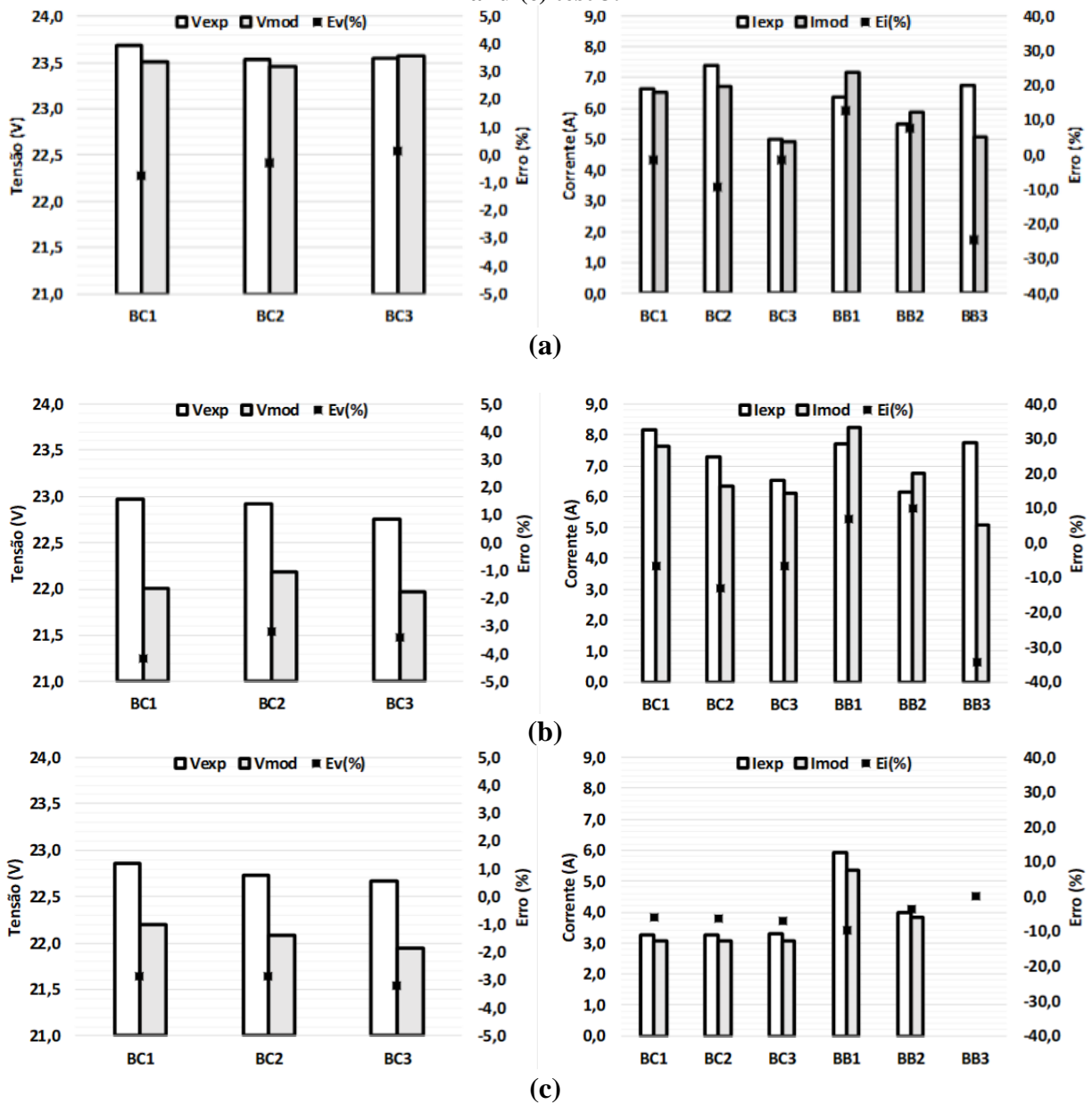

## 3.8 Dynamic simulation: system modeling on Simulink environment

Dynamic simulation comprises the integration of the model blocks developed in items 3.2 to 3.6. As indicated in the previous items, features from the SimPowerSystems library were used for the DCDN simulation. This library has specific features for power system simulation, such as controllable voltage and current source blocks, resistive, inductive and capacitive branches, among others.

Because the elements in this library are not native to the Simulink platform, a specific initialization mechanism has previously been made to allow the simulation to run. In this initial step, the equivalent state space model of the electric circuit to be simulated is obtained, characterizing the topology and the system inputs and outputs. For example, the implemented voltage and current sources are characterized as inputs, and the measurement points as outputs. From this state space model is built a model that can be interpreted by Simulink.

Thus, the simulation is performed in two stages:

1- execution of native Simulink blocks (charge controller algorithms, PVG models, BB, etc.)
2- the implementation of the grid implemented with Simpowersystems (load flow, voltages and currents at measuring points, input value update, etc.)

Simpowersystems blocks execution must occur with a multiple sampling period of native execution. It was decided to use the discrete simulation mode and fixed sampling rate, as this way it is possible to have greater control over the execution flow, which many models use, especially the battery model.

Importantly, the choice of sampling time is determined by different factors. Most important is the simulation time. Given the complexity of the modeled system, choosing a very short sampling time makes simulation impractical for long periods. However, choosing too high a sampling time can cause simulation errors, leading to instability situations.

Validation of dynamic simulation is presented in Chapter 4, comparing simulation results with measurements on an operating day.

# 4. CASE STUDIES OF DIFFERENT OPERATING SITUATIONS

## 4.1 Test under normal operating conditions

The results presented below correspond to the DCDN operation for 10/23/2018. The measured values of global irradiance in the inclined plane (approximately 10 °) and ambient temperature are shown in Figure 4.1. The measurements obtained are average values at 5-minute intervals. As observed in the irradiance chart, it was a very sunny day, with irradiance values above 1000 W/m² around 1 pm, but with considerable cloudiness in the afternoon. On this day, the solar energy available was HFS = 6.2 kWh/m².

Figure 4.1 – Measured irradiance and temperature data for 10/23/2018.

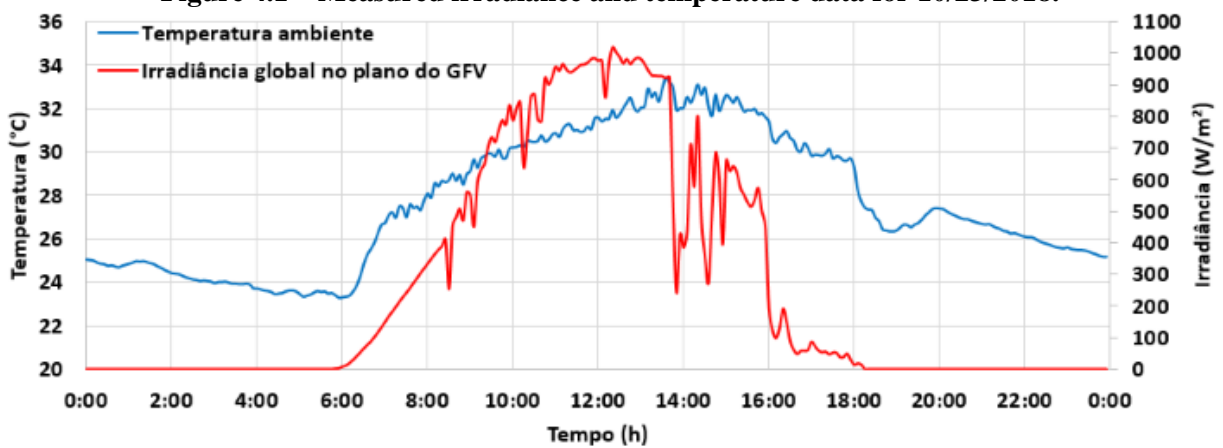

The measured voltage and current of the battery banks are illustrated in Figure 4.2 and Figure 4.3 respectively. These data were obtained by the monitoring system developed for the DCDN, and measurements are stored at 1-minute intervals. The standard sense of current measurement at the BBs corresponds to the reference adopted as in Figure 3.5, where positive current indicates battery charging.

Figure 4.2 - Voltage in the battery banks measured on 10/23/2018.

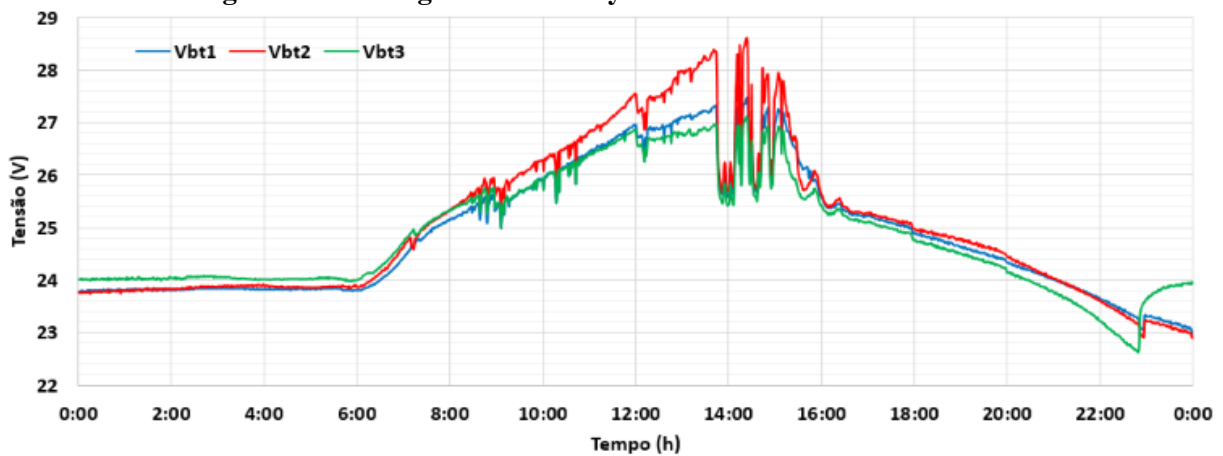

**Figure 4.3 - Current in the battery banks measured on 10/23/2018.**

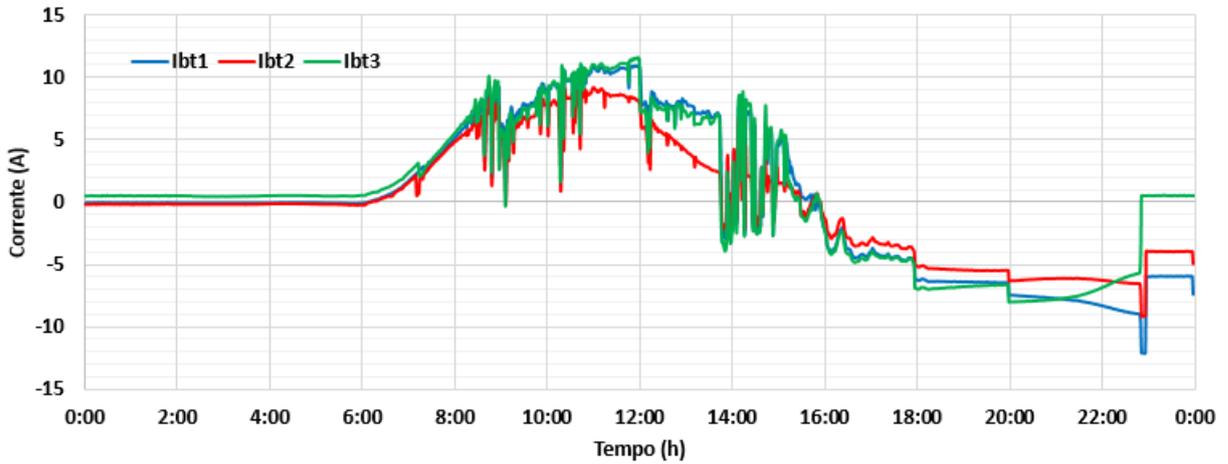

Figure 4.4 shows the measured values of power generated by each DCDN's PVG with data also stored at 1-minute intervals.

**Figure 4.4 - Power of PVGs measured on 10/23/2018.**

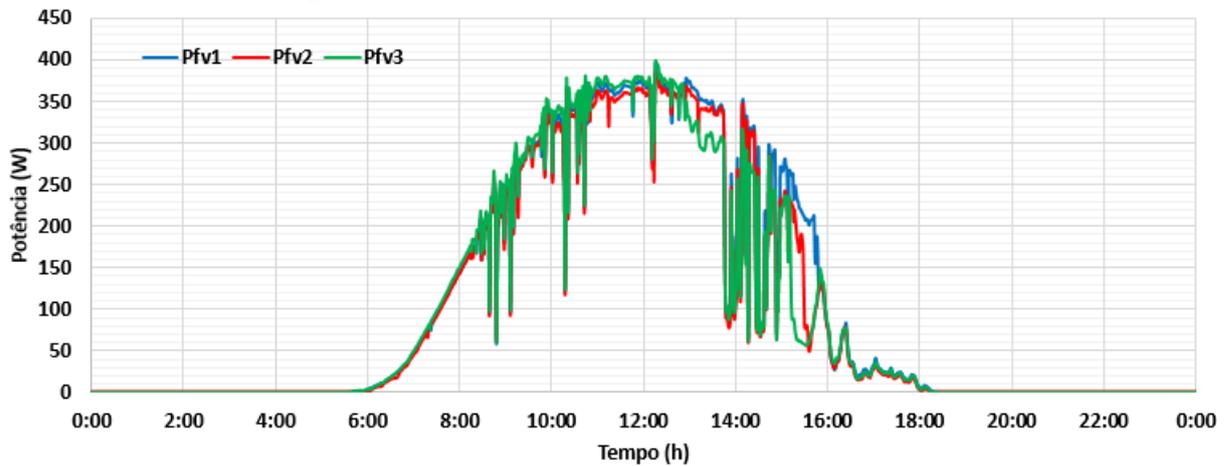

The measured voltage and current of the load banks are illustrated in Figure 4.5 and Figure 4.6, respectively. Data were also stored at 1-minute intervals.

**Figure 4.5- Voltage in the load banks measured on 10/23/2018.**

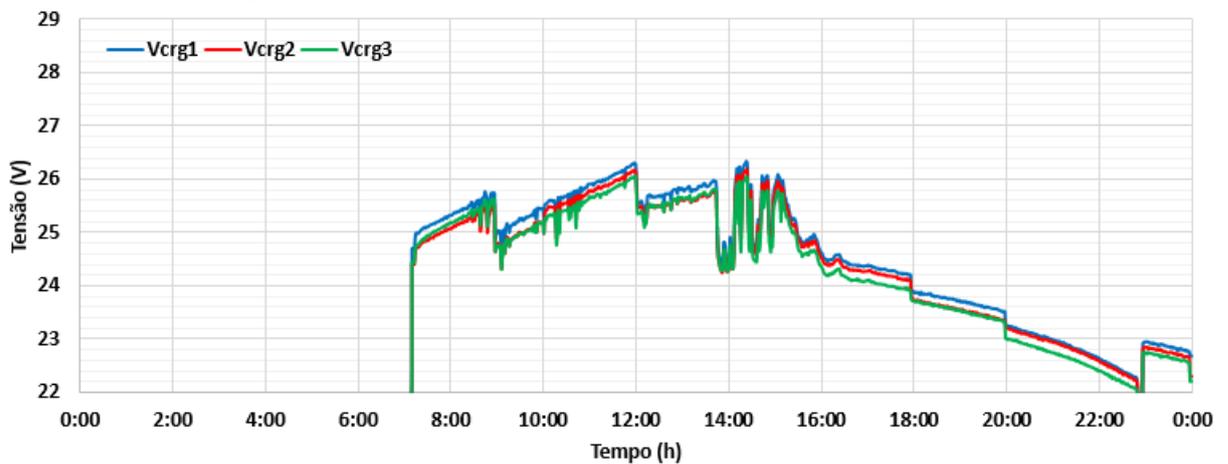

Figure 4.6- Current in the load banks measured on 10/23/2018.

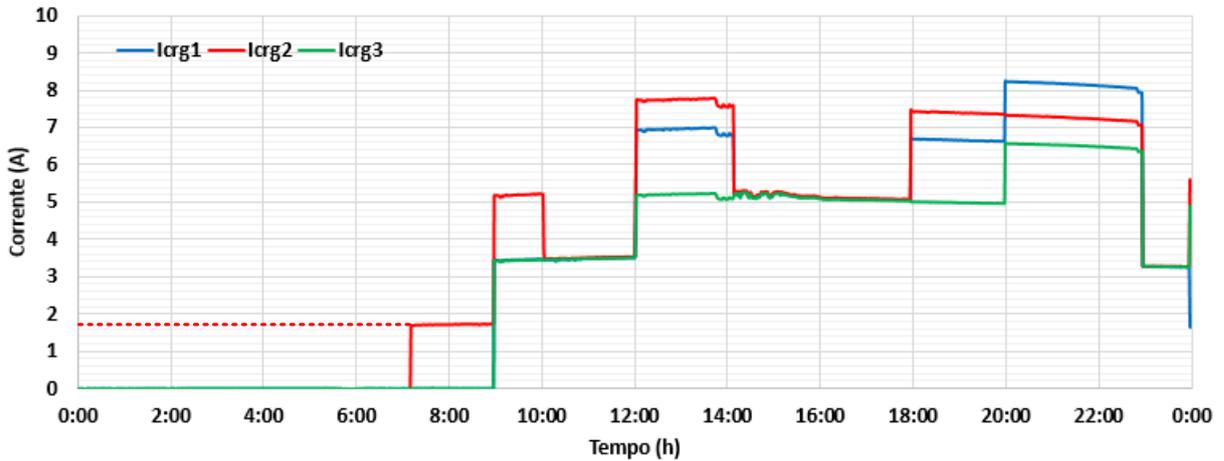

Initially, the three GSSs were disconnected from the DCDN due to the load cut-out performed by the charge controller to prevent deep discharge. The red dotted line in Figure 4.6 indicates the current that should be consumed according to the predetermined load curve for this bank, however, due to the disconnection of the three GSSs, the DCDN was completely de-energized.

At around 7:10 am, with the start of PV generation and voltage increase in the BBs, the three GSSs were reconnected to the grid by their respective charge controllers. Figure 4.7 shows in detail the voltages in the battery banks at the time of reconnection. Note that this reconnection does not occur simultaneously for all three GSSs, however, within 15 minutes, all GSSs were connected to the grid. The voltage drop that occurs at the time of reconnection is due to the discharge current of the batteries, so the most pronounced voltage drop is for the first reconnected BB (GSS2), since GSS2 is solely responsible for servicing the loads until the next GSS reconnects. However, this initial difference in voltages is soon suppressed by connecting all GSSs to the network.

Figure 4.7 - Detail of the variation of voltages at the BBs at the time of reconnection to the DCDN.

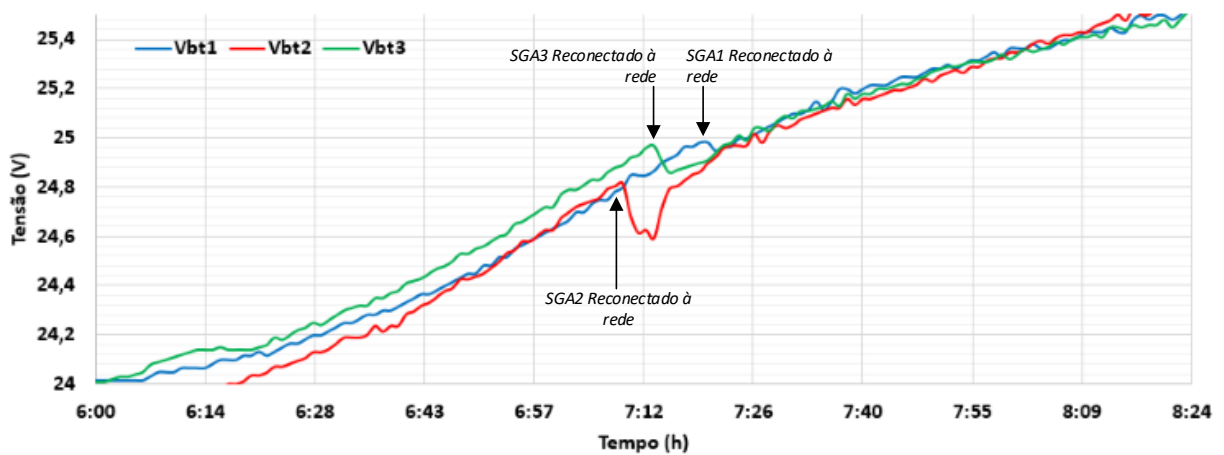

From 9 am to 2 pm there is an increase in the voltage difference of BB2 compared to the other battery banks, as indicated in the detail presented in Figure 4.8. The higher voltage of the BB2 is an indication that this bank has a state of charge higher than the others due to its lower capacity (as seen in charge and discharge bench tests), thus this bank charges faster. Given the higher voltage of BB2, the GSS2 is the one that will supply higher current to meet the loads. Therefore, part of the power delivered by the PVG2 is injected into the grid, and the excess is used to charge the battery bank. This justifies the lower BB2 charging current over the period shown in Figure 4.8 as the power provided by all PVGs is close, but the current injected into the DCDN is higher in the GSS2.

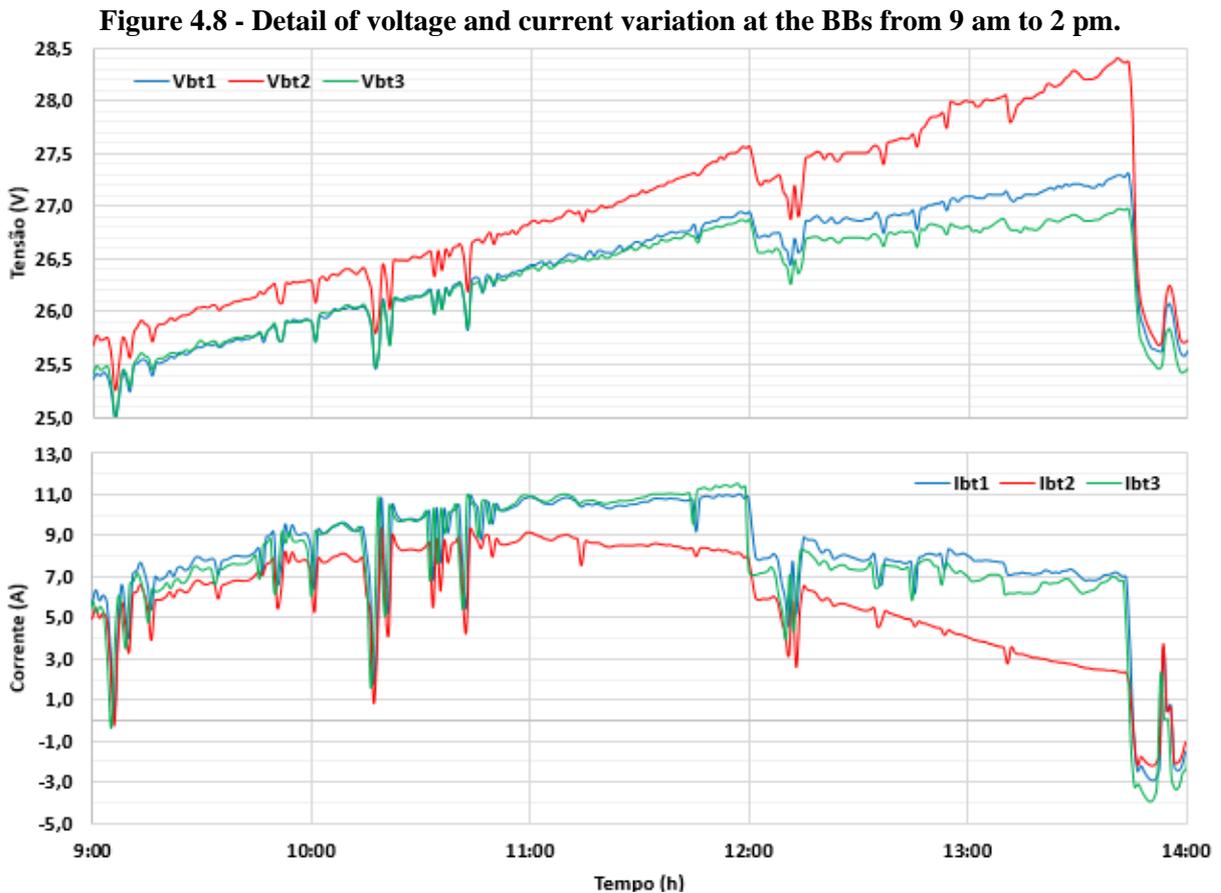

Figure 4.8 - Detail of voltage and current variation at the BBs from 9 am to 2 pm.

From 4 pm onwards, PV generation is no longer sufficient to meet the loads, requiring complementation from the BBs. With the end of the solar resource on this day, around 6 pm, the loads were met exclusively by the BBs. At about 11 pm, the voltage at the BB3 reaches the charge controller disconnect voltage limit, which removes the GSS3 from the grid. From this moment on, loads are serviced only by GSS1 and GSS2.

There is a large variation in the voltage in the battery and load banks throughout the day, and these variations are even more evident due to the variation in solar resource availability and load demand. As the loads used behave as constant impedances, there is a variation in the current consumed as a function of the voltage variations in the DCDN main bus, however this variation is not significant.

From the point of view of the end use of the loads, it was observed variation in the lighting capacity of the incandescent lamps as well as in the rotation speed of the fans and, consequently, in their ventilation capacity. Nevertheless, throughout the monitored period, no defect was observed in this equipment due to this variation in voltage.

### 4.1.1 Evaluation in terms of energy

Figure 4.9 shows the integrated power values over the day 23/10/2018 for the PVGs, BCs and BBs. The total energy produced by the three photovoltaic generators present in the DCDN is very similar, since all are operating under similar conditions of irradiance and ambient temperature. PVG1 produced slightly above the others, about $E_{PVG,1}$ = 2.41 kWh, while PVG2 and PVG3 produced $E_{PVG,2}$ = $E_{PVG,3}$ = 2.29 kWh each. Considering the nominal power of each PVG $P_{MP,STC}$ = 0.477 kWp, as shown in Table 2.2, one can calculate the productivity of each PVG, $Y_{FV}$, according to Equation (4.1). PVG1 yield on this day was 5.02 kWh/kWp and PVG2 and PVG3 yield was 4.80 kWh/kWp.

$$Y_{FV} = \frac{E_{GFV,n}}{P_{MP,stc}} \tag{4.1}$$

**Figure 4.9 - Energy (a) generated by the PVGs, (b) consumed by the BCs and (c) stored / extracted from the BBs.**

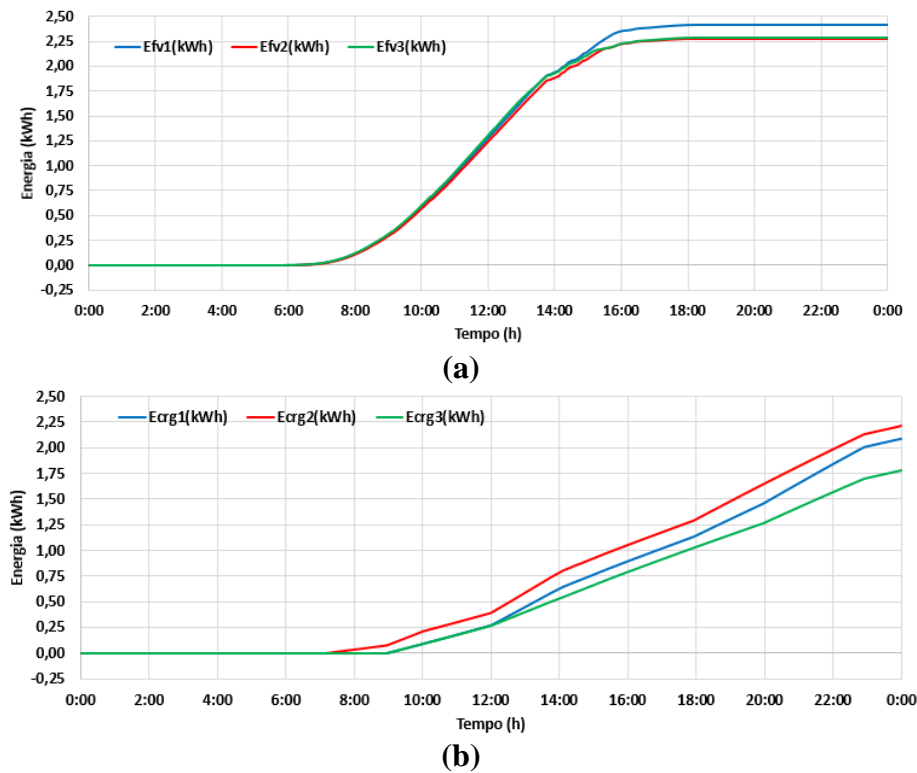

(a)

(b)

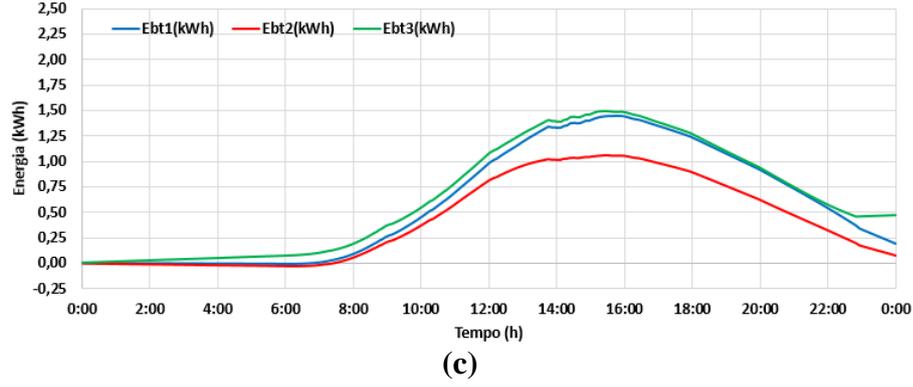

**(c)**

As shown in Figure 4.9b, the energy consumed by the load banks was 2.08 kWh, 2.22 kWh and 1.78 kWh for BC1, BC2 and BC3, respectively. This different daily consumption reflects the independent load curve implemented in each LB. In relation to the BBs, BB2 was the one that received the least energy from PVG throughout the day, as GSS2 was the one that contributed the most to the supply of loads, as previously observed. BB3 is the one that accumulated the most energy on this day, and GSS3 was the one that contributed the least to the supply of loads, being even more pronounced by the cut-out of the grid at the end of the day by its charge controller. Despite the GSS3 load shedding, the energy balance was positive in all BBs, which ended the day with more accumulated energy than they started. At the end of the day, the net energy balance was 0.19 kWh, 0.08 kWh and 0.47 kWh for BB1, BB2 and BB3, respectively.

The positive energy balance of the battery banks means that each BB can be interpreted as a load on the DCDN in terms of daily energy balance. Also in terms of daily energy balance, PV generation can be considered as only energy source in the system, so the energy balance in the DCDN can be equated as:

$$E_{GFV} = E_{BB} + E_{BC} + E_{loss} \qquad (4.2)$$

Where $E_{PVG}$ = 6.98 kWh is the sum of the energies produced by each PVG, $E_{BC}$ = 6.08 kWh is the sum of the energy consumed by each load bank and $E_{BB}$ = 0.74 kWh is the sum of the energy balance in every BB. This gives $E_{loss}$ = 0.16 kWh, which corresponds to the various sources of energy losses in the distribution (joule effect losses in the cables and connections) and in the charge controllers.

Considering that the main purpose of the distribution network is to meet the loads, the energy efficiency of supply of the DCDN can be evaluated through Equation (4.3).

$$\eta_{NRCC}(\%) = \left(1 - \frac{E_{PERDAS}}{E_{BC}}\right) \times 100 \qquad (4.3)$$

For this day, a load supply efficiency of 97.36% was obtained.

### 4.1.2 Comparison with simulation results

Figure 4.10 shows the simulation results of voltage and current in the battery banks, as modeled in Chapter 3. Figure 4.11 presents the simulation results for voltage and current in the load banks, while Figure 4.12 shows the values obtained in simulation for the PV generators output power. For comparison purposes, the measured values of the corresponding quantities in each figure are also presented, all for the day 23/10/2018.

**Figure 4.10 - Comparison of simulated and measured voltage and current values in (a) BB1, (b) BB2 and (c) BB3.**

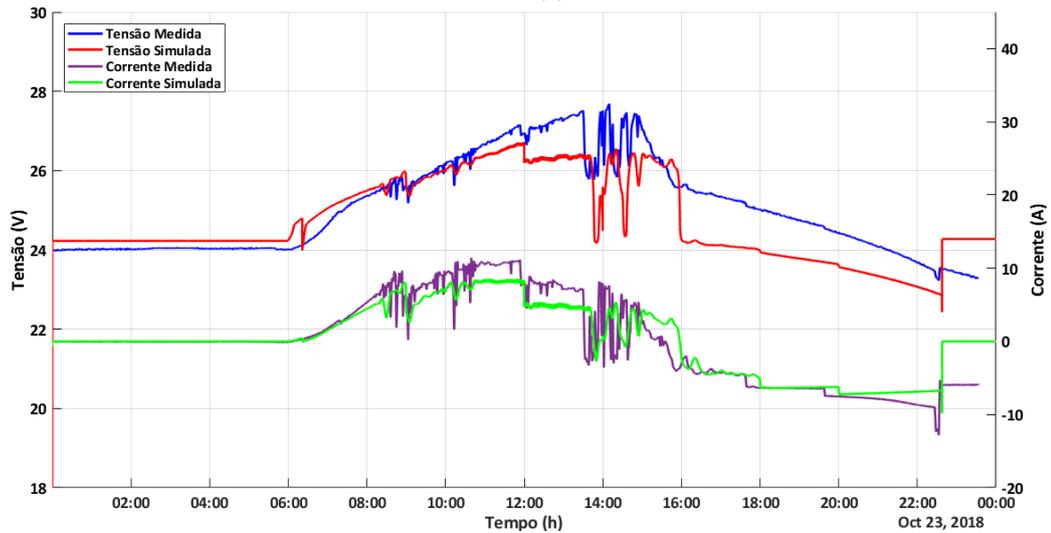

(a)

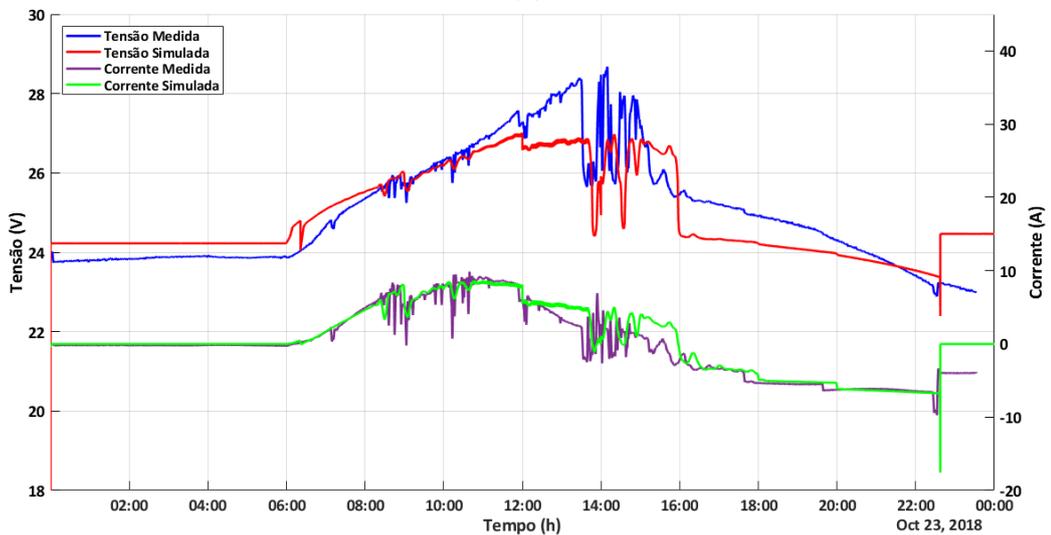

(b)

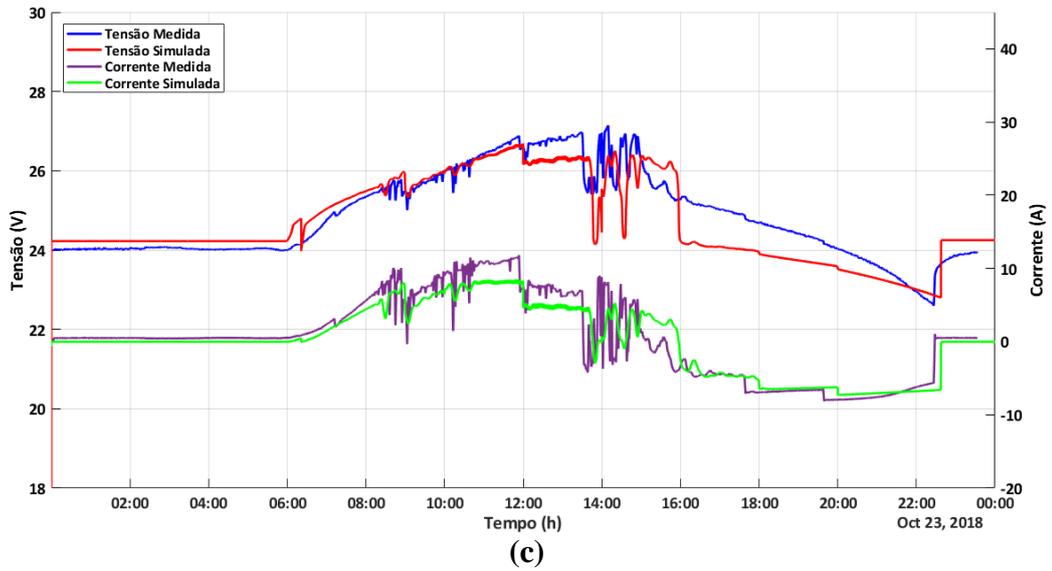

(c)

**Figure 4.11 - Comparison of simulated and measured voltage and current values in (a) BC1, (b) BC2 and (c) BC3.**

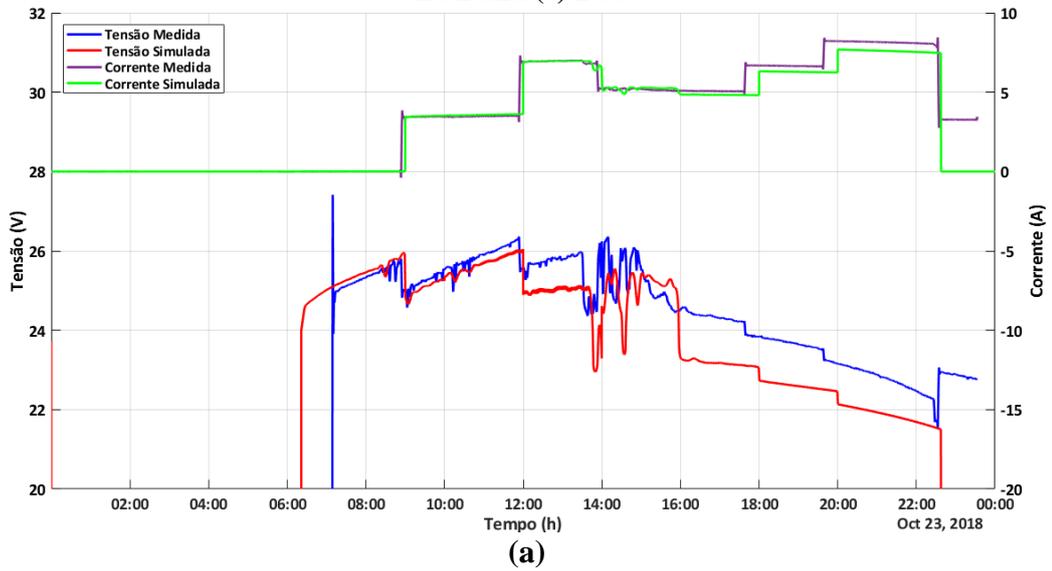

(a)

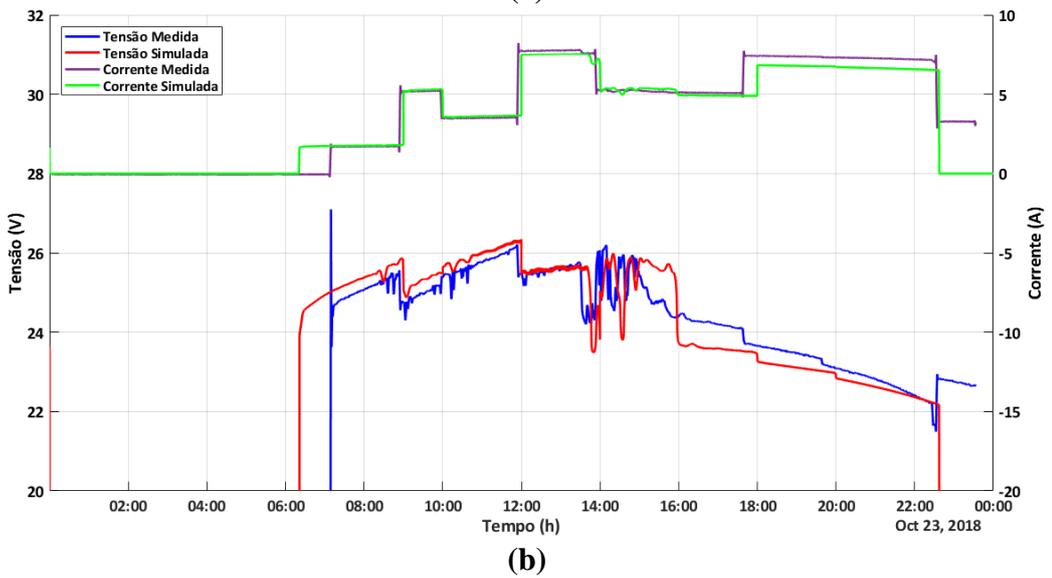

(b)

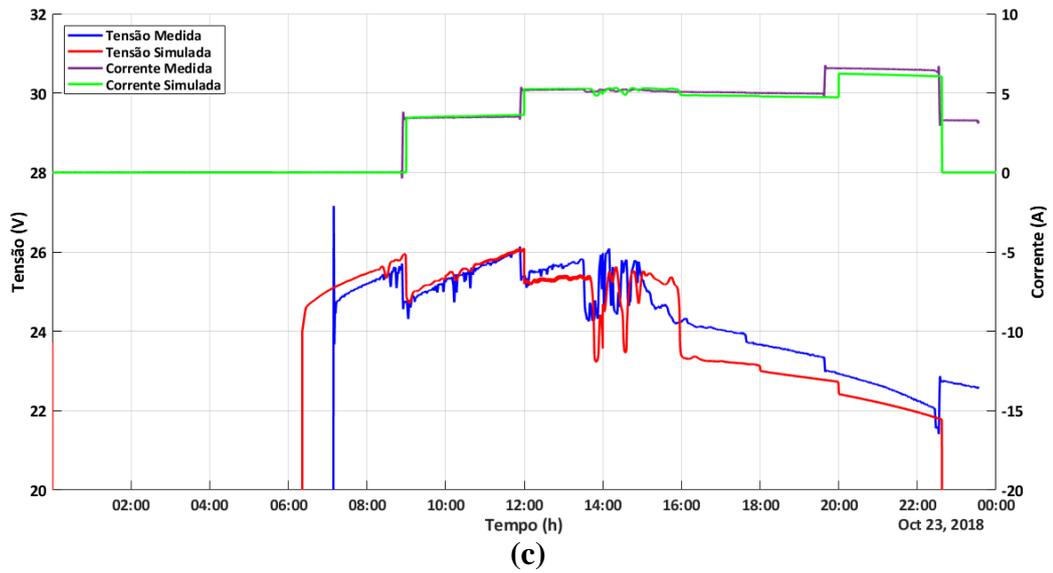

(c)

**Figure 4.12 - Comparison of simulated and measured power values in (a) PVG1, (b) PVG2 and (c) PVG3.**

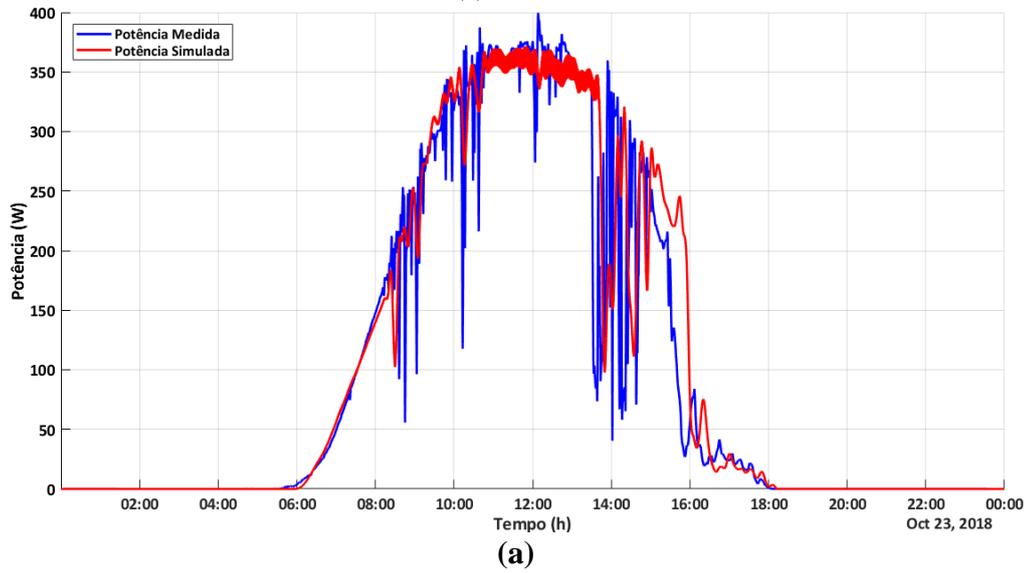

(a)

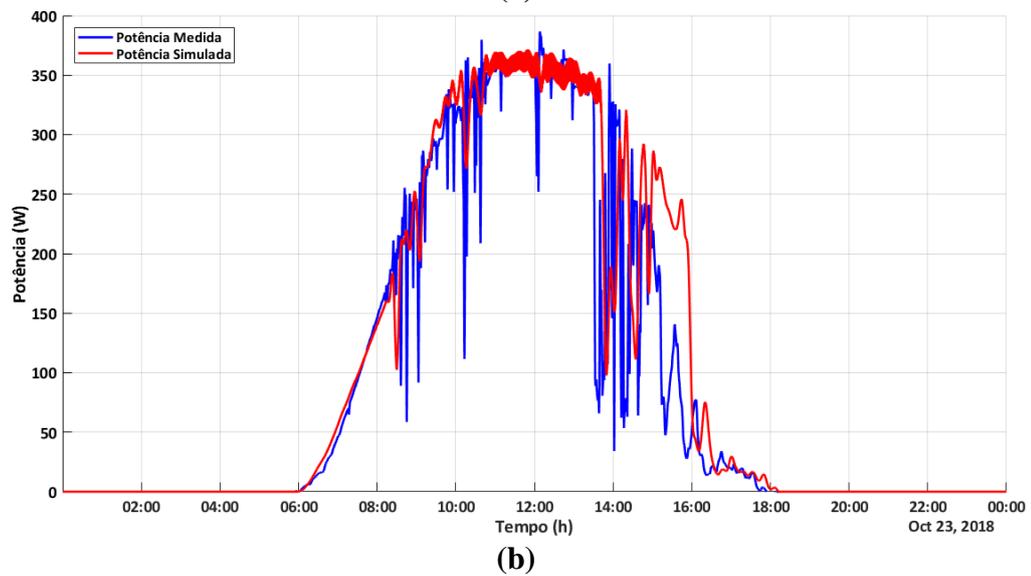

(b)

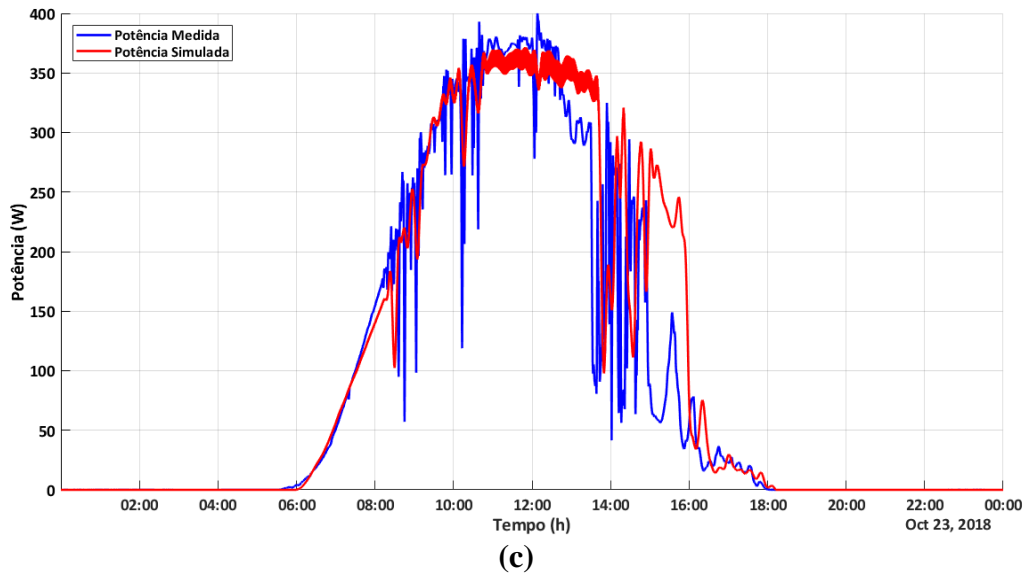
**(c)**

Regarding the simulated voltage values in the battery bank, it is noticed that although the behavior profile throughout the day is similar, there is a considerable difference compared to the measured values, especially in the region of higher battery charging. Another behavior to note is that in the simulation for BB1 and BB2 there is a load cut-out by the charge controller at the end of the day, bringing the current of the BB to zero. Regarding the monitored data, this load cut-out condition occurred only for BB3. Since in the simulation all BBs are removed from the grid by the charge controller action, the grid became completely de-energized, turning off all BCs, which did not actually occur according to measurement data.

Regarding the generation profile of the PVGs, there is a good reproduction of the power produced by the PV generation, with the largest differences observed between 3 pm and 4 pm, in which the results and simulation for PVG2 and PVG3 presented values much higher than those obtained from measurement data. This is due to the shading that occurs in the afternoon on the generators, as can be seen in Figure 4.13. Shading occurs from right to left, with PVG3 being the most affected, followed by PVG2. As shading effects were not considered in the model used for PVGs, power reduction is not observed.

**Figure 4.13- Occurrence of shading in the PVGs.**

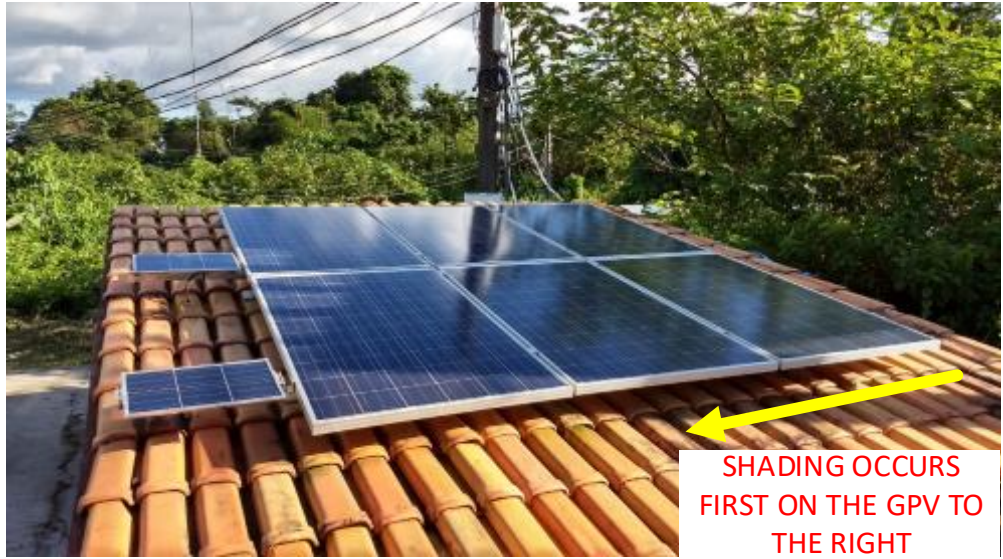

SHADING OCCURS FIRST ON THE GPV TO THE RIGHT

Figure 4.14 shows the accumulated energy values throughout the day for simulation, corresponding to the energies produced by the PVGs, consumed by the BCs and stored in the BBs.

**Figure 4.14- Simulated values of energy (a) generated by the PVGs, (b) consumed by the BCs and (c) stored / extracted from the BBs**

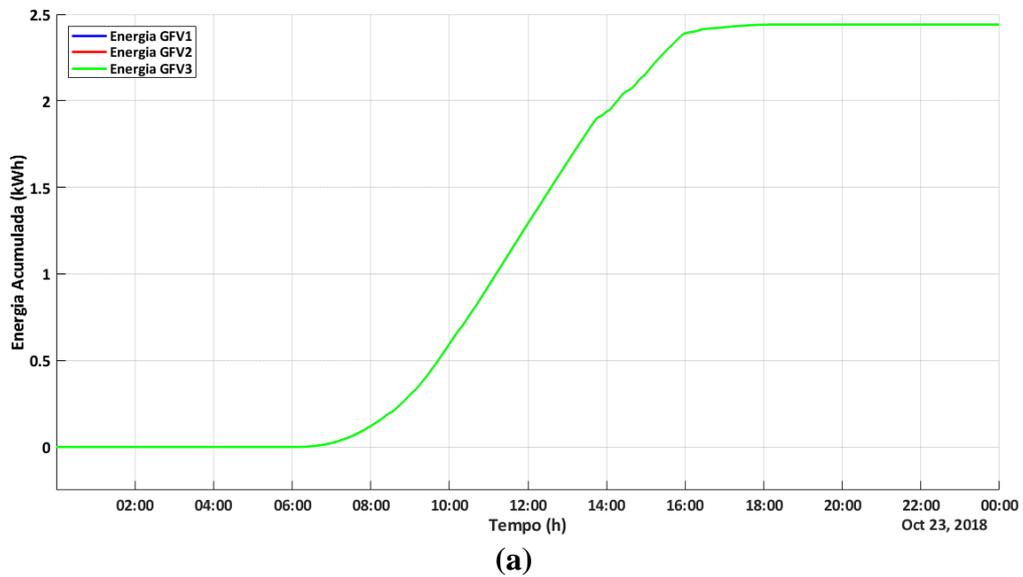

(a)

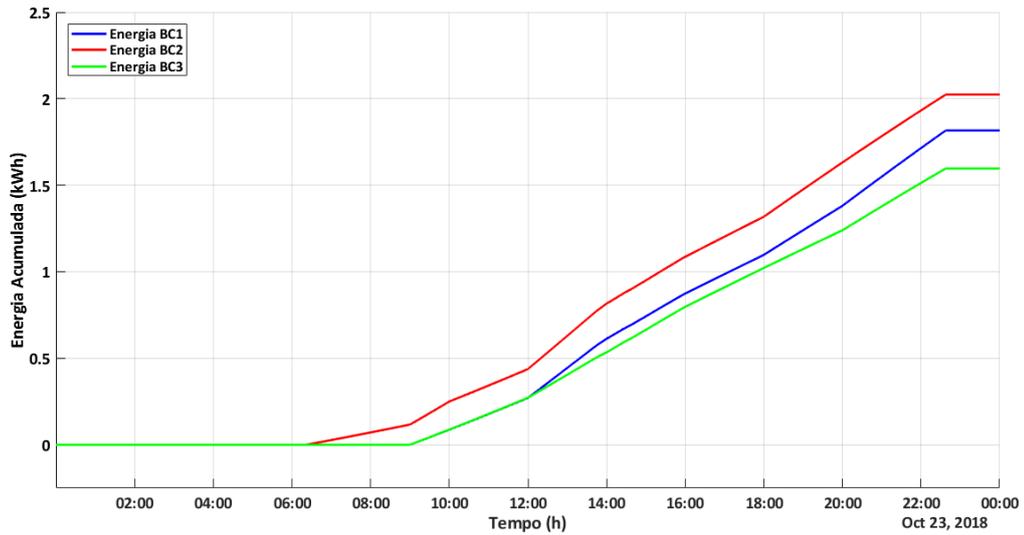

**(b)**

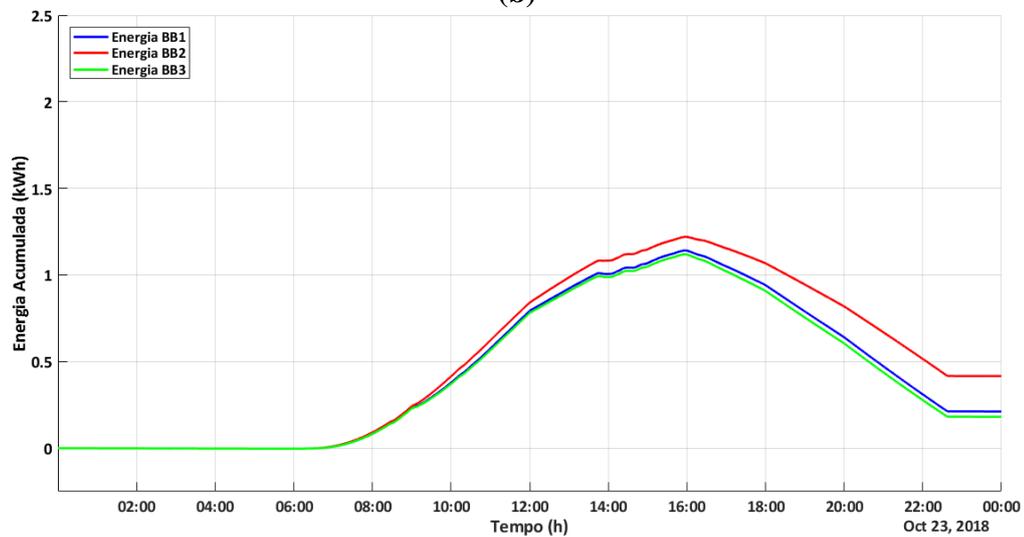

**(c)**

In terms of energy, the values obtained in simulation were penalized by two effects previously observed: the non-consideration of shading in PVGs and the load cut-out by all GSSs. Thus, in simulation, the total energy produced by the PVGs, $E_{PVG,SIM} = 7.32$ kWh is higher than the measured value (6.98 kWh). The simulated total energy consumed by the loads is lower in the simulation: $E_{BC,SIM} = 5.44$ kWh ($E_{BC} = 6.08$ kWh). In simulation, the total energy injected into the batteries was: $E_{BB,SIM} = 0.81$ kWh ($E_{BB} = 0.74$ kWh). Regarding losses, a total value of 1.07 kWh was observed in simulation, much higher than the measured value of 0.16 kWh.

As no generation cutoff occurred by any GSS, the simulation resulted in the same total energy produced by the three PVGs, since the simulation considered all operating under the same irradiance and temperature, so the three energy curves of the PVGs are superimposed in Figure 4.14a. In the event of generation limitation by the charge controller, due to the charging of the battery bank, it would be possible to observe differences between the energy produced by the PVGs.

## 4.2 Charge controller's operation without battery

As indicated in the charge controller characterization session, a contingency situation may occur where a battery bank is lost. An DCDN test was performed in an operation considering the loss of BB2. In this test, the battery bank of the GSS2 was disconnected following the procedure shown in Figure 4.15. At the end of the procedure, the charge controller is connected only to the PVG and to the DCDN main bus.

To monitor the test, a portable 4-channel oscilloscope model 190-204, manufactured by Fluke, was used. The connection diagram of the measuring channels is shown in Figure 4.16, monitoring the voltage and current of the PVG and the BB at the points of connection to the charge controller.

**Figure 4.15- Procedure for charge controller operation test without the BB.**

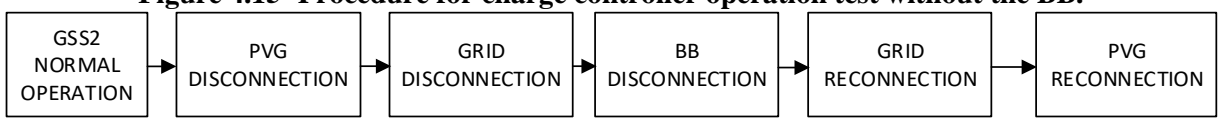

**Figure 4.16 - Instrumentation connection diagram for testing the charge controller without the BB.**

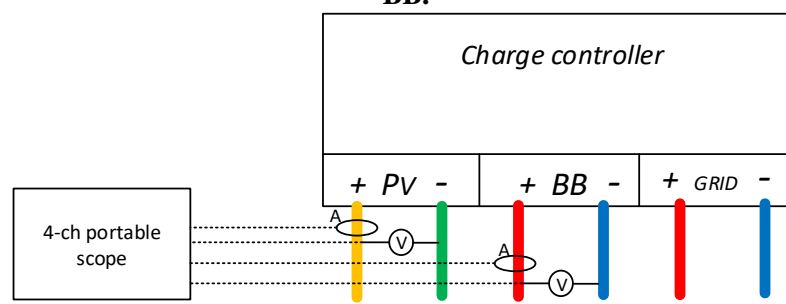

Figure 4.17 shows the measurements obtained during the test. At 11:03 am the PVG2 was disconnected, and then at 11:05 am the charge controller was disconnected from the grid and the BB circuit breaker was turned off, completely de-energizing the charge controller. At about 11:07 am the grid was reconnected, recovering the voltage at the BB connection point - even though the circuit breaker for battery connection remains off. Then, PVG2 was reconnected, which began to inject current, raising the voltage of the DCDN. Thus, it can be concluded that even after disconnecting the battery bank there is still power from the PV generation, which is being injected directly into the DCDN.

**Figure 4.17 - Operation test of the charge controller without the BB.**

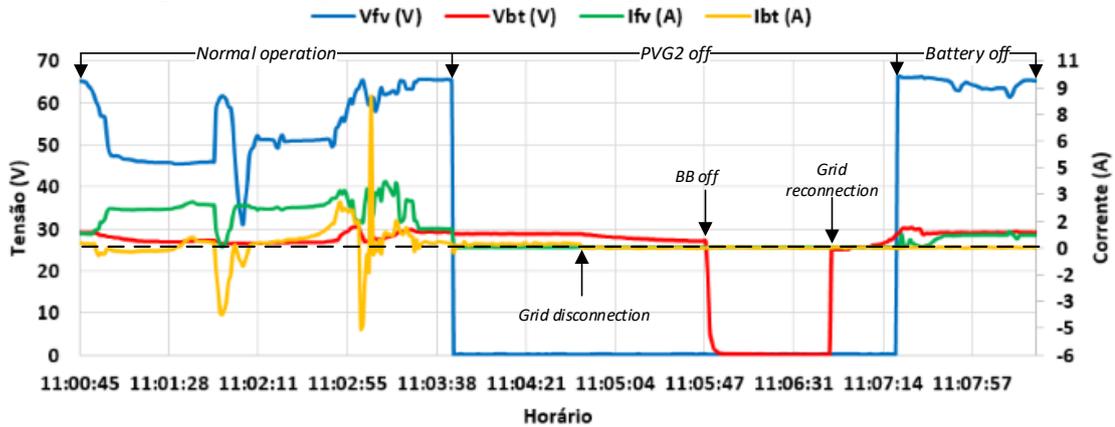

This ability of the charge controller to operate without a BB, provided a voltage reference from the DCDN, goes beyond the possibility of continuity of supply as it allows for greater modularity and ease in expanding grid's generation capacity. For example, a non-storage generating system can be installed so that the power of the PVG is injected directly into the grid. This setting may be important if the system load curve is in high demand at times of higher PV generation. This way one can expand the generation capacity without the need to expand the storage capacity.

## 4.3 Reverse power flow on the charge controller

Another possibility raised during the charge controller characterization process regards the charging of the battery bank through the DCDN. This can happen in two situations: if there is a PV generation surplus in a particular GSS, the spare power can be used to charge the battery bank of another GSS by the NDCC; also, in the occurrence of two BBs with different charging states, one BB can charge the other until it reaches the equilibrium between the voltages.

To verify the occurrence of these situations, a test was performed in DCDN operation considering the loss of PVG2. The procedure adopted in this test is shown in Figure 4.18. As pointed out in the procedure, BB2 is first discharged, while BB3 is kept under fully charged. After BB2 partial discharge, GSS3 is reconnected to the grid without its PVG, and then PVG is connected. The instrumentation used in this test was the same as presented in the previous item, keeping the same monitored variables.

**Figure 4.18- Procedure for testing reverse power flow operation on the charge controller.**

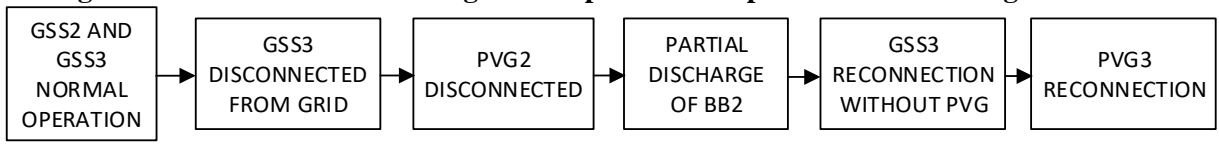

Figure 4.19 shows the measurements obtained during the test. At around 11:29 am PVG2 and GSS3 were disconnected from the grid, GSS1 remained disconnected throughout the test. Thus, only BB2 was responsible for servicing the loads, discharging with an approximate current of 7.5 A. Shortly after 11:36 am, GSS3 reconnected to the grid - keeping PVG2 and PVG3 disconnected; it is noticed that there is a reduction in the discharge current of BB2, due to the contribution of GSS3. At 11:38 am PVG3 reconnected, which started to inject power into the grid, generating a power surplus that was used to charge the BB associated with GSS2. Over the monitored period, it was reached a maximum BB2 charging current via NDCC of around 12.5 A.

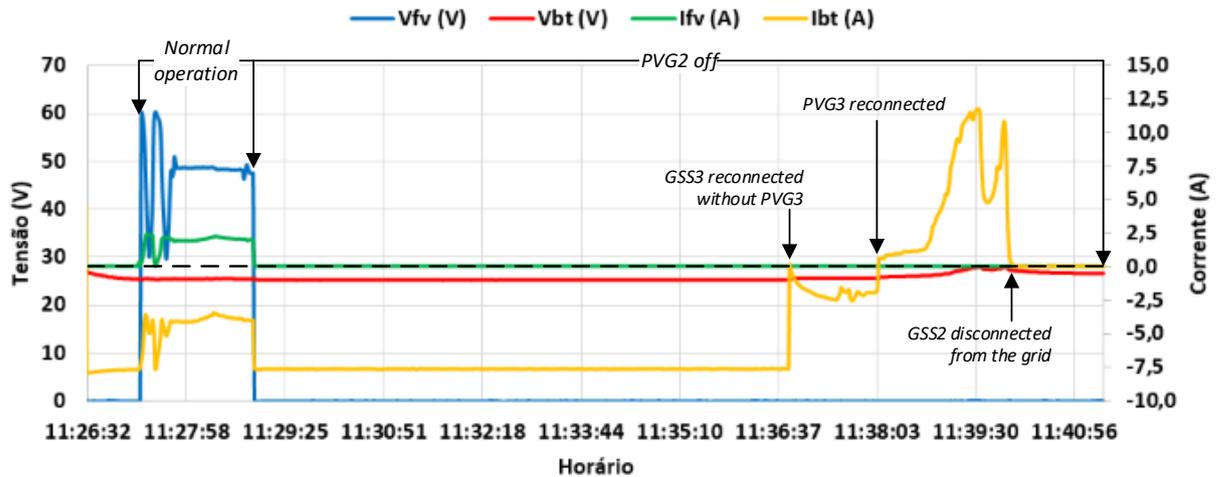

Figure 4.19- Charge controller reverse power flow operation test.

It is noted that there are benefits from the standpoint of power quality in the grid associated with this behavior of charging the BB via the DCDN. As one BB can charge another BB, voltages at the BBs tend to be equal, especially under light load conditions. In addition, with the possibility of using a PVG elsewhere on the grid to charge the BB, there is a reduction in the PVG limitation that can occur if the BB of an GSS is fully charged. Thus, with this configuration, a discharged BB behaves like a load on the grid.

On the other hand, in terms of energy efficiency, this behavior may not be the most recommended. Since the charge and discharge efficiency of batteries is not unitary, using the energy available in one BB to charge another BB is a waste of energy. In addition, losses in the distribution grid can be considerable depending on the distances between the BBs and can be avoided if the BB is charged only by the PVG associated with the respective GSS.

Given the short distances of the DCDN implemented in the laboratory, the losses associated with this behavior are insignificant, even more in normal operation, so that the difference between the voltages of the BBs is not significant. Thus, the charging current between the BBs is reduced and consequently the joule losses in the distribution grid are reduced. Therefore, it is concluded that this behavior of the charge controller is appropriate and beneficial to system operation.

# CONCLUSION

The DC nanogrid developed and installed in the laboratory of the Group of Studies and Development of Energy Alternatives - GEDAE/UFPA was adequate to meet the proposed loads, provided that the limitations of the storage system and photovoltaic generation are respected.

From the energy point of view, for a typical day of operation, a load servicing efficiency of over 97% was achieved. That is, less than 3% of the energy produced on this day was dissipated in the distribution grid and charge controllers (in the DC/DC conversion).

Regarding the power quality in the load attendance, it has been primarily verified the quality of voltage. During a typical day of operation, the voltage ranged from 26.45 V to 22 V (1.10 p.u. to 0.92 p.u., given the nominal 24 V operating voltage). This variation is characteristic of the configuration used to form the DCDN: the connection of the battery banks in parallel at different points of the grid, without any active bus voltage regulation. Thus, the voltage on the DC bus is a function of the voltage in the battery banks, which varies throughout the day with the state of charge, temperature, grid loading conditions and PV generation.

The variation in voltage observed in the load bars directly influences the operation of the loads used: a noticeable change in the lighting capacity of incandescent lamps and the ventilation capacity of the fans. However, during the entire operation period of the DCDN, there were no problems of burning of this equipment due to this variation.

It has been experimentally verified that the topology of the implemented charge controllers allows the occurrence of two unusual situations that can have a positive impact on the operation of the DCDN: the operation of a PV generator without necessarily being associated with a battery bank (operation without system storage) and the charging of a battery bank through the DCDN, called reverse flow in the charge controller (operation without generation system).

Regarding the protection devices, it was observed the tripping of DC circuit breakers in short circuits in the load banks and at the charge controller. However, during the entire DCDN operation period, two circuit breakers had to be replaced after their operation, as they did not reset after the fault clearance. It is important to highlight, however, that the implemented circuit breakers are of low cost compared to others available in the domestic market of traditional manufacturers. Also, they are not certified by the responsible agencies in the country.

From the static modeling developed, it was possible to attest the grid capacity to meet the loads in various charging conditions and voltage in the battery banks, so that in all tests performed the voltages in the three load banks obtained through simulation were close to the experimental values (difference less than 5%).

With the developed dynamic modeling, it is possible to predict the behavior of the DCDN throughout the day, considering the variations in solar resource availability and load bank demand. The accuracy of the simulation, however, has shortcomings associated mainly with the modeling of battery banks. With refinement in obtaining the specific parameters for each battery bank, better results should be obtained in the simulation.

One of the major difficulties in the development of this work was associated with the implementation of the measurement system. The large number of voltage and current measuring points that must be monitored simultaneously throughout the day has made the development of the measuring system quite complex. In addition, as the monitored voltage and current values are relatively low in magnitude, small reading errors in the sampling of electrical quantities may represent significant differences in the evaluation of system operation. For example, a measurement error in the order of 100 mV (less than 0.5%, considering the nominal voltage of 24 V), may represent a difference in the actuation or not of a charge controller. In this sense, it is considered that the improvement of the instrumentation used in the acquisition of DCDN operation data is essential for a better monitoring of the operation and to continue the work that has been developed.

It is important to highlight the difficulties associated with the use of battery banks with compromised service life, which penalized the use of load banks with higher consumption profiles. In addition, differences in discharge curves between batteries in the same bank made storage system modeling difficult. The bench characterization of the charge and discharge curves of the battery banks was important to attest the high state of degradation of the batteries used.

The experimental structure developed in this work is an important tool for studies of DC nanogrids, allowing the use of commercial equipment for grid development. Several more in-depth studies can be carried out, such as evaluating power quality in meeting different loads, evaluating protective devices and evaluating converter topologies.

Based on what was developed in this work, the following themes are proposed for continuity and further research in the area:
- Improve the load flow methodology to consider PV generation;
- Improve the measurement quality and reliability;
- Evaluate other types and technologies of batteries;
- Evaluate other voltage levels on the main DC bus;

- Use other common loads such as freezer, TV, acai machine;
- Investigate the effect of circulating currents on the charge controller used, as well as other effects associated with the parallelism of the grid converters.

## PUBLICATIONS RELATED TO THIS DISSERTATION

**Publication in national magazine**

**TORRES, P. F.**; VIEIRA FILHO, J. A. A. ; CHAAR JUNIOR, V. L. ; ARAUJO, L. F. ; WILLIAMSON, S. J. ; GALHARDO, M. A. B. ; PINHO, J. T. ; MACÊDO, W. N.; "Simulação de fluxo de carga em uma microrrede em corrente contínua alimentada por sistemas fotovoltaicos e baterias para atendimento de pequenas comunidades isoladas". In. Revista Fotovolt. Mar./2019.

**Publication in conference proceedings**

**TORRES, P. F.**; VIEIRA FILHO, J. A. A. ; CHAAR JUNIOR, V. L. ; ARAUJO, L. F. ; WILLIAMSON, S. J. ; GALHARDO, M. A. B. ; PINHO, J. T. ; MACÊDO, Wilson N. . LOAD FLOW SIMULATION OF A LOW-VOLTAGE PV-BATTERY BASED DC MICROGRID TO SUPPLY SMALL ISOLATED COMMUNITIES. In: **European PV Solar Energy Conference and Exhibition**, **2018**, Brussels. EU PVSEC Proceedings, 2018. v. 35. p. 1636-1640.

**TORRES, P. F.**; VIEIRA FILHO, J. A. A. ; WILLIAMSON, S. J. ; PINHO, J. T. ; GALHARDO, M. A. B. ; MACÊDO, Wilson N. . Concepção de estrutura laboratorial para realização de estudos em microrrede em corrente contínua de baixa tensão. In: Congresso Brasileiro de Energia Solar, 2018, Gramado. VII Congresso Brasileiro de Energia Solar - CBENS 2018, 2018. **Proceedings of the VII Brazilian conference of Solar Energy**. 2018.

**Summary paper accepted for presentation in conference**

**TORRES, P. F.**; COSTA, T. O.; ARAÚJO, L. F.; VIEIRA FILHO, J. A. A. ; WILLIAMSON, S. J.; MACÊDO, W. N.**.** Solar Photovoltaic-based DC Nanogrid Testing under Real-World Operating Conditions. **10th International Conference on Power Electronics, Machines and Drives.** 2020.

# ANNEX I

Solar Simulator test report of a PV module used in the DCDN.

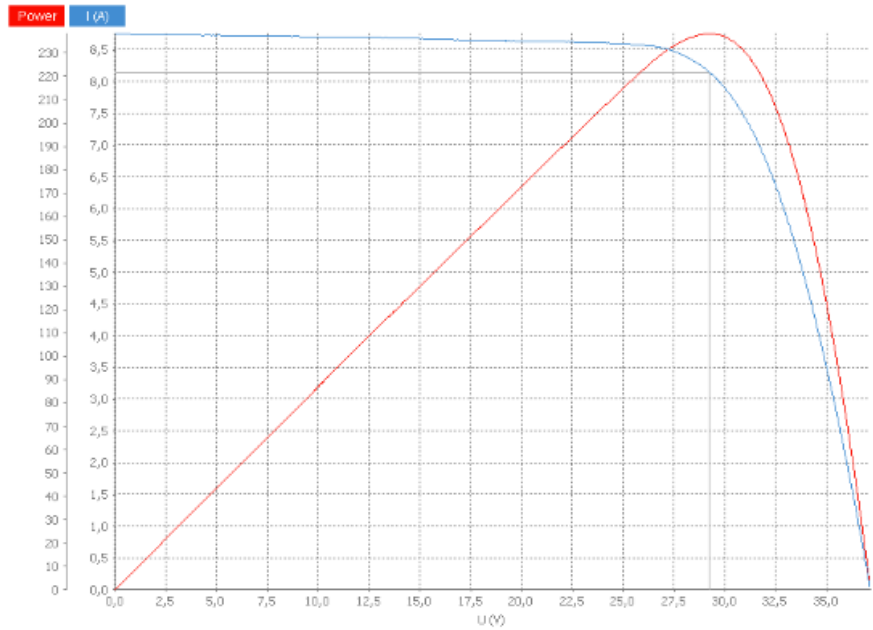

| Operator | SOLAR TESTE | Measurement | 2019/06/24 17.11.54 |
| PASAN Tester version | R2.4.0 / 2014/05/08 16:09:55 2.4.0 (9695) | | |

**YL245p-29b**

| Manufacturer | Yingli Solar | Type | policistralin |
| Serial number | 124305021064 | Configuration DUT | Module |
| Single cell area | 240.25 cm² | DUT area | 16335.00 cm² |
| Cells in series | 60 | Cells in parallel | 1 |

**Mono + th** — Irradiance Channel 1

| Serial number | 370 | | |
| Sensitivity | 128.700 mV/(kW/m²) | Temperature coefficient | 0.00 %/°C |

**direto-YL245p-29b** — Irradiance Channel 1

| Monitor cell temperature | 25.07 °C | Fill factor | 73.10 % |
| DUT temperature | 25.78 °C | Cell efficiency | 16.53 % |
| Compensated | 25.00 °C | DUT efficiency | 14.59 % |
| Gavg | 999.90 W/m² | | |
| GstdDev | 0.25 W/m² | | |
| Compensated Irradiance | 1000.00 W/m² | | |
| Regression linear for Voc | 37.208 V | | |
| Linear regression Isc | 8.760 A | | |
| Regression linear for | 0.625 Ω | | |
| Regression linear for | 176.423 Ω | | |
| Maximum power | 238.249 W | | |
| Voltage at Maximum | 29.221 V | | |
| Current at Maximum | 8.153 A | | |

# ANNEX II

Solar Simulator test report of the reference PV module used in the DCDN for irradiance sensing.

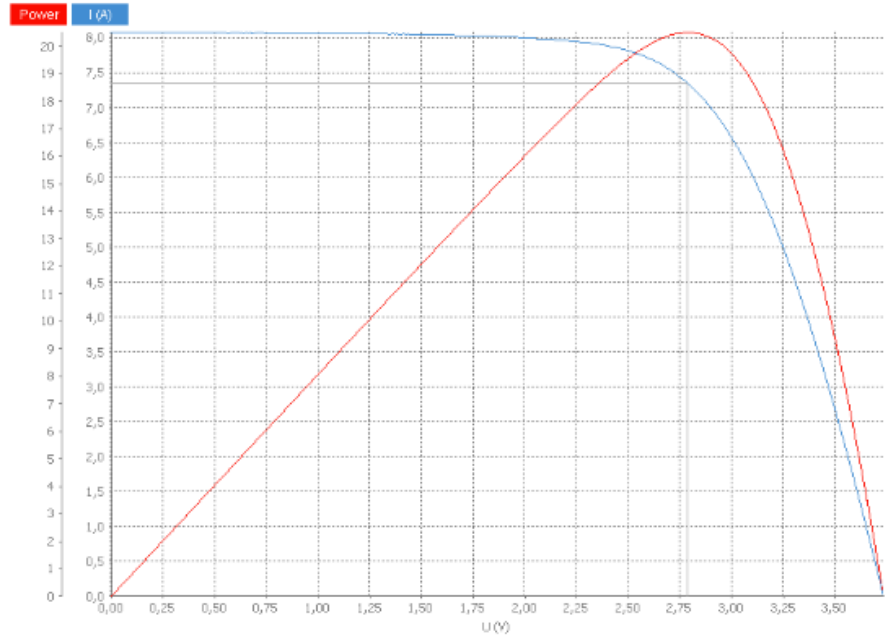